\documentclass[%
 reprint,
nofootinbib,
 amsmath,amssymb,
 aps,
]{revtex4-2}

\usepackage[T1]{fontenc}

\usepackage{graphicx}
\usepackage{dcolumn}
\usepackage{bm}
\usepackage{hyperref}

\usepackage{slashed}
\usepackage{tikz-feynman}
\usepackage{feynmp}
\tikzset{square/.style={draw, fill=black, shape=rectangle, minimum size=3pt, inner sep=0pt}}
\usepackage{subcaption}
\usepackage{caption}
\usepackage{multirow}
\usepackage{tabularx} 
\usepackage{breqn}
\usepackage{blkarray}
\usepackage{amsfonts} 
\usepackage{mathtools}
\usepackage{csquotes}
\usepackage{float}
\usepackage{color} 
\usepackage[english]{babel}
\usepackage{rotating}
\newcommand{\ket}[1]{\left| #1 \right\rangle}

\begin{document}


\title{Comprehensive study of hidden charm pentaquarks \\ with an improved unitarization method}

\author{E.E. Garcia-Gonzales$^1$}
\author{V.K. Magas$^{1,2}$}
\author{A. Ramos$^{1,2}$}
\affiliation{
	$^1$ Departament de Física Quàntica i Astrofísica (FQA)$\,,$ \\
	Universitat de Barcelona (UB)$\,,$ c. Martí i Franqués$\,,$ 1$\,,$ 08028 Barcelona$\,,$ Spain\\
    $^2$ Institut de Ciències del Cosmos (ICCUB)$\,,$\\  
    Universitat de Barcelona (UB)$\,,$ c. Martí i Franqués$\,,$ 1$\,,$ 08028 Barcelona$\,,$ Spain
}
\date{\today}

\begin{abstract}
    This work investigates dynamically generated hidden-charm baryon resonances arising from meson-baryon interactions. Using the local hidden gauge formalism, we model the interaction via t-channel vector meson exchange and unitarize the amplitude using the Bethe-Salpeter equation. To address regularization issues, we propose a novel ``hybrid loop function'' scheme that eliminates the unphysical poles ---common artifacts in cutoff or dimensional regularization--- while keeping
    the predictions of physical states. Consequently, the model successfully reproduces six experimentally  known hidden-charm pentaquarks as well as earlier theoretical results, and predicts new states in the $S=-1, I=1$ sector.

\end{abstract}


\maketitle


\section{Introduction}

The history of two-body molecular hadron states began with the discovery of the deuteron in 1931, the first known bound state composed of two hadrons – specifically one proton and one neutron. Years later, with the discovery of new hadrons, such as the pions and kaons in 1947, other bound systems besides the deuteron were expected. In 1959, the existence of a meson baryon molecule, the $\Lambda(1405)$, was predicted \cite{Dalitz:1959dn} and not too much later it was observed experimentally in 1961 \cite{Alston:1961zzd}.

With the development of the quark model in the early 1960s, it became clear that hadrons are not elementary particles, but complex structures composed of quarks. The quark models provided a simple way to classify hadrons according to their quark content and successfully described the properties of the hadrons in the SU(3) flavor multiplets.
Within this framework, the $\Lambda(1405)$ was interpreted as an excited 3-quark ($uds$) state, however its mass was systematically predicted to be too large \cite{Isgur:1978xj, Capstick:1986ter}. This initiated a long discussion that lasted for decades about the 3-quark or molecular nature of the $\Lambda(1405)$ \cite{Dobson:1972jib, Siegel:1988rq, Mueller-Groeling:1990uxr, Glozman:1995fu, PDG2010}. 

With the advent of effective field theories \cite{Gasser:1983yg,Weinberg:1991um}, which build the interactions between hadrons in the low energy regime respecting the symmetries of QCD, the use of Unitarized Chiral Perturbation Theory \cite{Kaiser:1995eg,Oset:1997it,Oller:2000ma,Hyodo:2011ur,Feijoo19} emerged as a powerful approach to generate molecules. It is based on the solution of the Bethe-Salpeter equation non-perturbatively using a particular regularization scheme. Within this molecular picture, the $\Lambda(1405)$ was found to have a two-pole structure \cite{Oller:2000fj,Jido:2003cb,Magas:2005vu}, opening the door to disentangle its nature and potentially resolving the decades-long debate. It was not until 2016 that the Particle Data Group acknowledged that the $\Lambda(1405)$ is most likely a hadronic molecule rather than a three-quark system \cite{PDG2016}, although further data is needed to precisely determine its two-pole structure (for a historical, experimental and theoretical review see \cite{Mai:2020ltx}).

It is clear now that more complex structures beyond the predictions of the classical quark model exist. A new charmonium state was discovered by Belle in 2003~\cite{Belle:2003} and named $\chi_{c_1}(3872)$ (also known as $X(3872)$).  It was found to be narrow, with a mass right below the $D^0 \bar{D}^{*0}$ threshold, suggesting the relevance of this meson pair in the $X(3872)$ dynamics.  Although the $X(3872)$ is proposed to be a dynamically generated state from hadron interactions, its structure as a compact state (i.e. a tetraquark) or a molecular state remains unknown.
More recently, the LHCb Collaboration observed a new candidate for a tetraquark state, named $T_{cc}(3875)^+$, in the $D^0 D^0 \pi^+ $ mass spectrum~\cite{LHCb:Tcc,Aijj2022_Tcc}. The novelty of this state lies in its open-charm truly exotic structure, as it contains two charm quarks, in contrast to the $X(3872)$, which contains a $c\bar{c}$ pair. The mass of the $T_{cc}(3875)^+$ was found to be near the $D^{*+}D^0$ and $D^{*0}D^+$ thresholds, around $3875 \, \text{MeV}$, and its width was measured to be small, $\Gamma = 410 \, \text{keV}$. The closeness to the $D^*D$ channel threshold enforces the possible molecular interpretation. Several theoretical studies have explored the nature of this structure within a coupled-channel framework~\cite{Du:2021zzh,Feijoo:2021ppq,Meng:2021jnw,Albaladejo:2021vln}.

As an example of exotic baryons, a few pentaquark states have been discovered in the hidden charm sector over the past decade~\cite{Aaij2015, Aaij2019, Aaij2020, Aaij2022, Belle:2025pey}: $P_{c\bar{c}}(4312)^+$, $P_{c\bar{c}}(4380)^+$, $P_{c\bar{c}}(4440)^+$, $P_{c\bar{c}}(4457)^+$, $P_{c\bar{c}s}(4338)^0$ and $P_{c\bar{c}s}(4459)^0$. 
Evidence for a structure in $J/\psi\, p$ and $J/\psi \,\bar{p}$ invariant mass distributions from the decay $B^0_s \to J/\psi \,p\,\bar{p}$, having a mass of $4337^{+7}_{-4}{}^{+2}_{-2}$~MeV and a width of $29^{+26}_{-12}{}^{+14}_{-14}$~MeV, has also been seen \cite{Aaij2021}.
To describe these exotic baryons ---and also 
the exotic mesons like $X(3872)$ and $T_{cc}(3875)^+$--- as hadron molecules, theoretical frameworks expanding the SU(3) flavor symmetry to include heavy quarks in meson-baryon interactions have been developed. In this work, we focus on two such models.  The model derived in ~\cite{Lutz,Molina:2010,Gloria} employs the local hidden gauge approach (LHGA) assuming SU(4) flavor symmetry (hereafter referred to as LHGA-SU(4)), while the model derived in~\cite{Debastiani} is also based on the LHGA, but explicitly uses baryon flavor wave functions without assuming SU(4) symmetry (hereafter referred to LHGA-WF). An important advantage of the LHGA-WF model is the possibility of taking baryons with $J^P = \frac{3}{2}^+$ into consideration, whereas only  $J^P = 1/2^+$ baryons are considered in the LHGA-SU(4) framework.
More recently, further studies have predicted hidden charm pentaquarks with strangeness $S=-2$ within the LHGA-SU(4) model~\cite{Valera23}, a sector later reanalyzed within the LHGA-WF method \cite{Roca}. Although both works predicted similar masses of the discovered resonances, the widths obtained in~\cite{Valera23} were substantially larger than those found in~\cite{Roca}.

In this work we present an extensive study of hidden charm pentaquarks dynamically generated from meson-baryon interactions derived from t-channel vector meson exchange, covering all strangeness and isospin sectors. To this end, we employ both the LHGA-SU(4) and LHGA-WF interaction models and compare the obtained results. The scattering amplitude is obtained by solving the Bethe-Salpeter coupled-channel equations within the framework of Unitarized Chiral Perturbation Theory. The unitarization procedure requires regularization of the meson-baryon loop function, which is typically done using either dimensional regularization or cut-off regularization. However, both regularization schemes introduce unphysical structures in the scattering amplitude, tied to the appearance of what we refer to as ``fake poles'' in the complex plane, as they are generated by repulsive interactions. We propose a new hybrid loop function that, by combining both dimensional regularization and cut-off regularization schemes, avoids the unphysical issues and permits disentangling the physical states naturally.

This work is organized as follows. In Section~\ref{formalism}, we describe the two models used to compute the meson-baryon interactions. We also introduce and describe the hybrid loop function. In Section~\ref{Gdr_co_hy_comp}, we study the dynamically generated states in all strangeness, $S$, and isospin, $I$, sectors using the LHGA-SU(4) model and explore the results obtained with the three different loop function regularization schemes: dimensional, cut-off and hybrid. We discuss the inconveniences of using the dimensional and cut-off schemes and show the advantages of the hybrid approach. In Section~\ref{LHGA-WF_results}, we employ the LHGA-WF interaction and the hybrid loop function to study the same $(S,I)$ sectors and, when applicable, we discuss the differences with respect to the LHGA-SU(4) results.  Finally, to support the reliability of the hybrid regularization method, we compare our findings with those predicted by other theoretical studies in Section~\ref{model_comp} and with the experimentally observed pentaquark states in Section~\ref{data_comp}. The conclusions of our work are summarized in Section~\ref{conclusions}.

\section{Formalism}
\label{formalism}
\subsection{LHGA-SU(4) and LHGA-WF models}
\label{models}

\begin{figure}[h!]
    \includegraphics[width=0.5\linewidth]{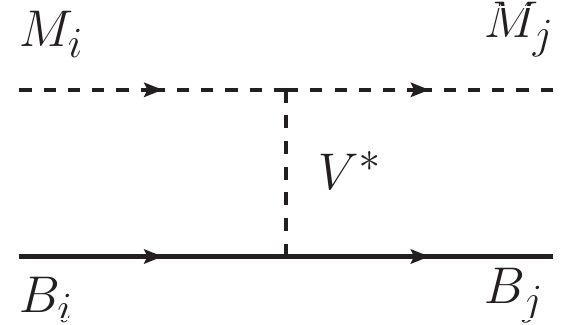}
       \caption{The t-channel of the meson-baryon interaction in the LHGA.}
    \label{t-channel}
\end{figure}

The LHGA-SU(4) formalism is developed in Ref.~\cite{Lutz}, where the interaction proceeds via the exchange of vector mesons, as illustrated in Fig.~\ref{t-channel}. 
The vector-pseudoscalar-pseudoscalar (${VPP}$) and vector-baryon-baryon (${VBB}$) vertices are given by
\begin{eqnarray}
        &&\!\!\!\!\!\! \!\!\! \!\! \!\!\mathcal{L}_{VPP} = i g\, Tr\left(\left[\partial_\mu \Phi_{[16]}, \Phi_{[16]} \right]V^\mu_{[16]} \right), \label{L_VPP}\\
        &&\!\!\!\!\!\!\!\!\! \!\!\!\! \mathcal{L}_{VBB} = \frac{g}{2} \!\!\! \!\!\sum_{i,j,k,l = 1}^4 \!\!\! \overline{B}_{ijk}^{[20]} \gamma^{\mu} \left( V_{\mu,l}^{[16],k} B^{ijl}_{[20]} + 2V_{\mu,l}^{[16],j} B^{ilk}_{[20]}\! \right).
\end{eqnarray}
The constant $g$ is the universal coupling, related to the pion decay constant $f_\pi$ and the representative mass of the light vector mesons $m_V$ from the nonet: 
\begin{equation}
g = \frac{m_V}{2f_{\pi}}.
\label{KSFR}
\end{equation}
The fields $\Phi_{[16]}$ and $V_{[16]}^{\mu}$ represent the pseudoscalar and vector meson SU(4) 16-plets, respectively, and the tensor $B^{ijk}$ corresponds to the 20-plet $\frac{1}{2}^+$- baryon field. Their explicit matrix elements are
\begin{widetext}
\begin{equation}
    \Phi_{[16]} = 
    \begin{pmatrix}
        \frac{1}{\sqrt{2}}\pi^0 + \frac{1}{\sqrt{3}}\eta + \frac{1}{\sqrt{6}}\eta' & \pi^+ & K^{+} & \bar{D}^{0} \\
        \pi^- & -\frac{1}{\sqrt{2}}\pi^0 + \frac{1}{\sqrt{3}}\eta + \frac{1}{\sqrt{6}}\eta' & K^{0} & D^{-} \\
        K^{-} & \bar{K}^{0} & \frac{1}{\sqrt{3}}\eta + \sqrt{\frac{2}{3}}\eta' & D_s^{-} \\
        D^{0} & D^{+} & D_s^{+} & \eta_c
    \end{pmatrix},
    \label{P_mat}
\end{equation}
\begin{equation}
    V^{\mu}_{[16]} = 
    \begin{pmatrix}
        \frac{1}{\sqrt{2}}\rho^0 + \frac{1}{\sqrt{2}}\omega & \rho^+ & K^{*+} & \bar{D}^{*0} \\
        \rho^- & -\frac{1}{\sqrt{2}}\rho^0 + \frac{1}{\sqrt{2}}\omega & K^{*0} & \bar{D}^{*-} \\
        K^{*-} & \bar{K}^{*0} & \phi & D_s^{*-} \\
        D^{*0} & D^{*+} & D_s^{*+} & J/\psi
    \end{pmatrix}
    ^\mu.
    \label{V_mat}
\end{equation}
\begin{equation}
    \begin{matrix*}[l]
        B_{[20]}^{121} = p, &  B_{[20]}^{122} = n, & B_{[20]}^{132} = \frac{1}{\sqrt{2}} \Sigma^0 - \frac{1}{\sqrt{6}} \Lambda, \\ \\
        B_{[20]}^{213} = \frac{2}{\sqrt{6}}\Lambda, & B_{[20]}^{231} = \frac{1}{\sqrt{2}} \Sigma^0 + \frac{1}{\sqrt{6}} \Lambda, &  B_{[20]}^{232} = \Sigma^-, \\ \\
        B_{[20]}^{233} = \Xi^-, & B_{[20]}^{311} = \Sigma^+, & B_{[20]}^{313} = \Xi^0, \\ \\
        B_{[20]}^{141} = -\Sigma_c^{++} , & B_{[20]}^{142} = \frac{1}{\sqrt{2}}\Sigma_c^+ + \frac{1}{\sqrt{6}}\Lambda_c, & B_{[20]}^{143} = \frac{1}{\sqrt{2}} {\Xi'}_c^+ - \frac{1}{\sqrt{6}} \Xi_c^+, \\ \\
        B_{[20]}^{241} = \frac{1}{\sqrt{2}}\Sigma_c^+ - \frac{1}{\sqrt{6}}\Lambda_c, & B_{[20]}^{242} = \Sigma_c^0, & B_{[20]}^{243} = \frac{1}{\sqrt{2}} {\Xi'}_c^0 + \frac{1}{\sqrt{6}} \Xi_c^0, \\ \\
        B_{[20]}^{341} = \frac{1}{\sqrt{2}} {\Xi'}_c^+ + \frac{1}{\sqrt{6}} \Xi_c^+, & B_{[20]}^{342} = \frac{1}{\sqrt{2}} {\Xi'}_c^0 - \frac{1}{\sqrt{6}} \Xi_c^0, & B_{[20]}^{343} = \Omega_c, \\ \\
        B_{[20]}^{124} = \frac{2}{\sqrt{6}} \Lambda_c^0, & B_{[20]}^{234} = \frac{2}{\sqrt{6}}\Xi_c^0, & B_{[20]}^{314} = \frac{2}{\sqrt{6}}\Xi_c^+, \\ \\
        B_{[20]}^{144} = \Xi_{cc}^{++}, & B_{[20]}^{244} = -\Xi_{cc}^+, & B_{[20]}^{344} = \Omega_{cc}.
    \end{matrix*}
\end{equation}
\end{widetext}
Following Refs.~\cite{Lutz, Gloria}, the {\it s}-wave scattering kernel in this model reads:
\begin{multline}
    V_{ij} = g^2 \sum_{v} C_{ij}^v\, \overline{u}(p_j) \gamma^\mu u(p_i) \frac{1}{t-m_v^2} \\
    \times \left[ (k_i + k_j)_\mu - \frac{k_i^2 - k_j^2}{m_v^2}(k_i - k_j)_\mu \right] \ ,
    \label{LHGA_kernel1}
\end{multline}
where the subindices $i$ and $j$ refer to the initial and final channels, $k_i$ ($k_j$) is the momentum of the initial (final) meson, $t=(k_i-k_j)^2= (p_j-p_i)^2$ is the four momentum transfer, with $p_i$ ($p_j$) being the momentum of the initial (final) baryon, $m_v$ is the mass of the exchanged vector meson, and $C_{ij}^v$ is the corresponding transition coefficient. For low momentum transfers ($t/m_v^2 \rightarrow 0$) the kernel simplifies to:
\begin{equation}
    V_{ij} = - g^2 \, \overline{u}(p_j) \gamma^\mu u(p_i) (k_{j,\mu} + {k}_{i,\mu}) \sum_v \frac{C_{ij}^v}{m_v^2} \ .
 \label{LHGA_kernel22}
\end{equation}
In the summation $\sum_v C_{ij}^v/m_v^2 $, one may assume that all the light vector mesons have the same mass, $m_V$ (related to the universal coupling $g$ in Eq.~\eqref{KSFR}). For transitions involving charmed vector mesons, we introduce correction factors, $\kappa_c$ and $\kappa_{cc}$, for $D^*$ and $J/\psi$, respectively, defined as~\cite{Gloria}:
\begin{equation}
    \begin{gathered}
    k_c = \left(\frac{m_V}{m_D} \right)^2 \approx \frac{1}{4}, \\
    k_{cc} = \left(\frac{m_V}{m_{J/\psi}} \right)^2 \approx \frac{1}{9} \ .
\end{gathered}
\label{kappas}
\end{equation}
In this way, one can factor $m_V^2$ out of the summation in Eq.~\eqref{LHGA_kernel22}, which then simplifies to 
\begin{equation}
    V_{ij} = -C_{ij} \frac{1}{4f_\pi^2}
     \, \overline{u}(p_j) \gamma^\mu u(p_i) (k_{j,\mu} + {k}_{i,\mu}) \ ,
    \label{LHGA_kernel2}
\end{equation}
an expression that reproduces the Weinberg–Tomozawa term of the lowest order chiral SU(3) meson-baryon Lagrangian, but now the kernel coefficients $C_{ij}$ encode the information of SU(4) flavor symmetry and its breaking introduced by the factors $\kappa_c$ and $\kappa_{cc}$.  Working out the Dirac algebra of \eqref{LHGA_kernel2} and retaining only the $s$-wave terms, one obtains
\begin{equation}
    V_{ij} = -C_{ij} \frac{1}{4f_\pi^2} \left(2\sqrt{s} - M_i - M_j \right) \sqrt{\frac{E_i + M_i}{2M_i}} \sqrt{\frac{E_j+M_j}{2M_j}} \ ,
    \label{V_SU4_swave}
\end{equation}
where $E_i (E_j)$ and $M_i (M_j)$ are the energy and mass, respectively, of the baryon in channel $i (j)$.

The LHGA-WF scheme developed in Ref.~\cite{Debastiani} also builds the meson-baryon interaction from a t-channel vector meson exchange, as shown in Fig.~\ref{t-channel}. The upper $VPP$ vertex is again given by $\mathcal{L}_{VPP}$ in Eq.~\eqref{L_VPP} but, in contrast to the LHGA-SU(4) model, the lower vertex is built from:
\begin{equation}
    \left< \phi_j\chi_j|g \ q\overline{q}(V) | \phi_i\chi_i \right> \ ,
    \label{VBB_debastiani}
\end{equation}
where $q \overline{q}(V)$ denotes the flavor wave function of the exchanged vector meson and $\ket{\phi \chi}$ stands for the baryon wave function with flavor and spin degrees of freedom. Taking into account that the diagonal transitions, which are the most important for the model, are dominated by the exchange of the light vector mesons, the LHGA-WF scheme assumes the $c$ quark to act merely as a spectator in the transition. Consequently, the baryon wave functions do not incorporate SU(4) symmetry: in constructing the flavor part of the baryon wave function the heavy quark is singled out and only the light quarks are symmetrized (or antisymmetrized) ensuring that the  spin-flavor wave function is symmetric. In this way one is effectively using SU(3) symmetry. A complete list of the wave functions needed in the present work can be found in Ref.~\cite{Yu:2018yxl}. 
It is important to emphasize that this method naturally allows for the inclusion of $\frac{3}{2}^+$ baryons, since their spin-flavor structure is explicitly incorporated in Eq.~\eqref{VBB_debastiani}.


After multiplication of the meson and baryon vertices, one obtains an expression analogous to Eq.~\eqref{V_SU4_swave}, but replacing the $C_{ij}$ coefficients by $D_{ij}$ ones.
Appendices \ref{app:B} and \ref{app:C} collect the $C_{ij}$ and $D_{ij}$ kernel coefficients for all the sectors studied in this work.
Note that, up to a possible sign tied to phase conventions of the states, the $C_{ij}$ and $D_{ij}$ coefficients are the same, except for transitions connecting channels with an $\eta_c$ meson with channels with a $\bar{D}$ or $\bar{D}_s$ meson.  These transitions involve the exchange of a charmed vector meson and their size is a factor $\sqrt{3}$ higher in the LHGA-SU(4) model. 
This difference is a direct consequence of heavy quarks being treated as spectators in the LHGA-WF model, so that only the light $u,d,s$ quarks of the baryon wave functions are symmetrized. Hence, transitions not involving the participation of the charm quark (i.e. those that are mediated by light vector mesons) give rise to coefficients having the same size in both models, as only the common subjacent SU(3) symmetry plays a role. However, when the charm quark of the initial $\eta_c$ meson ends up in the final charmed baryon it is no longer an spectator. The symmetrization of this charm quark with the other two in the final charmed baryon, an assumption formally embedded in the SU(4) symmetry of the LHGA-SH(4) model, implicitly implies the existence of three distinct contributions in the baryon wave-function, instead of the sole charm-unsymmetrized one of the LWGA-WF model. Together with the required $1/\sqrt{3}$ normalization factor, the three contributions end up producing a LHGA-SU(4) transition amplitude which is a factor $\sqrt{3}$ larger in size than the LHGA-WF one.

The vector meson-baryon (VB) interactions are also obtained from a t-channel vector meson exchange formalism in the LHGA. In this case the $VPP$ vertex of Eq.~\eqref{L_VPP} is replaced by the the $VVV$ one~\cite{Gloria, Bando, OsetRamos10}:
\begin{equation}
    \mathcal{L}_{VVV} = i g\, Tr\left(\left[V_{[16]}^{\nu}, \partial_\mu V_{[16],\nu} \right]V^\mu_{[16]} \right)
    \label{L_VVV}
\end{equation}
For the baryon vertex, we use $\mathcal{L}_{VBB}$ in Eq.~\eqref{L_VPP} within the LHGA-SU(4) model or Eq.~\eqref{VBB_debastiani} within the LHGA-WF method. In both cases, one can see that the resulting interaction kernels are equal to that obtained for PB interactions in Eq.~\eqref{V_SU4_swave}, but multiplied by a factor related to the polarizations of the initial and final vector mesons, $\epsilon \cdot \epsilon'$.

\subsection{Unitarized Chiral Perturbation Theory (UChPT)} \label{UChPT}
\begin{figure}[h!]
      \includegraphics[width=\linewidth]{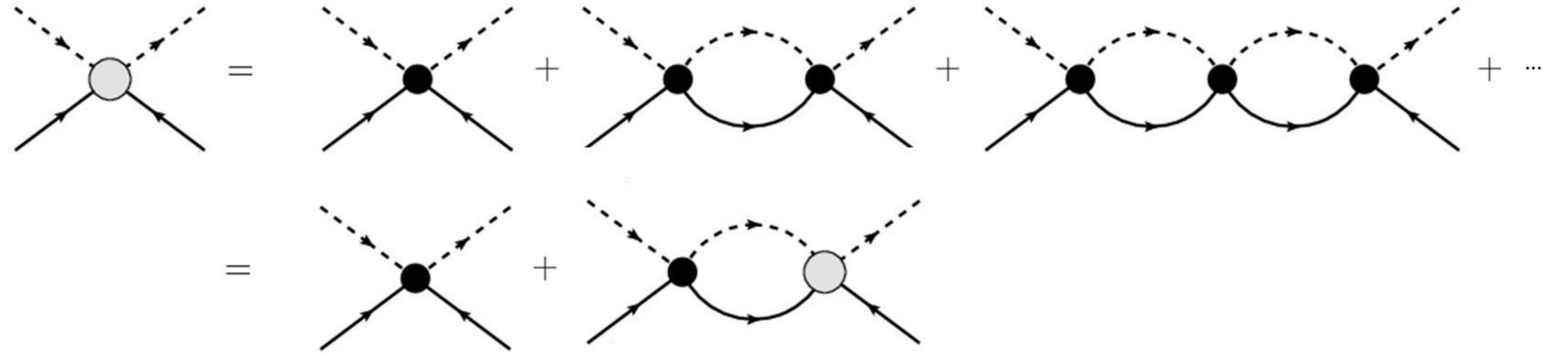}
    \caption{Diagrammatic representation of the Bethe-Salpeter equation. The gray circle corresponds to the scattering amplitude $T$, the black circles represent the interaction kernel $V$ and the internal lines correspond to the meson-baryon propagator $G$.}
    \label{MB_BSdiagram}
\end{figure}

All the information about a scattering process is contained in the scattering amplitude or T-matrix. The unitarized T-matrix can be obtained from the meson-baryon interaction kernel by solving the Bethe-Salpeter equation in coupled channels, which is shown diagrammatically in Fig.~\ref{MB_BSdiagram}.

Considering the on-shell contribution of the interaction kernel, it can be factorized and the Bethe-Salpeter equation reduces to a set of algebraic equations \cite{Oller:1997ti}, which, in matrix form, reads:
\begin{equation}
    T= V+  V G T \quad \Rightarrow \quad T
    =\left( 1 - V G \right)^{-1} V\,,
    \label{factorized_LS}
\end{equation}
where $G$ is a diagonal matrix containing the loop function for all  channels $k$ in the given sector
\begin{equation}
    G_k(\sqrt{s})= {\rm i} \int \frac{d^4q_k}{(2\pi)^4} \frac{2M_k}{(P-q_k)^2 - M_k^2 +i\epsilon} \frac{1}{q_k^2 - m_k^2 +i\epsilon} \ .
    \label{loop_function}
\end{equation}
The real part of this last integral diverges logarithmically, so a regularization scheme must be employed to compute the loop function. Different methods will be discussed in Section~\ref{sect:Ghy}. The imaginary part  reads
\begin{equation}
    \text{Im}\left(G_k(\sqrt{s})\right) = - \frac{2M_k}{8\pi}\frac{q_{\text{cm}}}{\sqrt{s}} \theta\left(s - (M_k + m_k)^2\right),
    \label{ImG_oller}
\end{equation}
where $q_{\text{cm}}$ is the center-of-mass momentum:
\begin{equation}
q_{\text{cm}} = \frac{1}{2\sqrt{s}}\sqrt{[s-(M_k + m_k)^2][s-(M_k - m_k)^2]}.
\label{qcm}
\end{equation}

The so called dynamically generated states, appear in the T-matrix when the condition~$det(1-VG)=0$ is satisfied. These poles usually appear in the complex plane, so the T-matrix needs to be analytically continued to complex energies. Close to the pole, the T-matrix can be approximated as,
\begin{equation}
    T_{ij}(\sqrt{s}) \approx \frac{g_i g_j}{\sqrt{s} - M_R + i \frac{\Gamma_R}{2}},
    \label{breit-wigner}
\end{equation}
from which we can extract the mass, $M_R$, and the width, $\Gamma_R$, of the resonance or bound state, as well as its coupling strength to the different channels, $g_i$.

Regarding the computation of the scattering amplitude in VB interactions, when performing the resummation of the loops in the Bethe-Salpeter equations, we only need to consider the transverse vector mesons to find poles in the T-matrix~\cite{Roca2005}. In this context, knowing that the sum over all the polarizations of the internal vector mesons is
\begin{equation}
    \sum_{\text{pol}} \epsilon_i \epsilon_j = \delta_{ij} + \frac{q_i q_j}{M_V^2},
\end{equation}
we can neglect the term $\sim \vec{q}\,^2/M_V^2$ (in line with previous low energy approximations). 
Thus,  the formalisms of PB and VB interactions are equivalent, so the VB channels in each strangeness-isospin sector are the same as the PB channels, with the replacements
\begin{equation}
    \eta_c \rightarrow J/\psi \hspace{1cm} \bar{D} \rightarrow \bar{D}^* \hspace{1cm} \bar{D}_s \rightarrow \bar{D}_s^* \ .
\end{equation}

\subsection{Loop function regularization} \label{sect:Ghy}
In this section we discuss different methods to regularize the real part of the integral defining the
meson-baryon loop function in Eq.~\eqref{loop_function}, which diverges logarithmically. The cut-off scheme limits the integration over the spatial components at a given value $q_{\max}$:
\begin{equation}
    G_k^{\text{CO}} = 2M_k \int^{q_{\max}} \frac{q^2 dq}{4\pi^2} \frac{(\omega_k+E_k)}{\omega_k E_k} \\
    \frac{1}{s-(\omega_k+E_k)^2 + i \epsilon},
    \label{G_co_integral}
\end{equation}
where $\omega_k(\vec{q}\,) = \sqrt{m_k^2 + \vec{q}\,^2}$, $E_k(\vec{q}\,) = \sqrt{M_k^2 + \vec{q}\,^2}$. Within this approach, the loop function can be evaluated analytically~\cite{OllerOset99}:
\begin{widetext}
\begin{multline}
    G^{\text{CO}}_k = \frac{2M_k}{32\pi^2} \left\{ -\frac{\Delta}{s} \log \left( \frac{m_k^2}{M_k^2}\right) + \frac{\nu}{s} \left[\log \left(\frac{s-\Delta+\nu\sqrt{1+\frac{m_k^2}{q^2_{\max}}}}{-s+\Delta+\nu\sqrt{1+\frac{m_k^2}{q^2_{\max}}}}\right) + \log \left(\frac{s+\Delta+\nu\sqrt{1+\frac{M_k^2}{q^2_{\max}}}}{-s-\Delta+\nu\sqrt{1+\frac{M_k^2}{q^2_{\max}}}}\right) \right] \right. \\
    \left. + 2\frac{\Delta}{s} \log \left(\frac{1+\sqrt{1+\frac{m_k^2}{q^2_{\max}}}}{1+\sqrt{1+\frac{M_k^2}{q^2_{\max}}}}\right) - 2 \, \log \left[ \left( 1+\sqrt{1+\frac{m_k^2}{q^2_{\max}}} \right)\left(1+\sqrt{1+\frac{M_k^2}{q^2_{\max}}}\right) \right] + \log \left( \frac{m_k^2 M_k^2}{q^4_{\max}}\right) \right\},
    \label{G_co}
\end{multline}
\end{widetext}
where $\Delta = M_k^2 - m_k^2$, and $\nu = 2\sqrt{s}\, q_{\text{cm}}$.

Alternatively, one can employ the dimensional regularization scheme:
\begin{widetext}
\begin{multline}
    G_k^{\text{DR}} = \frac{2M_k}{(4\pi)^2} \Bigg\{a_k(\mu) + \ln \left(\frac{M_k^2}{\mu^2}\right) + \frac{m_k^2 - M_k^2+s}{2s} \ln\left(\frac{m_k^2}{M_k^2}\right) + \frac{q_{\text{cm}}}{\sqrt{s}} \left[ \ln(s-(M_k^2 - m_k^2) + 2\sqrt{s}\,q_{\text{cm}}) \right. \\
     \left. + \ln(s+(M_k^2 - m_k^2) + 2\sqrt{s}\,q_{\text{cm}}) - \ln(-s+(M_k^2 - m_k^2) + 2\sqrt{s}\,q_{\text{cm}}) - \ln(-s-(M_k^2 - m_k^2) + 2\sqrt{s\,}q_{\text{cm}})\right]\Bigg\},
    \label{G_dr}
\end{multline}
\end{widetext}
where $a_k(\mu)$ are the subtraction constants, which determine the finite part of the integral at the regularization scale $\mu$. In Eq.~\eqref{G_co} and~\eqref{G_dr} both the cut-off and subtraction constants are free parameters that can be determined by fitting the model to data.

\begin{figure}
    \includegraphics[width=\linewidth]{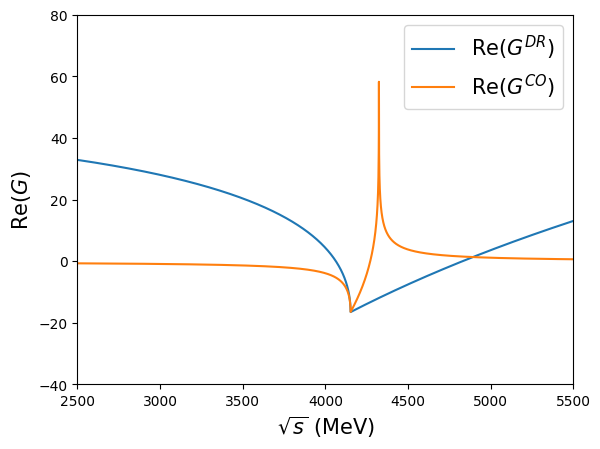}
    \caption{Typical behavior of the real part of the loop function in the dimensional (blue line) and cut-off (orange line) regularization schemes.}
    \label{Gdr_Gco_demo}
\end{figure}

Although both schemes yield similar amplitudes close to the energy at which the two loops are forced to be equal (usually, the channel threshold), at further away energies there might appear some structures which should not be associated to any physical resonance or bound state. Recall that poles appear when the T-matrix diverges, that is when $\det(1-VG) = 0$ or when $V^{-1} \sim \text{Re}\,G$ if the pole is strongly coupled to one channel (uncoupled approach). If the latter relation is satisfied when the potential $V$ is positive, the pole generated is fake in the sense that it can not be associated to a genuine state, since there is no physical justification for a repulsive interaction to produce a resonance or bound state. In Fig.~\ref{Gdr_Gco_demo} we show the typical behavior of the real part of the loop function in the dimensional (blue line) and cut-off (orange line) regularization schemes. As we can see, $\text{Re}\,G^{\text{DR}}$ may become positive below the channel threshold, hence potentially giving rise to fake poles in that region of energies. Likewise, $\text{Re}\,G^{\text{CO}}$ acquires large positive values at energies where $q_{\text{cm}}$ is close to $q_{\max}$ (where a divergence appears), a region where fake poles in the scattering amplitude may then appear.

Thus, in both dimensional and cut-off regularization schemes one needs to check whether the dynamically generated states are genuine or fake. Such a study implies performing a quite involved analysis, and the attempts to make a choice in a simplified, more qualitative, way may lead to erroneous results. Some particular examples will be presented and discussed later in this work. 

In the present work we introduce an improved unitarization scheme that avoids the appearance of non-genuine poles. We propose a hybrid loop function, $G^{\text{HY}}$, that preserves only the well-behaved parts of $\text{Re}\,G^{\text{CO}}$ and $\text{Re}\,G^{\text{DR}}$. It is defined as follows:
\begin{equation}
\begin{aligned}
\text{Re}\,G_k^{\text{HY}}(\sqrt{s}) &= \left\{ 
\begin{array}{rcl} 
\text{Re}\,G^{\text{CO}}_k(\sqrt{s}) \ \ \ \text{if} \ \ \ \sqrt{s} < M_k + m_k  \\[1mm]
\text{Re}\,G^{\text{DR}}_k(\sqrt{s}) \ \ \ \text{if} \ \ \ \sqrt{s} \ge M_k + m_k
\end{array} 
\right. \\[2mm]
\text{Im}\,G_k^{\text{HY}}(\sqrt{s})  &= \text{Im}\,G^{\text{DR}}_k(\sqrt{s}) 
\end{aligned}
\label{Ghy}
\end{equation}
with the condition that $\text{Re}\,G^{\text{CO}}$ and $\text{Re}\,G^{\text{DR}}$ match at the channel threshold. Consequently, the subtraction constants $a_k(\mu)$ are no longer independent parameters but are connected to the value of $q_{\max}$ via the following relation:
\begin{equation}
    a_k(\mu) = \frac{16\pi^2}{2M_k}(G^{\text{CO}}_k(q_{\max}) - G_k^{\text{DR}}(\mu, a_k = 0)) \ ,
\label{subs_constants}
\end{equation}
for a given value of the regularization scale parameter, which we choose $\mu = 1000 \, \text{MeV}$.
In this work we will present results for $q_{\max}$ = $ 600 \, \text{MeV}$ and $800 \, \text{MeV}$ (following the values used in Refs.~\cite{Roca, Feijoo19}), as a measurement of  uncertainties. 

Note that the hybrid prescription assigns to $\text{Re}\,G^{\text{HY}}$ the negative value of 
 $\text{Re}\,G^{\text{CO}}$ for energies below threshold, where potential bound states might occur. For energies above the threshold, $\text{Re}\,G^{\text{HY}}$ takes the value of $\text{Re}\,G^{\text{DR}}$, which is negative in an extended energy region and does not show the divergent structure of the cut-off loop function.
As we will show in the next section, this hybrid loop function naturally eliminates the fake poles generated when $G^{\text{CO}}$ and $G^{\text{DR}}$ are used individually, while leaving the physical poles almost unmodified.


Let us note that the hybrid prescription of the loop does not satisfy a dispersion relation relating its real and imaginary parts. In this sense, it must be viewed within a phenomenological perspective as a practical way of avoiding spurious structures in the scattering amplitude. 

\section{Results}
\subsection{PB and VB interactions within the LHGA-SU(4) model employing $G^{\text{DR}}$, $G^{\text{CO}}$ and $G^{\text{HY}}$} 
\label{Gdr_co_hy_comp}

In this section, we perform a complete study of the structures which appear in the scattering amplitude obtained in meson-baryon interactions for all possible values of strangeness $S$ and isospin $I$ in the hidden charm sector. We present the results obtained using the hybrid loop function of Eq.~\eqref{Ghy}, and compare them with those obtained using the cut-off one of Eq.~\eqref{G_co} and the dimensional regularization one of~ Eq.\eqref{G_dr}. The kernel coefficients employed in this part of the analysis are taken from the LGHA-SU(4) model developed in~\cite{Lutz} and listed in Appendix \ref{app:B}. All the scattering amplitude plots shown for each sector are computed with $q_{\max} = 600 \, \text{MeV.}$

\subsubsection{Sector $S=0, \, I=1/2$} \label{sec:S0I1}
\begin{figure*}
    \includegraphics[width=\linewidth]{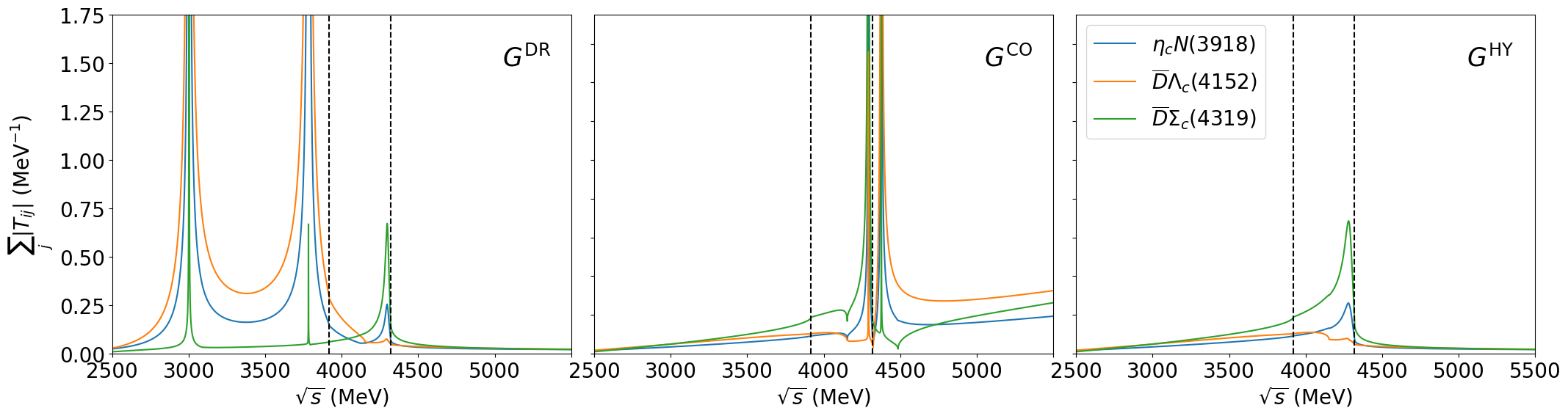}
    \caption{From left to right, the panels illustrate the scattering amplitude obtained with $G^{\text{DR}}$, $G^{\text{CO}}$ and $G^{\text{HY}}$ for the PB interaction in the $S=0, \, I=1/2$ sector. All three plots share the same y-axis scale and color legend. Each line represents the sum of the moduli of the transition amplitudes to the i-th final channel. Vertical black dashed lines indicate the thresholds of the lightest and heaviest channels.}
    \label{fig_PB_S0I1}
\end{figure*}

We start discussing the results for the  PB interaction in the sector with strangeness 0 and isospin $1/2$, where three channels are coupled: $\eta_c N$, $\bar{D} \Lambda_c$ and $\bar{D} \Sigma_c$.
In Fig.~\ref{fig_PB_S0I1} 
we represent $\sum_{j} |T_{ij}|$ for all different $i$ channels in different colors as functions of $\sqrt{s}$ within the range $\sqrt{s} \in [2500, 5500] \, \text{MeV}$. 
The vertical black dashed lines indicate the thresholds of the lightest and heaviest channels. The left, middle and right panels correspond to the results obtained employing, respectively, the $G^{\text{DR}}$, 
$G^{\text{CO}}$ and $G^{\text{HY}}$ prescription for the loop function.

When the $G^{\text{DR}}$ loop is employed, the T-matrix presents three peaks. Two of them appear below the lightest channel threshold and would be classified as bound states, while the third one would correspond to a resonance, since it is generated above the lightest channel threshold. However, the two bound states correspond to what we call fake poles, since they are generated by a repulsive potential, as we will discuss below. In the case of the $G^{\text{CO}}$ loop, one observes two peaks in the T-matrix but, again, as we justify below, the one above the higher threshold is a fake one. Finally, the T-matrix obtained with $G^{\text{HY}}$ leads to a single peaked structure that corresponds to a genuine resonance.

\begin{figure*}[ht!]
     \begin{subfigure}[b]{0.49\textwidth}
         \includegraphics[width=\linewidth]{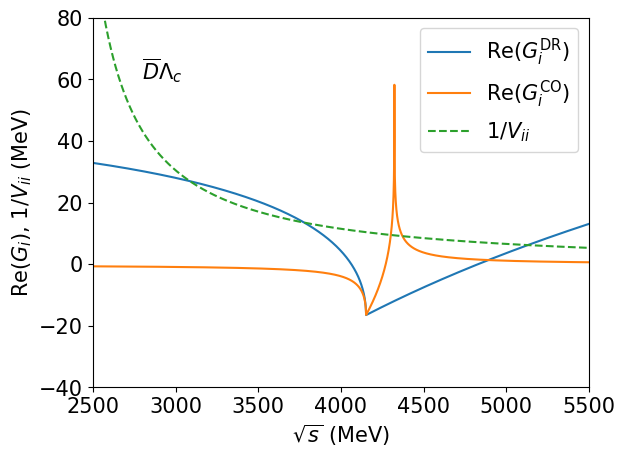}
         \caption{}
         \label{G-V_S0I1_a}
     \end{subfigure}
     \hfill
     \begin{subfigure}[b]{0.49\textwidth}
         \includegraphics[width=\linewidth]{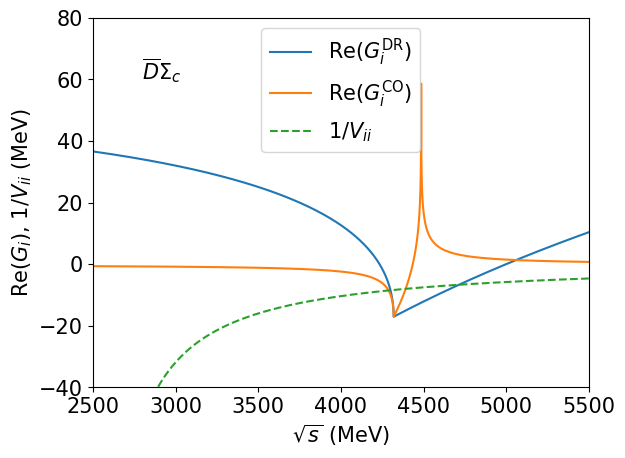}
         \caption{}
         \label{G-V_S0I1_b}
     \end{subfigure}
    \caption{
    Real part of the loop function in the dimensional (blue solid line) and cut-off (orange solid line) regularization scheme. The inverse potential $1/V_{ii}$ is also plotted. The subindex i stands for the corresponding channel:  $\bar{D}\Lambda_c$ [panel (a)] and $\bar{D}\Sigma_c$ [panel (b)].
    }
    \label{G-V_S0I1}
\end{figure*}

To understand the origin of the fake poles, it is instructive to analyze the problem within an uncoupled approach, where structures in the T-matrix for real values of $\sqrt{s}$ appear when $\text{Re}\ G_k = V_k^{-1}$ as long as the size of $\text{Im}\ G_k$ is small.
In Fig.~\ref{G-V_S0I1} we show the inverse potential with dashed green lines for the $\bar{D}  \Lambda_c$ and $\bar{D} \Sigma_c$ channels in the left and right panels, respectively.  We have omitted the figure of the $\eta_c N$ channel because the corresponding kernel coefficient is zero (see Table~\ref{tab:C_S0I1} in Appendix \ref{app:B}). Each panel also shows the corresponding real part of the loop function $G^{\text{DR}}$ with a solid blue line and that of $G^{\text{CO}}$ with a solid orange line. 

For the $\bar{D} \Lambda_c$ channel, we find in Fig.~\ref{G-V_S0I1_a} two values of $\sqrt{s}$ --- $3087  \, \text{MeV}$ and $3779 \, \text{MeV}$ ---at which $\text{Re}\ G^{\text{DR}}$ and $1/V$ intersect, both below the channel threshold. These correspond to the structures found in the left panel of Fig.~\ref{fig_PB_S0I1} at $\sqrt{s} = 3002 \, \text{MeV}$ and $\sqrt{s} = 3782 \, \text{MeV}$, which couple strongly to $\bar{D}\Lambda_c$ (orange line). The differences are attributed to coupled-channel effects, which
shift the pole positions obtained in the uncoupled approach. Similarly, in Fig.~\ref{G-V_S0I1_a}, we observe that the inverse potential crosses $\text{Re}\ G^{\text{CO}}$ of $\bar{D} \Lambda_c$ at two points around the spike. The two intersections would give rise to two poles in the scattering amplitude in the complex energy plane, although they appear merged as one asymmetric structure in the real axis above the $\bar{D}\Sigma_c$ threshold, coupling strongly to $\bar{D} \Lambda_c$, as seen in the middle panel of Fig.~\ref{fig_PB_S0I1}. 

Now we come to the important observation: the interaction of the $\bar{D} \Lambda_c$ channel is repulsive at the points where $1/V$ intersects $\text{Re}\ G^{\text{DR}}$ and $\text{Re}\ G^{\text{CO}}$. Therefore, it should not generate bound states or resonances. Consequently,  the two peaks below the lowest threshold in the amplitude generated by $G^{\text{DR}}$ or the structure above the highest threshold obtained with $G^{\text{CO}}$ do not correspond to physical poles and must be disregarded.

Once the nonphysical structures are eliminated, the only genuine peak remaining in the amplitudes obtained either with $G^{\text{DR}}$ or $G^{\text{CO}}$ couples strongly to the $\bar{D} \Sigma_c$ channel. In an uncoupled picture, this structure is generated by the intersection of the inverse potential with $\text{Re}\,G^{\text{DR}}$ or $\text{Re}\,G^{\text{CO}}$ just below the channel threshold, as seen in the right panel of Fig.~\ref{G-V_S0I1_b}. 

Note that both in Fig.~\ref{G-V_S0I1_a} and Fig.~\ref{G-V_S0I1_b} there are also some intersections above the channel threshold, which in principle would give rise to resonances. However, due to the sizable imaginary part of the loop at the crossing energy, the poles lie far away from the real axis in the complex plane and do not stand out as a peak of the scattering amplitude. 

The scattering amplitude obtained with $G^{\text{HY}}$, presented in the right panel of Fig.~\ref{fig_PB_S0I1}, naturally eliminates the artificial structures observed with the previous regularization schemes and only generates the physical pole, thus substantially simplifying  the analysis of the obtained results. 

\begin{table}
    \begin{tabular}{c|c|c|cc}
 \hline\hline
         \multicolumn{5}{c}{ }  \\[-3mm]
        \multicolumn{5}{c}{PB interaction ($J^P = \frac{1}{2}^-$) in the $(S,I) = (0,\frac{1}{2})$ sector}  \\
            \multicolumn{5}{c}{ }  \\[-3mm]  \hline \hline
           & & & & \\[-3mm] 
        & $G^{\text{DR}}$ & $G^{\text{CO}}$ & \multicolumn{2}{c}{$G^{\text{HY}}$} \\ \hline
        $q_{\text{max}}$ (MeV) & ~~~~600~~~~&~~~~600~~~~ & ~~~~~600~~~~~ & ~~~800~~~
        \\ \hline
        $M$ (MeV) & 4299& 4309 & 4283 & 4191 \\
        $\Gamma$ (MeV) & 22 & 57 & 29 & 54 \\ \hline
        & $|g_i|$& $|g_i|$& $|g_i|$& $|g_i|$ \\
        $\eta_c N (3918)$ & 0.89 & 1.08 & 1.16 & 1.77 \\
        $\bar{D} \Lambda_c (4152)$& 0.19 & 0.40 & 0.24 & 0.32 \\
        $\bar{D} \Sigma_c(4319)$ & 2.28 & 3.06 & 2.98 & 4.40
    \end{tabular}
    \begin{tabular}{c|c|c|cc}
            \hline\hline
                     \multicolumn{5}{c}{ }  \\[-3mm]
        \multicolumn{5}{c}{VB interaction ($J^P = \frac{1}{2}^-, \frac{3}{2}^-$) in the $(S,I) = (0,\frac{1}{2})$ sector}  \\
                             \multicolumn{5}{c}{ }  \\[-3mm]
                             \hline \hline 
                                        & & & & \\[-3mm] 
        & $G^{\text{DR}}$ & $G^{\text{CO}}$ & \multicolumn{2}{c}{$G^{\text{HY}}$} \\ \hline
                $q_{\text{max}}$ (MeV) & ~~~~600~~~~&~~~~600~~~~ & ~~~~~600~~~~~ & ~~~800~~~
                \\ \hline
        M(MeV) & 4439 & 4450 & 4422 & 4326 \\
        $\Gamma$(MeV) & 22& 61 & 30 & 55 \\ \hline
        & $|g_i|$ & $|g_i|$& $|g_i|$ & $|g_i|$\\
        $J/\psi N (4035)$ &0.88 & 1.11 & 1.18 & 1.80 \\
        $\bar{D}^* \Lambda_c (4293)$& 0.19& 0.39 & 0.25 & 0.33 \\
        $\bar{D}^*  \Sigma_c(4460)$ & 2.31 & 3.18& 3.07 & 4.54 \\
    \hline\hline
    \end{tabular}    
    \caption{Masses, widths and coupling strengths of the genuine poles found in the $S=0, \, I=1/2$ sector with the three regularization schemes, $G^{\text{DR}}$, $G^{\text{CO}}$ and $G^{\text{HY}}$.}
    \label{tab_S0I1}
\end{table}

The position of the pole, its width and the coupling
strengths to the different channels, obtained for the three regularization schemes, are collected in the upper half of Table~\ref{tab_S0I1}. The resonance has a mass of around 4300~MeV, the differences between regularization schemes being small, and couples predominantly to $\bar{D} \Sigma_c$ and substantially to
$\eta_c N$.
The widths are also relatively similar for the DR and HY schemes, 22~MeV and 29~MeV, respectively, but its size doubles in the case of the CO scheme.
This can be understood as follows.
The poles are found in the second Riemann sheet, which is defined by employing the following loop function:
\begin{equation}
    G_k^{II}(\sqrt{s})= G_k(\sqrt{s})+{\rm i}\frac{2 M_k}{4\pi\sqrt{s}}q_k(\sqrt{s}) \ ,
    \label{eq:2ndRiemann}
\end{equation}
if $\text{Re}\ \sqrt{s}> M_k+m_k$.  In the real energy axis,
the above expression amounts to changing the sign of the momentum $q_k$, i.e. taking the negative prescription of the square root defining this momentum, or, equivalently, changing the sign of the original $\text{Im}\ G_k$ by adding twice its opposite value. This is the case for the DR or HY schemes.  However, in the CO scheme, the analytical formula of Eq.~(\ref{G_co}) makes $\text{Im}\ G_k^{\text{CO}}$ to be zero for energies larger than $\sqrt{M_k^2+q_{\max}^2}+\sqrt{m_k^2+q_{\max}^2}$ and then the prescription of Eq.~(\ref{eq:2ndRiemann}) transform a zero-valued $\text{Im}\ G_k^{\text{CO}}$ into a $\text{Im}\ G_k^{II, \text{CO}}$ with a value {\it twice} as large as that acquired by the DR or HY schemes. This explains the wider states obtained in the CO regularization scheme.

In Table~\ref{tab_S0I1}, we also include the results for $q_{\max} = 800 \, \text{MeV}$ for the HY scheme,  as an estimate of the uncertainty in the pole properties. With this new value of the cut-off parameter, the mass shifts downwards by approximately $90 \, \text{MeV}$, and the width increases by about $25 \, \text{MeV}$. 

In the case of the VB interaction in this sector, we observe the same systematics as in the PB case but the structures are shifted to higher energies. This is expected since the kernel coefficients are identical to those of the PB case, with the following substitutions in this sector:
$\eta_c \rightarrow J/\psi$, $\bar{D} \rightarrow \bar{D}\,^*$.
The properties of the pole computed for $q_{\max} = 600\, \text{MeV}$ within the different regularization schemes are collected in the bottom half of Table~\ref{tab_S0I1}, where we also show the HY results for a cut-off value of $q_{\max} = 800\, \text{MeV}$. In this VB case, the states are spin-degenerate, with $J^P = 1/2^-$ and $J^{P} = 3/2^-$, since no spin-dependent term is included in the interaction potential of Eq.~\eqref{V_SU4_swave}.
Note that the mass difference of about 140~MeV between the states found for the VB or PB interactions corresponds, essentially, to the difference between the thresholds $\bar{D}^* \Sigma_c$ and  $\bar{D} \Sigma_c$ states as they are, respectively, the channels to  which the PB and VB states couple most.

\subsubsection{Sector  $S=-1, \, I=0$} \label{sec:S1I1}

\begin{figure*}[ht!]
    \includegraphics[width=\linewidth]{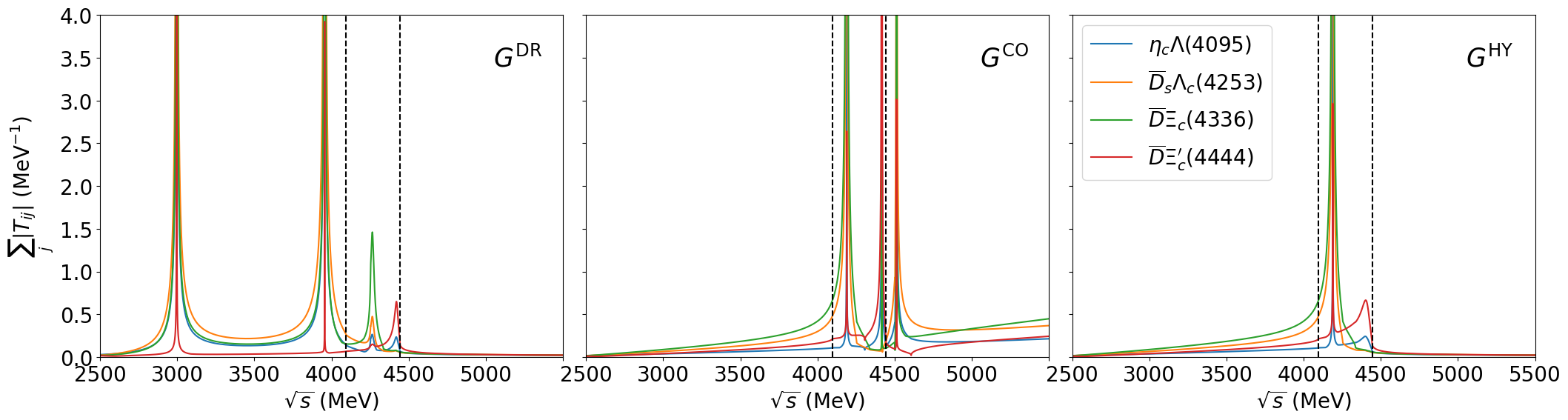}
    \caption{From left to right, the panels illustrate the scattering amplitude results obtained with $G^{\text{DR}}$, $G^{\text{CO}}$ and $G^{\text{HY}}$ for the PB interaction in the $S=-1, \, I=0$ sector. All three plots share the same y-axis scale and color legend. Each line represents the sum of the moduli of the transition amplitudes to the i-th final channel. Vertical black dashed lines indicate the thresholds of the lightest and heaviest channels.}
    \label{fig_PB_S1I1}
\end{figure*}

The PB interaction in the $S= -1, \, I=0$ sector involves four heavy channels: $\eta_c\Lambda$, $\bar{D}_s \Lambda_c$, $\bar{D} \Xi_c$ and $\bar{D} \Xi_c'$. When using the $G^{\text{HY}}$ prescription, two resonances are found in the scattering amplitude,  as can be seen in the right panel of Fig.~\ref{fig_PB_S1I1}. The number of structures is larger when we use $G^{\text{DR}}$ (left panel) or $G^{\text{CO}}$ (midle panel), and an additional study is required to determine
their origin.

\begin{figure*}[ht!]
     \begin{subfigure}[b]{0.32\textwidth}
         \includegraphics[width=\linewidth]{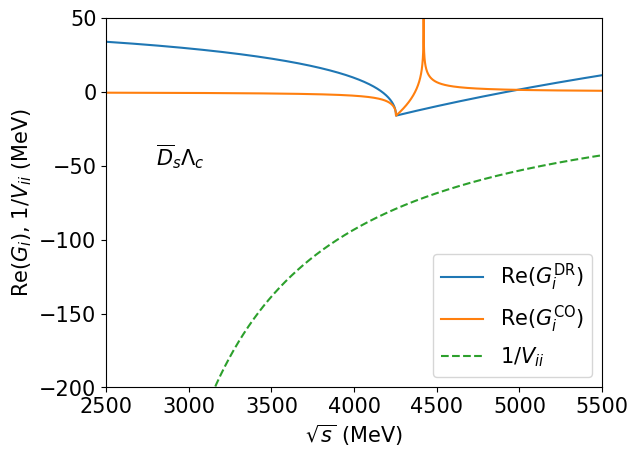}
         \caption{}
         \label{G-V_S1I1_a}
     \end{subfigure}
     \hfill
     \begin{subfigure}[b]{0.32\textwidth}
         \includegraphics[width=\linewidth]{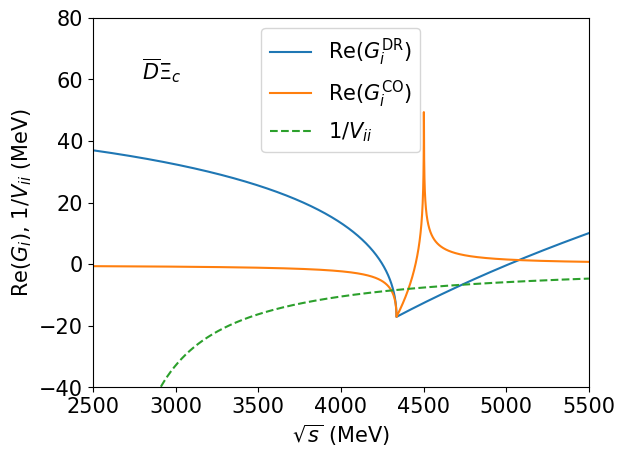}
         \caption{}
         \label{G-V_S1I1_b}
     \end{subfigure}
     \hfill
     \begin{subfigure}[b]{0.32\textwidth}
         \includegraphics[width=\linewidth]{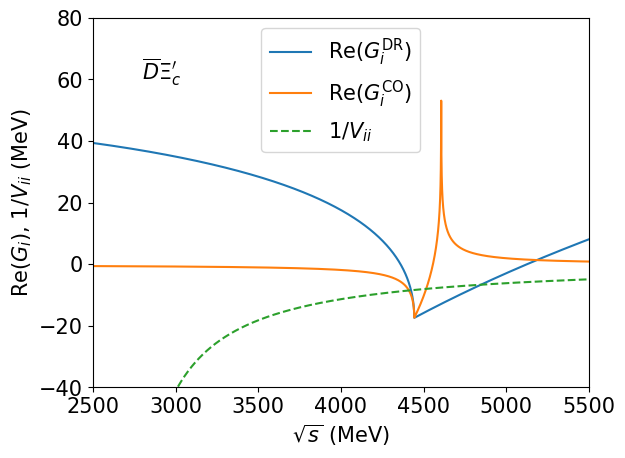}
         \caption{}
         \label{G-V_S1I1_c}
     \end{subfigure}
    \caption{
    Real part of the loop function in the dimensional (blue solid line) and cut-off (orange solid line) regularization scheme. The inverse potential $1/V_{ii}$ is also plotted. The subindex i stands for the corresponding channel:  $\bar{D}_s\Lambda_c$ [panel (a)], $\bar{D}\Xi_c$ [panel (b)] and $\bar{D} \Xi_c'$ [panel (c)].}
    \label{G-V_S1I1}
\end{figure*}

This sector is particularly interesting because the uncoupled approach fails to explain the existence of the two bound states that appear around 3000~MeV and 4000~MeV in the T-matrix computed with $G^{\text{DR}}$ 
or the structure above the $\bar{D}\Xi_c'$ threshold in the T-matrix computed with $G^{\text{CO}}$, which all couple strongly to $\bar{D}_S\Lambda_c$ (orange line Fig.~\ref{fig_PB_S1I1}). In  Fig.~\ref{G-V_S1I1} we display the real parts of the DR and CO loop functions and the inverse potential for the relevant channels in this sector ($\bar{D}_s \Lambda_c$, $\bar{D} \Xi_c$ and $\bar{D} \Xi_c'$). Indeed, for the $\bar{D}\Lambda_c$ case (left panel) there are no intersections between the inverse of the potential (dashed green line) and $\text{Re}\,G^{\text{DR}}$ (solid blue line) or $\text{Re}\,G^{\text{CO}}$ (orange line).
Consequently, the above mentioned structures are due to the coupled-channel dynamics, as we show in the following.


We analyze the problem by means of an effective single-channel interaction which takes into account the coupling to other channels. In the case of two channels, labeled as 1 and 2, we obtain the following expression from Eq.~\eqref{factorized_LS}:
\begin{widetext}
\begin{gather}
    \begin{pmatrix}
        T_{11} & T_{12} \\
        T_{21} & T_{22}
    \end{pmatrix}
    =
    \begin{pmatrix}
        V_{11} & V_{12} \\
        V_{21} & V_{22}
    \end{pmatrix}
    +
    \begin{pmatrix}
        V_{11} & V_{12} \\
        V_{21} & V_{22}
    \end{pmatrix}
    \begin{pmatrix}
        G_1 & 0 \\
        0 & G_2
    \end{pmatrix}
    \begin{pmatrix}
        T_{11} & T_{12} \\
        T_{21} & T_{22}
    \end{pmatrix}.
\end{gather}
\end{widetext}
The solution for $T_{11}$ is:
\begin{equation}
    T_{11} = \frac{V_{11} + V_{12} G_2(1-V_{22}G_2)^{-1} V_{21}}{1 -(V_{11} + V_{12} G_2(1-V_{22}G_2)^{-1} V_{21})G_1} \ ,
\label{eq:Teff}
\end{equation}
and, defining $V^{\text{eff}}_{11}$ as
\begin{equation}
 V^{\text{eff}}_{11} = V_{11} + V_{12} G_2(1-V_{22}G_2)^{-1} V_{21},
 \label{eq:Veff}
\end{equation}
we recover the one-channel type expression:
\begin{equation}
    T_{11} = \frac{V^{\text{eff}}_{11}}{1 - V^{\text{eff}}_{11}G_1} \ .
\end{equation}
We compute the effective potential for the coupled channels $\bar{D}_s\Lambda_c$ and $\bar{D} \Xi_c$ because their coupling coefficient is significantly larger than the other channel couplings (see Table~\ref{tab:C_S1I1} in Appendix \ref{app:B}). In Fig.~\ref{Veff_S1I1}, we show the real part of the loop function of the $\bar{D}_s \Lambda_c$ channel employing DR regularization (solid blue line) or the CO scheme (solid orange line). The figure also shows the inverse of the effective potential, obtained from the kernels involving the $\bar{D}_s \Lambda_c$ and $\bar{D} \Xi_c$ channels [See Eq.~(\ref{eq:Veff})], using $G^{\text{DR}}$ ($1/V_{\text{eff}}^{\text{DR}}$, dashed green line) or $G^{\text{CO}}$  ($1/V_{\text{eff}}^{\text{CO}}$, dotted red line). The two intersections between the dashed green line and $\text{Re}\,G^{\text{DR}}$ below the $\bar{D}_s\Lambda_c$ threshold, at around 3000~MeV and 4000~MeV, correspond to the peaks observed in the amplitude below the $\eta_c \Lambda$ threshold in the left panel of Fig.~\ref{fig_PB_S1I1} and they must be considered as non-physical, since the effective interaction is repulsive in this energy region.  These two lines also intersect above the $\bar{D}_s\Lambda_c$ threshold, where the effective potential is negative. The crossing right above the threshold is the origin of the genuine state represented by the third peak in the left panel of Fig.~\ref{fig_PB_S1I1}, and that occurring much further up in energy (beyond 5000~MeV) leaves no signature in the scattering amplitude at real energies as the corresponding pole lies far away in the complex plane. Moving to the results for the CO scheme, we observe that the inverse effective potential (dotted red line) crosses $\text{Re}\,G^{\text{CO}}$ (solid orange line) below the $\bar{D}_s\Lambda_c$ threshold, in a region where the potential is negative, hence giving rise to the genuine state seen right above the lowest threshold in the middle panel of Fig.~\ref{fig_PB_S1I1}. There is also a double intersection through the spike of the CO loop around 4400~MeV when the potential is positive, giving rise to a non-physical state represented by the asymmetrical peak above the highest threshold in the middle panel of Fig.~\ref{fig_PB_S1I1}.
We recall that Fig.~\ref{Veff_S1I1}, which has served us for the characterization of a single genuine state from the DR and CO scattering amplitudes of Fig.~\ref{fig_PB_S1I1}, replaces the uncoupled Figs.~\ref{G-V_S1I1_a} and \ref{G-V_S1I1_b}. The characterization of the remaining peak is done by inspecting Fig.~\ref{G-V_S1I1_c}, where a crossing of the inverse of the $\bar{D} \Xi^\prime_c$ potential with $\text{Re}\,G^{\text{DR}}$ or $\text{Re}\,G^{\text{CO}}$ is seen right below threshold and corresponds to the physical state seen around that energy in Fig.~\ref{fig_PB_S1I1}. The crossing higher up in energy (either with the DR or CO loop) does not leave a visible signal in the T-matrix.
Having eliminated the non-physical poles, both DR and CO schemes give rise to two genuine states that are precisely the ones that come out straightforwardly in the HY scheme.

\begin{figure}[h!]
         \includegraphics[width=\linewidth]{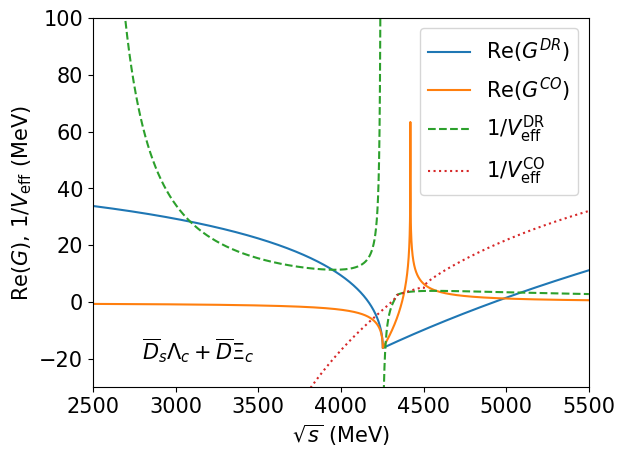}
    \caption{
    Effective potential constructed with the channels $\bar{D}_s \Lambda_c$ and $\bar{D} \Xi_c$  employing the DR loop (dashed green line) or the CO one (dotted red line). The real part of the loop functions for the $\bar{D}_s \Lambda_c$ channel, $\text{Re}\,G^{\text{DR}}$ (blue line) and $\text{Re}\,G^{\text{CO}}$ (orange line), are also shown.}
    \label{Veff_S1I1}
\end{figure}
\begin{table}[h!]
    \begin{tabular}{c|cc|cc|cc|cc}
    \hline \hline
            \multicolumn{9}{c}{ }\\[-3mm]
        \multicolumn{9}{c}{PB interaction ($J^P = \frac{1}{2}^-$) in the $(S,I) = (-1,0)$ sector}  \\
                 \multicolumn{9}{c}{ }\\[-3mm] 
                 \hline \hline
                 &&&&&\multicolumn{4}{c}{} \\[-3mm]
        & \multicolumn{2}{c|}{$G^{\text{DR}}$} & \multicolumn{2}{c|}{$G^{\text{CO}}$} & \multicolumn{4}{c}{$G^{\text{HY}}$} \\ \hline
        $q_{\text{max}}$ (MeV) & \multicolumn{2}{c|}{600} & \multicolumn{2}{c|}{600} & \multicolumn{2}{c|}{600} & \multicolumn{2}{c}{800}  \\ \hline
        M (MeV) & 4265 & 4422 & 4189 & 4435 & 4189 & 4405 & 4037 & 4304 \\
        $\Gamma$ (MeV) & 20 & 23 & 0.73& 62 & 0.29 & 31 & 
        $\sim0$ & 54 \\ \hline
        & $|g_i|$ & $|g_i|$ & $|g_i|$ & $|g_i|$& $|g_i|$ & $|g_i|$ & $|g_i|$ & $|g_i|$ \\
        $\eta_c \Lambda (4095)$ & 0.60 & 0.86 & 0.21& 1.05 & 0.16 & 1.14 & 0.10 & 1.78 \\
        $\bar{D}_s \Lambda_c (4253)$& 0.93 & 0.15& 3.24 & 0.25 & 2.38 & 0.20 & 3.29 & 0.24 \\
        $\bar{D} \Xi_c(4336)$ & 3.20& 0.13 &5.05 & 0.32 & 3.68 & 0.17 & 4.73 & 0.48 \\
        $\bar{D} \Xi_c'(4444)$ &0.20 & 2.27 & 0.11& 3.04 & 0.10 & 3.01 & 0.03 & 4.50 \\
    \end{tabular}
    \begin{tabular}{c|cc|cc|cc|cc}
    \hline \hline
                \multicolumn{9}{c}{ }\\[-3mm]
        \multicolumn{9}{c}{VB interaction ($J^P = \frac{1}{2}^-, \frac{3}{2}^-$) in the $(S,I) = (-1,0)$ sector}   \\
                 \multicolumn{9}{c}{ }\\[-3mm] 
                 \hline \hline
            &&&&&\multicolumn{4}{c}{} \\[-3mm]
        & \multicolumn{2}{c|}{$G^{\text{DR}}$} & \multicolumn{2}{c|}{$G^{\text{CO}}$} & \multicolumn{4}{c}{$G^{\text{HY}}$} \\ \hline
        $q_{\text{max}}$ (MeV) & \multicolumn{2}{c|}{600} & \multicolumn{2}{c|}{600} & \multicolumn{2}{c|}{600} & \multicolumn{2}{c}{800} \\ \hline
        M (MeV) & 4409 & 4563 & 4328 & 4575 & 4328 & 4544 & ~4170~ & 4439 \\
        $\Gamma$ (MeV) & 19 & 23 & 0.74 & 65 & 0.29 & 31 & $\sim 0$ & 56 \\ \hline
        & $|g_i|$ & $|g_i|$ & $|g_i|$ & $|g_i|$& $|g_i|$ & $|g_i|$ & $|g_i|$ & $|g_i|$ \\
        $J/\psi \Lambda (4212)$ & 0.60&0.85&0.20&1.08& 0.15 & 1.15 & 0.01 & 1.81 \\
        $\bar{D}_s^* \Lambda_c (4397)$& 0.89&0.15&3.40&0.26& 2.44 & 0.20 & 3.38 & 0.24\\
        $\bar{D}^* \Xi_c(4477)$ & 3.18&0.13&5.29&0.33& 3.77 & 0.18 & 4.85 & 0.53 \\
        $\bar{D}^* \Xi_c'(4585)$ & 0.20&2.29&0.12&3.16& 0.11 & 3.10 & 0.04 & 4.64 \\
    \hline \hline
    \end{tabular}
    \caption{Masses, widths and coupling strengths of the genuine poles found in the $S=-1, \, I=0$ sector with the three regularization schemes, $G^{\text{DR}}$, $G^{\text{CO}}$ and $G^{\text{HY}}$.}
    \label{tab_S1I1}
\end{table}

\begin{figure*}[ht!]
    \includegraphics[width=\linewidth]{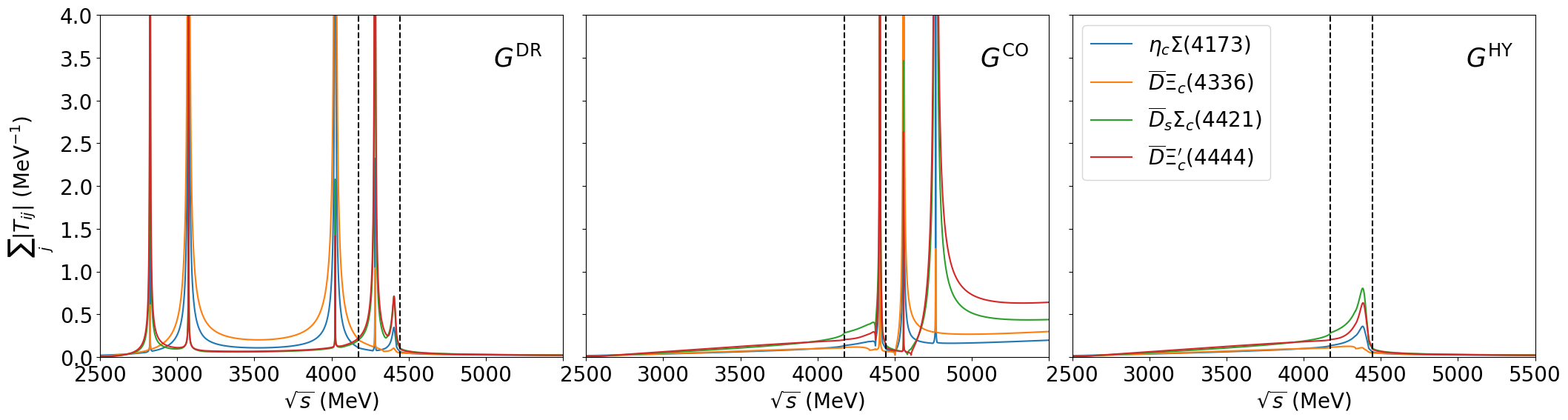}
    \caption{From left to right, the panels illustrate the scattering amplitude results obtained with $G^{\text{DR}}$, $G^{\text{CO}}$ and $G^{\text{HY}}$ for the PB interaction in the $S=-1, \, I=1$ sector. All three plots share the same y-axis scale and color legend. Each line represents the sum of the moduli of the transition amplitudes to the i-th final channel. Vertical black dashed lines indicate the thresholds of the lightest and heaviest channels.}
    \label{fig_PB_S1I2}
\end{figure*}

The properties of the PB states obtained in the $S=-1, I=0$ sector employing a cut-off $q_{\max} = 600 \, \text{MeV}$ for the three regularization schemes are summarized in the upper part of Table~\ref{tab_S1I1}, together with the results computed for $q_{\max} = 800 \, \text{MeV}$ in the case of the HY scheme. All cases predict a lower energy pole, coupling strongly to $\bar{D} \Xi_c$ and sizably to $\bar{D}_s \Lambda_c$. In the case of the HY and CO regularization schemes, this pole appears at 4189~MeV and has a small width (0.29~MeV for HY and slightly larger, 0.73~MeV, for CO). This state appears at 4265~MeV in the DR regularization scheme and, since it lies above the threshold of the $\bar{D}_s \Lambda_c$ channel, to which it couples substantially, its width is much larger, amounting to 20~MeV.
The higher energy pole appears in the range 4400--4435~MeV for the different regularization schemes, couples strongly to $\bar{D} \Xi^\prime_c$ and decays mainly into  $\eta_c\Lambda$ because of its sizable coupling to this channel. This explains the relatively large width obtained for this state, of the order of 20--30~MeV in the case of the HY and DR schemes and, for the reasons explained in the previous section, twice this value in the case of CO.
Employing a cut-off $q_{\max} = 800 \, \text{MeV}$ in the HY scheme, the first and second resonances lower their energies by about 150~MeV and 100~MeV respectively. The higher resonance acquires a much higher width, in spite of the loss of phase space, because of the increased value of its coupling to the $\eta_c\Lambda$ channel.

Our results for the VB states in this sector follow the same trends as the PB ones and are all gathered in the lower half of Table~\ref{tab_S1I1}.

\subsubsection{Sector $S=-1, \, I=1$} \label{sec:S1I2}

\begin{figure*}[ht!]
     \begin{subfigure}[b]{0.32\textwidth}
         \includegraphics[width=\linewidth]{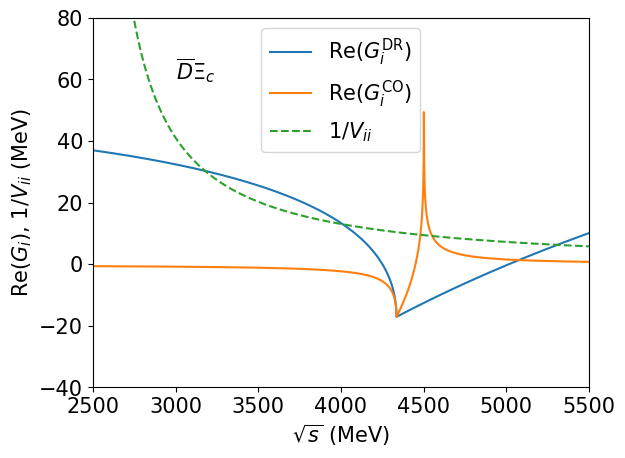}
         \caption{}
         \label{G-V_S1I2_a}
     \end{subfigure}
     \hfill
     \begin{subfigure}[b]{0.32\textwidth}
         \includegraphics[width=\linewidth]{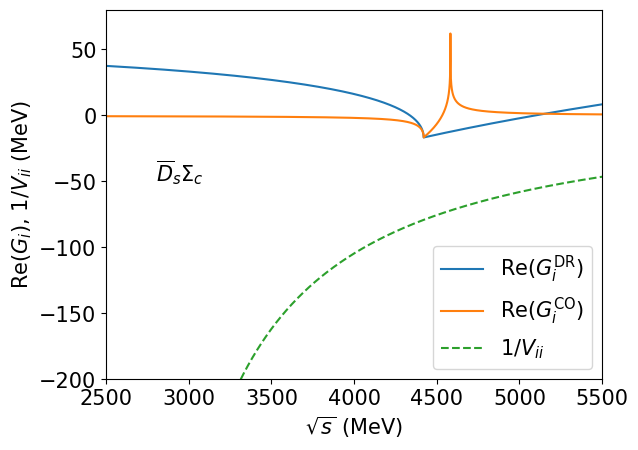}
         \caption{}
         \label{G-V_S1I2_b}
     \end{subfigure}
     \hfill
     \begin{subfigure}[b]{0.32\textwidth}
         \includegraphics[width=\linewidth]{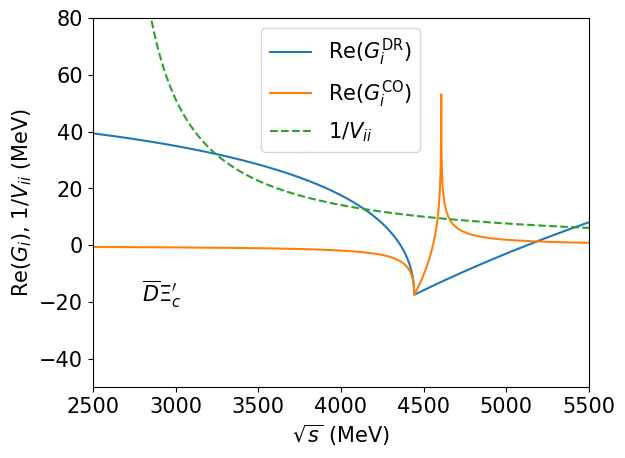}
         \caption{}
         \label{G-V_S1I2_c}
     \end{subfigure}
    \caption{Real part of the loop function in the dimensional (blue solid line) and cut-off (orange solid line) regularization scheme. The inverse potential $1/V_{ii}$ is also plotted. The subindex i stands for the corresponding channel:  $\bar{D}\Xi_c$ [panel (a)], $\bar{D}_s\Sigma_c$ [panel (b)] and $\bar{D} \Xi_c'$ [panel (c)].}
    \label{G-V_S1I2}
\end{figure*}

\begin{figure}[h!]
         \includegraphics[width=\linewidth]{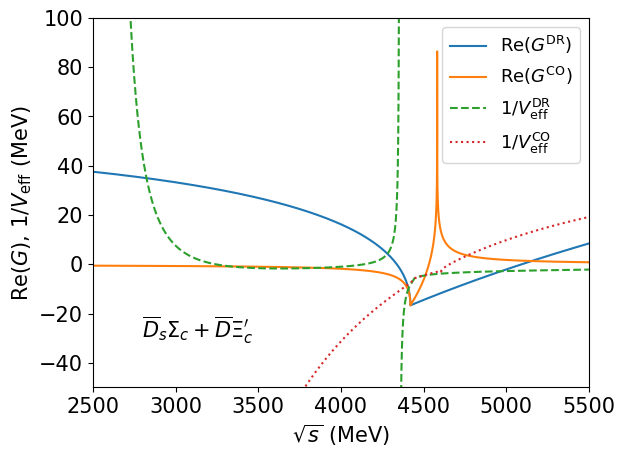}
    \caption{Effective potential constructed with the channels $\bar{D}_s \Sigma_c$ and $\bar{D} \Xi_c'$ employing the DR loop (dashed green line) or the CO one (dotted red line). The real part of the loop functions for the $\bar{D}_s \Sigma_c$ channel, $\text{Re}\,G^{\text{DR}}$ (blue line) and $\text{Re}\,G^{\text{CO}}$ (orange line), are also shown.}
    \label{Veff_S1I2}
\end{figure}

The PB interaction in the  $S=-1, \, I=1$ sector has four coupled-channels: $\eta_c\Sigma$, $\bar{D} \Xi_c$, $\bar{D}_s \Sigma_c$ and $\bar{D} \Xi_c'$. The corresponding T-matrix, shown in Fig.~\ref{fig_PB_S1I2}, has five structures when employing the $G^{\text{DR}}$ loop
(left panel), three structures in the case of $G^{\text{CO}}$ (middle panel) and only one for $G^{\text{HY}}$ (right panel). As in previous sections, the uncoupled approach helps in understanding the origin of the unphysical states obtained in the DR and CO schemes. In Fig.~\ref{G-V_S1I2} we represent $\text{Re}\,G^{\text{DR}}$ (solid blue line) and $\text{Re}\,G^{\text{CO}}$ (solid orange line), together with the inverse of the potential (dashed green line) for $\bar{D}\Xi_c$ (a), $\bar{D}_s\Sigma_c$ (b) and $\bar{D} \Xi_c'$ (c) channels. We observe two intersections of $\text{Re}\,G^{\text{DR}}$ with the inverse potential for the $\bar{D} \Xi_c$ channel, which are tied, after coupled channel effects, to the second and third peaks of the T-matrix in the DR scheme that couple strongly to $\bar{D} \Xi_c$. These structures are then unphysical since they have been generated by a repulsive interaction. In a totally similar manner, the first and fourth peaks of the DR T-matrix, which couple strongly to $\bar{D} \Xi_c'$, are fake states, since they are connected to the two intersections of $\text{Re}\,G^{\text{DR}}$ with the positive inverse potential of the $\bar{D} \Xi_c'$ channel. Turning to the CO case, we can also discard as physical states the second and third peaks of the CO T-matrix, as they are generated, respectively, from the crossing of the positive $\bar{D} \Xi_c$ and $\bar{D} \Xi_c'$ potentials with the spike of their corresponding loop functions. 

The origin of the remaining peak in the DR or CO scattering amplitudes right below the  $\bar{D} \Xi_c'$ threshold, which is the only one appearing in the HY regularization scheme, cannot be understood from an uncoupled analysis based on Fig.~\ref{G-V_S1I2}. Note that the non-zero diagonal kernel coefficients (see Table~\ref{tab:C_S1I2} in Appendix \ref{app:B}) correspond to  either a repulsive interaction or a very tiny attractive one.  However, the 
off-diagonal coefficient for the transition $\bar{D}_s \Sigma_c\rightarrow \bar{D} \Xi'_c $ is sizable and we construct the effective $\bar{D}_s \Sigma_c$ potential using Eq.~\eqref{eq:Veff} from this pair of coupled channels. Its inverse is represented in Fig.~\ref{Veff_S1I2}, when calculated with the $G^{\text{DR}}$ loop (dashed green line) or the $G^{\text{CO}}$ one (dotted red line). This figure should now replace the middle and right panels of Fig.~\ref{G-V_S1I2}.  We now observe two intersections of $\text{Re}\,G^{\text{DR}}$ with the positive branch of the corresponding inverse effective potential, generating the two unphysical states already found in the uncoupled analysis, and one intersection with the negative branch of the inverse potential right below the $\bar{D}_s \Sigma_c$ threshold explaining the physical peak in the DR T-matrix at that location. Similarly, for the CO case, $\text{Re}\,G^{\text{CO}}$ intersects the inverse potential when it is negative right below threshold, thus generating a genuine state, and there is a double crossing at energies around the position of the spike, where the effective potential is positive, thus leading to a wide asymetrical peak in the CO T-matrix corresponding to the unphysical structure already found in the uncoupled analysis. 
Consequently, only a valid physical pole below the $\bar{D}_s \Sigma_c$ threshold remains for the DR and CO schemes, which is the one directly obtained by the HY scheme without the need of an ulterior analysis to discard unphysical structures. Note that this state is generated from the coupled channel dynamics. 

\begin{table}[h!]
    \begin{tabular}{c|c|c|cc}
\hline \hline
         \multicolumn{5}{c}{ }\\[-3mm]
        \multicolumn{5}{c}{ PB interaction ($J^P = \frac{1}{2}^-$) in the $(S,I) = (-1, 1)$ sector }  \\ 
           \multicolumn{5}{c}{ }\\[-3mm]      
        \hline \hline
        &&&& \\[-3mm]
        & $G^{\text{DR}}$ & $G^{\text{CO}}$ & \multicolumn{2}{c}{$G^{\text{HY}}$} \\ \hline
        $q_{\text{max}}$ (MeV) & ~~~~~~600~~~~~~& ~~~~~~600~~~~~~ & ~~~~600~~~~ & 800 \\ \hline
        M (MeV) & 4406& 4413& 4386 & 4283\\
        $\Gamma$(MeV) & 19& 55& 27 & 37 \\ \hline
        & $|g_i|$ & $|g_i|$& $|g_i|$ & $|g_i|$ \\
        $\eta_c \Sigma (4173)$ & 0.82 & 1.05 & 1.13 & 1.64 \\
        $\bar{D} \Xi_c (4336)$& 0.19 & 0.43& 0.25 & 0.52 \\
        $\bar{D}_s \Sigma_c(4421)$ & 1.65& 2.45& 2.45& 3.66 \\
        $\bar{D} \Xi_c'(4444)$ & 1.62& 1.94& 1.92 & 2.65 \\
    \end{tabular}
    \begin{tabular}{c|c|c|cc}
    \hline \hline
             \multicolumn{5}{c}{ }\\[-3mm]
        \multicolumn{5}{c}{VB interaction ($J^P = \frac{1}{2}^-, \frac{3}{2}^-$) in the $(S,I) = (-1, 1)$ sector}   \\ 
           \multicolumn{5}{c}{ }\\[-3mm]      
        \hline \hline
        &&&& \\[-3mm]
        & $G^{\text{DR}}$ & $G^{\text{CO}}$ & \multicolumn{2}{c}{$G^{\text{HY}}$} \\ \hline
        $q_{\text{max}}$ (MeV) & ~~~~~~600~~~~~~& ~~~~~~600~~~~~~ & ~~~~600~~~~ & 800 \\ \hline
        M(MeV) & 4549 &4555& 4527 & 4419 \\
        $\Gamma$(MeV) & 20 &62& 28 & 40 \\ \hline
        & $|g_i|$& $|g_i|$& $|g_i|$& $|g_i|$ \\
        $J\psi \Sigma (4290)$ & 0.82 & 1.08 & 1.14 & 1.67 \\
        $\bar{D}^* \Xi_c (4477)$ & 0.19 &0.46& 0.27 & 0.56 \\
        $\bar{D}_s^* \Sigma_c(4564)$ & 1.69 &2.60& 2.54 & 3.77 \\
        $\bar{D}^* \Xi_c'(4585)$ & 1.62 &2.01& 1.96 & 2.72 \\
\hline \hline
    \end{tabular}
    \caption{Masses, widths and coupling strengths of the genuine poles found in the $S=-1, \, I=1$ sector with the three regularization schemes, $G^{\text{DR}}$, $G^{\text{CO}}$ and $G^{\text{HY}}$.}
    \label{tab_S1I2}    
\end{table}

The results for this sector are summarized in the upper half of Table~\ref{tab_S1I2}. The $S=-1$, $I=1$ resonance appears around 4400~MeV in all three regularization schemes, and couples strongly to the  $\bar{D}_s \Sigma_c$ and $\bar{D} \Xi_c^\prime$ channels. It also couples significantly to $\eta_c \Sigma$, one of its decay channels, giving rise to a sizable width, especially in the CO case. The mass of the resonance decreases by about 100 MeV and its width increases by 10 MeV when $q_{\max} = 800 \, \text{MeV}$ is employed in the HY scheme.

In the case of the VB interaction we find a completely analogous pattern and the properties of the only remaining physical state for the different regularization schemes are collected in the bottom half of Table~\ref{tab_S1I2}.

\subsubsection{Sector  $S=-2, \, I=1/2$} \label{sec:S2I1}
\begin{figure*}[ht!]
    \includegraphics[width=\linewidth]{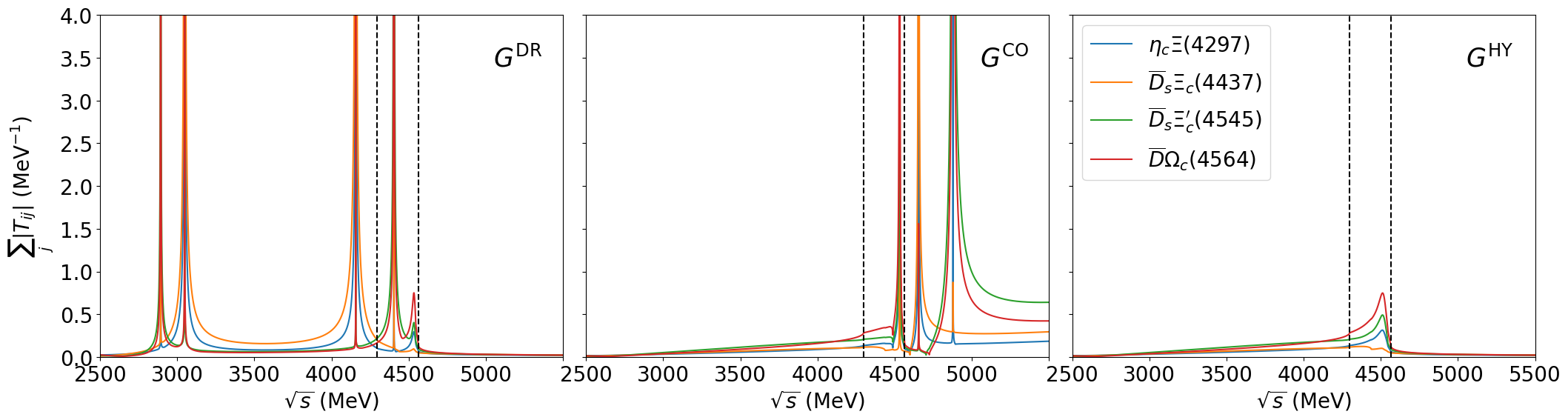}
    \caption{From left to right, the panels illustrate the scattering amplitude results obtained with $G^{\text{DR}}$, $G^{\text{CO}}$ and $G^{\text{HY}}$ for the PB interaction in the $S=-2, \, I=1/2$ sector. All three plots share the same y-axis scale and color legend. Each line represents the sum of the moduli of the transition amplitudes to the i-th final channel. Vertical black dashed lines indicate the thresholds of the lightest and heaviest channels.}
    \label{fig_PB_S2I1}
\end{figure*}

The coupled channels of the PB interaction in the $S=-2, \, I=1/2$ sector are $\eta_c\Xi$, $\bar{D}_s \Xi_c$, $\bar{D}_s \Xi_c'$, and $\bar{D} \Omega_c$. The corresponding T-matrix with the HY prescription presents a single structure, as seen in the right panel of Fig.~\ref{fig_PB_S2I1}, while the shapes of the scattering amplitudes with the DR (left panel) and CO schemes (middle panel) are more complex.

\begin{figure*}[ht!]
     \begin{subfigure}[b]{0.45\textwidth}
         \includegraphics[width=\linewidth]{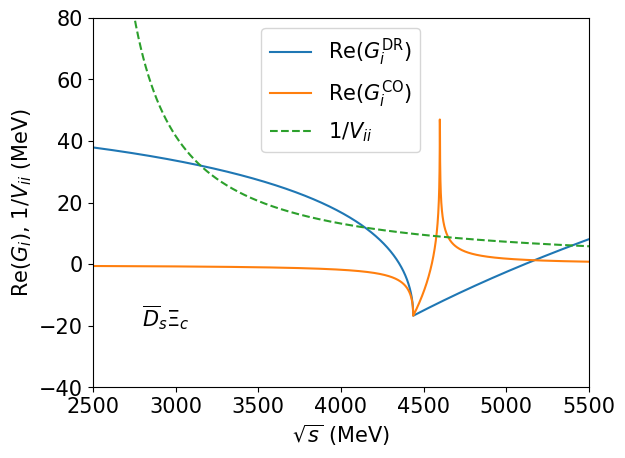}
         \caption{}
         \label{G-V_S2I1_a}
     \end{subfigure}
     \hfill
     \begin{subfigure}[b]{0.45\textwidth}
         \includegraphics[width=\linewidth]{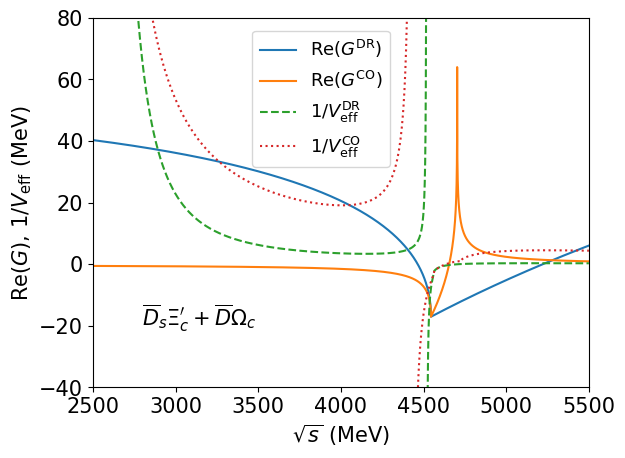}
         \caption{}
         \label{Veff_S2I1}
     \end{subfigure}
    \caption{(a): Real part of the loop function in the DR (blue solid line) and CO (orange solid line) regularization schemes, and the inverse potential in the $\bar{D}_s\Xi_c$ channel. (b): Real part of the loop function in the DR (blue solid line) and CO (orange solid line) schemes for the $\bar{D}_s\Xi_c'$ channel, and inverse $\bar{D}_s \Xi_c'-\bar{D} \Omega_c$ effective potential built with the DR loop (dashed green line) or the CO one (dotted red line).} 
    \label{G-V_S2I1}
\end{figure*}

Notice the similarity of the amplitudes in Fig.~\ref{fig_PB_S2I1} with those shown for the $S = -1, \, I=1$ sector in Fig.~\ref{fig_PB_S1I2}.
This is directly related to the structure of the kernel matrix (see Tables~\ref{tab:C_S1I2} and~\ref{tab:C_S2I1} in Appendix \ref{app:B}). The diagonal coefficients are equal with the following correspondences:
    $\eta_c \Sigma \rightarrow \eta_c \Xi$, 
    $\bar{D} \Xi_c \rightarrow \bar{D}_s \Xi_c$, 
    $\bar{D}_s \Sigma_c \rightarrow \bar{D} \Omega_c$, $\bar{D} \Xi'_c \rightarrow \bar{D}_s \Xi'_c$.
The off-diagonal are all small (proportional to the $\kappa_c$ factor), except for the $\sqrt{2}$ in the $S = -1, \, I=1$ sector, connecting the $\bar{D}_s \Sigma_c$  and $\bar{D} \Xi'_c$  channels, and the  $- \sqrt{2}$  in the $S = -2, \, I = 1/2$ sector, connecting the corresponding  
$\bar{D} \Omega_c $ and $\bar{D}_s \Xi'_c$ ones \footnote{The sign of the off-diagonal potential does not affect since it appears squared in the effective interaction built from the two coupled channels.}. 
\begin{table}[h!]
    \begin{tabular}{c|c|c|cc}
    \hline \hline
     \multicolumn{5}{c}{ }\\[-3mm]
        \multicolumn{5}{c}{PB interaction ($J^P = \frac{1}{2}^-$) in the $(S,I) = (-2,\frac{1}{2})$ sector}   \\ 
           \multicolumn{5}{c}{ }\\[-3mm]      
        \hline \hline
        &&&& \\[-3mm]
        & $G^{\text{DR}}$ & $G^{\text{CO}}$ & \multicolumn{2}{c}{$G^{\text{HY}}$} \\ \hline
        $q_{\text{max}}$ (MeV) & ~~~~600~~~~&~~~~600~~~~ & ~~~~600~~~~ & 800 \\ \hline
        M(MeV) & 4536&4544 & 4515 & 4406 \\
        $\Gamma$(MeV) & 21& 59&  33 & 41 \\ \hline
        & $|g_i|$& $|g_i|$& $|g_i|$& $|g_i|$ \\
        $\eta_c \Xi (4297)$ & 0.82& 1.02& 1.12 & 1.66 \\
        $\bar{D}_s \Xi_c (4437)$&0.20 &0.45 & 0.27 & 0.53 \\
        $\bar{D}_s \Xi_c' (4545)$ & 1.07& 1.70& 1.67 & 2.57 \\
        $\bar{D} \Omega_c(4564)$ & 2.01& 2.63& 2.60 & 3.70 \\
    \end{tabular}
    \begin{tabular}{c|c|c|cc}
    \hline \hline
     \multicolumn{5}{c}{ }\\[-3mm]
        \multicolumn{5}{c}{VB interaction ($J^P = \frac{1}{2}^-, \frac{3}{2}^-$) in the $(S,I) = (-2,\frac{1}{2})$ sector}   \\ 
           \multicolumn{5}{c}{ }\\[-3mm]      
        \hline \hline
        &&&& \\[-3mm]
        & $G^{\text{DR}}$ & $G^{\text{CO}}$ & \multicolumn{2}{c}{$G^{\text{HY}}$} \\ \hline
        $q_{\text{max}}$ (MeV) &~~~~600~~~~&~~~~600~~~~ & ~~~~600~~~~ & 800 \\ \hline
        M(MeV) & 4677&4686& 4654 & 4541 \\
        $\Gamma$(MeV) &22&65& 30 & 44 \\ \hline
        & $|g_i|$& $|g_i|$& $|g_i|$& $|g_i|$ \\
        $J\psi \Xi (4415)$ & 0.81 &1.05& 1.13 & 1.69 \\
        $\bar{D}_s^* \Xi_c (4581)$ & 0.20 &0.48& 0.28 & 0.59 \\
        $\bar{D}_s^* \Xi_c'(4688)$ & 1.10 &1.80& 1.74 & 2.65 \\
        $\bar{D}^* \Omega_c (4705)$ & 2.02 &2.72& 2.66 & 3.80 \\
     \hline \hline
    \end{tabular}
    \caption{Masses, widths and coupling strengths of the genuine poles found in the $S=-2, \, I=1/2$ sector with the three regularization schemes, $G^{\text{DR}}$, $G^{\text{CO}}$ and $G^{\text{HY}}$.}
    \label{tab_S2I1}
\end{table}
Therefore, in this  $S=-2, \, I=1/2$ sector we expect similar arguments to those used in the $S = -1, \, I = 1$ one and only one physical pole should remain in each prescription. 
For completeness, in the left panel of Fig.~\ref{G-V_S2I1} we plot the real parts of the DR and CO loop functions (blue and orange lines, respectively) for the $\bar{D}_s \Xi_c$ channel, together with the inverse of its potential. In the right panel, we present $\text{Re}\,G^{\text{DR}}$ and $\text{Re}\,G^{\text{CO}}$ for the $\bar{D}_s \Xi_c'$ channel, together with the inverse of the effective $V_{11}$ potential, built from the $\bar{D}_s \Xi'_c $ and $\bar{D} \Omega_c$ channels employing either the DR loop (dashed green line) or the CO one (dotted red line). Performing a similar analysis as for the $S = -1, \, I = 1$ sector, the presence of four and two fake poles for the DR and CO cases, respectively, can be clearly inferred from Fig.~\ref{G-V_S2I1}. Only one physical resonance remains and it is generated from a coupled channel mechanism between the $\bar{D}_s \Xi'_c $ and $\bar{D} \Omega_c$ channels, as already noticed in Ref.~\cite{Valera23}.
 The properties of this physical pole are summarized in the upper half of Table~\ref{tab_S2I1}. Completely analogous results are obtained for VB interactions and the results are presented in the bottom half of Table~\ref{tab_S2I1}.

\subsubsection{Summary of all sectors}


\begin{figure}[ht!]
         \includegraphics[width=\linewidth]{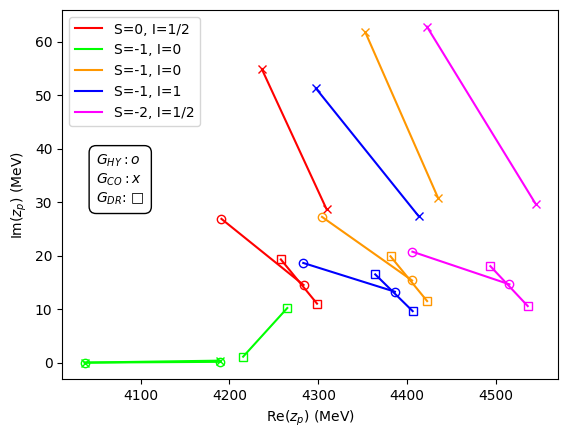}
         \caption{}
         \label{PB_ALL_POLES}
    \caption{Results of the pole positions for the PB  interaction. The marker is related to the type of regularization employed for the loop function and the color is related to the sector. Results with $q_{\max} = 600$ and $800 \, \text{MeV}$ are connected by a straight line of the same color. 
    }
    \label{fig:ALL_POLES}
\end{figure}

We end Section~\ref{Gdr_co_hy_comp} by collecting in Fig.~\ref{fig:ALL_POLES} the positions in the complex plane of all physical poles found with PB interactions.  We present the results obtained with three different markers, which are related to the scheme that was used to regularize the loop function: $G^{\text{HY}}$ (circles), $G^{\text{CO}}$ (crosses) and $G^{\text{DR}}$ (squares). The poles are computed using two different cut-off parameters, $q_{\max} = 600 \, \text{MeV}$  and $q_{\max} = 800 \, \text{MeV}$, and the results lie, respectively, in the rightmost and leftmost end of the straight line that joins them. The color of the line (or the color of the marker) indicates the $(S,I)$ sector in which the pole was found. 

First, we observe the CO states (crosses) to be much wider than those obtained for the other two regularization schemes. This is tied to an artifact in the definition of the second Riemann sheet, as explained before, and the CO width should simply be discarded. 
By comparing the three types of symbols vertically within each sector for a cut-off value of $q_{\max} = 600 \, \text{MeV}$ (rightmost end of the lines), we observe the HY prescription (circle) to produce a state with slightly lower mass than the two other prescriptions, and slightly wider than the DR one (square). This tendency is accentuated for $q_{\max} = 800 \, \text{MeV}$ (leftmost end of the lines). 
The lightest state in the $S=-1, \, I= 0$ sector (indicated with green lines in Fig.~\ref{PB_ALL_POLES}) deviates from this behavior. For a given $q_{\max}$ value, the HY and CO prescriptions predict an identical bound state, the reason being that it is located 
below the thresholds of the channels to which it couples most strongly ($\bar{D}_s \Lambda_c$ and $\bar{D} \Xi_c$), as shown by the coupling strengths in Table~\ref{tab_S1I1}. Consequently, $\text{Re}\,G^{\text{HY}}$ is effectively using $\text{Re}\,G^{\text{CO}}$ in these relevant channels. 
The corresponding states obtained with the DR scheme have a substantially higher masses, and in the case of $q_{max}=600$ MeV even higher than the $\bar{D}_s \Lambda_c$ threshold, a channel that is then open for decay, thus explaining the much higher width of this state.

\subsection{LHGA-WF model and hybrid regularization} \label{LHGA-WF_results}

\begin{figure*}[t!]
     ~\hfil \begin{subfigure}[b]{0.42\textwidth}
         \includegraphics[width=\textwidth]{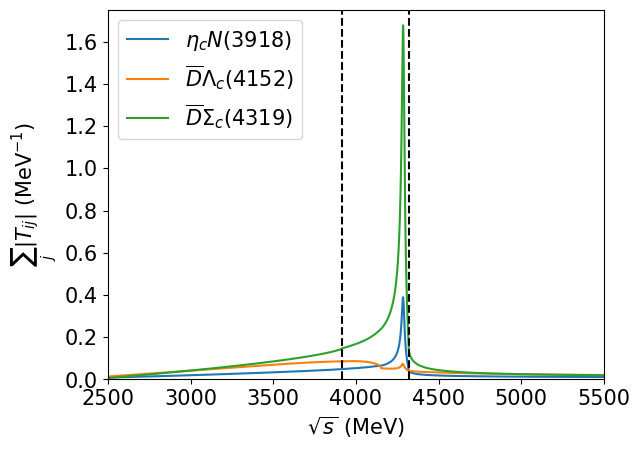}
          \caption{S=0, I=1/2}
     \end{subfigure}
     \hfill
     \begin{subfigure}[b]{0.42\textwidth}
         \includegraphics[width=\textwidth]{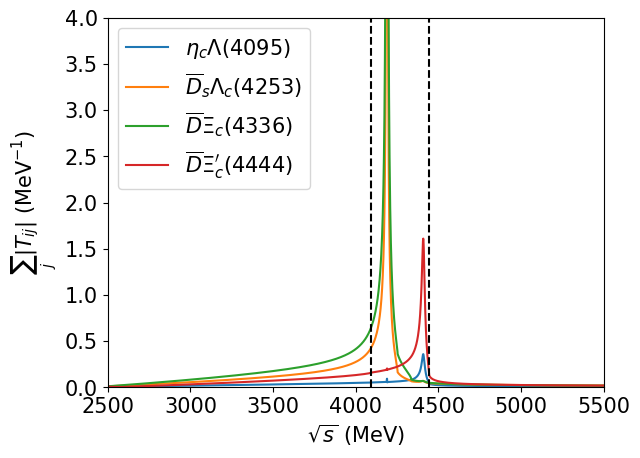}
          \caption{S=-1,I=0}
     \end{subfigure}
          \begin{subfigure}[b]{0.42\textwidth}
         \includegraphics[width=\textwidth]{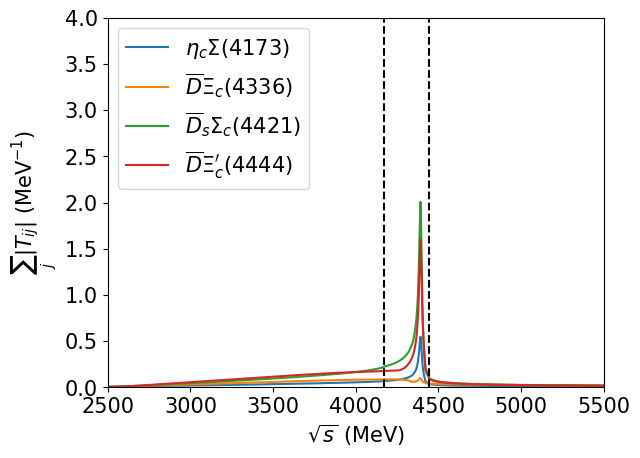}
          \caption{S=-1,I=1}
     \end{subfigure}
     \hfill
     \begin{subfigure}[b]{0.42\textwidth}
         \includegraphics[width=\textwidth]{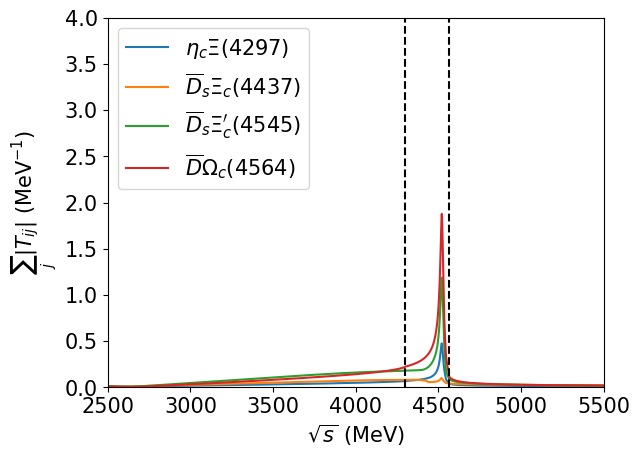}
          \caption{S=-2,I=1/2}
     \end{subfigure}
        \caption{PB scattering amplitudes for different strange and isospin sectors within the LHGA-WF method. Vertical black dashed lines indicate the thresholds of the lightest and heaviest channels.
        }
        \label{fig_WF_PB}
        \hfil~
\end{figure*}

In this section we review the hidden charm states dynamically generated in different strangeness and isospin sectors, employing the LHGA-WF method for the meson-baryon interaction and the hybrid scheme for the loop regularization. 
We will show results for PB and VB interactions in s-wave, involving $J^P = 1/2^+$ baryons, which will be compared to those in the previous section obtained with the LHGA-SU(4) interaction model. We will also report results for PB$^*$ and VB$^*$ interactions, involving $J^P = 3/2^+$ baryons, which can additionally be obtained within the LHGA-WF method and lead to the appearance of new pentaquarks.

All the figures of the scattering amplitudes that appear in this section  are obtained for $q_{\max} = 600 \, \text{MeV}$. However, the mass, width and coupling strengths of the states will be given for $q_{\max} = 600$~MeV and $800 \, \text{MeV}$, as a measurement of the theoretical uncertainty. 

We start presenting the results obtained with the PB interaction. The kernel coefficients within the LHGA-WF method are displayed in Tables \ref{tab:D_S0I1}, \ref{tab:D_S0I2}, \ref{tab:D_S1I1}, \ref{tab:D_S1I2}, \ref{tab:D_S2I1} and \ref{tab:D_S3I1} of Appendix \ref{app:B} for the different strangeness and isospin sectors. The T-matrix amplitudes are shown in Fig.~\ref{fig_WF_PB} for the sectors that present structures tied to dynamically generated states. The corresponding pole positions, widths and coupling strengths are displayed in the upper half of Tables~\ref{tab_WF-S0I1}, \ref{tab_WF-S1I1}, \ref{tab_WF-S1I2} and \ref{tab_WF-S2I1}.

\begin{table}[h!]
    \begin{tabular}{c|c|c}
    \hline \hline
            \multicolumn{3}{c}{}\\[-3mm] 
        \multicolumn{3}{c}{PB interaction ($J^P = \frac{1}{2}^-$) in the $(S,I) = (0,1/2)$ sector}    \\ 
           \multicolumn{3}{c}{ }\\[-3mm]      
        \hline \hline
        && \\[-3mm]
        $q_{\max}$ (MeV) & ~~~~~~~~~~600~~~~~~~~~~ & 800 \\ \hline
        M(MeV) & 4286 & 4211 \\
        $\Gamma$(MeV) & 9 & 17 \\ \hline
        & $|g_i|$& $|g_i|$ \\
        $\eta_c N (3918)$ & 0.66 & 0.96 \\
        $\bar{D} \Lambda_c (4152)$& 0.08 & 0.11 \\
        $\bar{D} \Sigma_c(4319)$ & 2.82 & 4.04 \\
    \end{tabular}
    \begin{tabular}{c|c|c}
    \hline \hline
                \multicolumn{3}{c}{}\\[-3mm] 
        \multicolumn{3}{c}{VB interaction ($J^P = \frac{1}{2}^-, \frac{3}{2}^-$) in the $(S,I) = (0,1/2)$ sector}    \\ 
           \multicolumn{3}{c}{ }\\[-3mm]      
        \hline \hline
        && \\[-3mm]
        $q_{\max}$ (MeV) & ~~~~~~~~~~600~~~~~~~~~~ & 800 \\ \hline
        M(MeV) & 4425 & 4347 \\
        $\Gamma$(MeV) & 10 & 17 \\ \hline
        & $|g_i|$& $|g_i|$ \\
        $J/\psi N (4035)$ & 0.67 & 0.97 \\
        $\bar{D}^* \Lambda_c (4293)$ & 0.08 & 0.11 \\
        $\bar{D}^* \Sigma_c(4460)$ & 2.91 & 4.17 \\
    \hline \hline
    \end{tabular}
    \caption{Masses, widths and coupling strengths of the poles found in the $S=0, \, I=1/2$ sector using LHGA-WF.}
        \label{tab_WF-S0I1}
\end{table}
\begin{table}[h!]
    \begin{tabular}{c|cc|cc}
    \hline\hline
               \multicolumn{5}{c}{ }\\[-3mm]   
        \multicolumn{5}{c}{PB interaction ($J^P = \frac{1}{2}^-$) in the $(S,I) = (-1,0)$ sector}  \\ 
           \multicolumn{5}{c}{ }\\[-3mm]      
        \hline \hline
        &\multicolumn{2}{c|}{ }&\multicolumn{2}{c}{ } \\[-3mm]
        $q_{\max}$ (MeV) & \multicolumn{2}{c|}{~~~~~~~~~~600~~~~~~~~~~} & \multicolumn{2}{c}{800}  \\ \hline
        M(MeV) & ~~~4189~~~ & 4409 & ~~~4037~~~ & 4332 \\
        $\Gamma$(MeV) & $\sim 0$ & 10 & $\sim 0$ & 17 \\ \hline
        & $|g_i|$ & $|g_i|$ & $|g_i|$ & $|g_i|$ \\
        $\eta_c \Lambda (4095)$ & 0.08 & 0.64& 0.06 & 0.94 \\
        $\bar{D}_s \Lambda_c (4253)$& 2.29 & 0.06 &3.28 & 0.08\\
        $\bar{D} \Xi_c(4336)$ & 3.58 & 0.06& 4.73& 0.11 \\
        $\bar{D} \Xi_c'(4444)$ & 0.05 & 2.82& 0.01& 4.04 \\
    \end{tabular}
    \begin{tabular}{c|cc|cc}
    \hline\hline
        \multicolumn{5}{c}{VB interaction ($J^P = \frac{1}{2}^-, \frac{3}{2}^-$) in the $(S,I) = (-1,0)$ sector}\\ 
           \multicolumn{5}{c}{ }\\[-3mm]      
        \hline \hline
        &\multicolumn{2}{c|}{ }&\multicolumn{2}{c}{ } \\[-3mm]
        $q_{\max}$ (MeV) & \multicolumn{2}{c|}{~~~~~~~~~~600~~~~~~~~~~} & \multicolumn{2}{c}{800}  \\ \hline
        M(MeV) & ~~~4328~~~& 4548 & ~~~~4171~~~~ & 4467\\
        $\Gamma$(MeV) & $\sim 0$ & 10 & $\sim 0$ & 17 \\ \hline
        & $|g_i|$ & $|g_i|$& $|g_i|$ & $|g_i|$ \\
        $J/\psi \Lambda (4212)$ & 0.08 & 0.65 & 0.05& 0.95 \\
        $\bar{D}_s^* \Lambda_c (4397)$& 2.25 & 0.07 & 3.38 & 0.08\\
        $\bar{D}^* \Xi_c(4477)$ & 3.50 & 0.06 & 4.86 & 0.12 \\
        $\bar{D}^* \Xi_c'(4585)$ & 0.04 & 2.92& 0.01 & 4.17 \\
    \hline\hline
    \end{tabular}
    \caption{Masses, widths and coupling strengths of the poles found in the $S=-1, \, I=0$ sector using LHGA-WF. }
    \label{tab_WF-S1I1}
\end{table}

\begin{table}[h!]
    \begin{tabular}{c|c|c}
        \hline \hline
          \multicolumn{3}{c}{ }\\[-3mm] 
        \multicolumn{3}{c}{PB interaction ($J^P = \frac{1}{2}^-$) in the $(S,I) = (-1,1)$ sector}  \\ 
           \multicolumn{3}{c}{ }\\[-3mm]      
        \hline \hline
        $q_{\max}$ (MeV) & ~~~~~~~~~~600~~~~~~~~~~ & 800 \\ \hline
        M(MeV) & 4392 & 4313\\
        $\Gamma$(MeV) & 8 & 13 \\ \hline
        & $|g_i|$& $|g_i|$ \\
        $\eta_c \Sigma (4173)$ & 0.62 & 0.90\\
        $\bar{D} \Xi_c (4336)$& 0.08& 0.15\\
        $\bar{D}_s \Sigma_c(4421)$ &2.25 & 3.32 \\
        $\bar{D} \Xi_c'(4444)$ & 1.79& 2.44 \\
    \end{tabular}
    \begin{tabular}{c|c|c}
            \hline \hline
                  \multicolumn{3}{c}{ }\\[-3mm] 
        \multicolumn{3}{c}{VB interaction ($J^P = \frac{1}{2}^-, \frac{3}{2}^-$) in the $(S,I) = (-1,1)$ sector}  \\ 
           \multicolumn{3}{c}{ }\\[-3mm]      
        \hline \hline
        $q_{\max}$ (MeV) & ~~~~~~~~~~~~600~~~~~~~~~~~~ & 800 \\ \hline
        M(MeV) & 4533 & 4451 \\
        $\Gamma$(MeV) & 9 & 14 \\ \hline
        & $|g_i|$& $|g_i|$ \\
        $J\psi \Sigma (4290)$ & 0.63 & 0.91 \\
        $\bar{D}^* \Xi_c (4477)$ & 0.09 & 0.16\\
        $\bar{D}_s^* \Sigma_c(4564)$ & 2.34 & 3.43\\
        $\bar{D}^* \Xi_c'(4585)$ & 1.83 & 2.50 \\
        \hline \hline
    \end{tabular}
    \caption{Masses, widths and coupling strengths of the poles found in the $S=-1, \, I=1$ sector using LHGA-WF.}
    \label{tab_WF-S1I2}
\end{table}
\begin{table}[h!]
    \begin{tabular}{c|c|c}
            \hline \hline
     \multicolumn{3}{c}{ }\\[-3mm]   
        \multicolumn{3}{c}{PB interaction ($J^P = \frac{1}{2}^-$) in the $(S,I) = (-2,\frac{1}{2})$ sector}  \\ 
           \multicolumn{3}{c}{ }\\[-3mm]      
        \hline \hline
        $q_{\max}$ (MeV) & ~~~~~~~~~~~600~~~~~~~~~~~ & 800 \\ \hline
        M(MeV) & 4521 & 4441\\
        $\Gamma$(MeV) & 9 & 15 \\ \hline
        & $|g_i|$& $|g_i|$ \\
        $\eta_c \Xi (4297)$ & 0.62 & 0.91 \\
        $\bar{D}_s \Xi_c (4437)$& 0.09 & 0.12 \\
        $\bar{D}_s \Xi_c' (4545)$ & 1.52 & 2.31 \\
        $\bar{D} \Omega_c(4564)$ & 2.41 & 3.37 \\
    \end{tabular}
    \begin{tabular}{c|c|c}
            \hline \hline
                       \multicolumn{3}{c}{ }\\[-3mm] 
        \multicolumn{3}{c}{VB interaction ($J^P = \frac{1}{2}^-, \frac{3}{2}^-$) in the $(S,I) = (-2,\frac{1}{2})$ sector}   \\ 
           \multicolumn{3}{c}{ }\\[-3mm]      
        \hline \hline
        $q_{\max}$ (MeV) & ~~~~~~~~~~~~600~~~~~~~~~~~~ & 800 \\ \hline
        M(MeV) & 4661 & 4577 \\
        $\Gamma$(MeV) & 10 & 15 \\ \hline
        & $|g_i|$& $|g_i|$ \\
        $J\psi \Xi (4415)$ & 0.63 & 0.92\\
        $\bar{D}_s^* \Xi_c (4581)$ & 0.09 & 0.15\\
        $\bar{D}_s^* \Xi_c'(4688)$ & 1.60 & 2.39\\
        $\bar{D}^* \Omega_c (4705)$& 2.48 & 3.45\\
                \hline \hline
    \end{tabular}
    \caption{Masses, widths and coupling strengths of the poles found in the $S=-2, \, I=1/2$ sector using LHGA-WF.}
    \label{tab_WF-S2I1}
\end{table}

Upon comparing with the amplitudes on the right panel of Figs.~
\ref{fig_PB_S0I1}, \ref{fig_PB_S1I1}, \ref{fig_PB_S1I2} and \ref{fig_PB_S2I1}, obtained with the LHGA-SU(4) model and the hybrid regularization, we immediately observe that the LHGA-WF states appear at practically the same energy but are much narrower. This systematics can be easily traced back to the kernel coefficients. On the one hand, the diagonal coefficients of Tables \ref{tab:C_S0I1}, \ref{tab:C_S0I2}, \ref{tab:C_S1I1}, \ref{tab:C_S1I2}, 
\ref{tab:C_S2I1} and \ref{tab:C_S3I1}, corresponding to the LWGA-SU(4) model, are the same as those in the corresponding tables for the LWGA-WF model. This explains why the resonances appear at similar energies in both models, for sectors that generate them via sizable attractive interactions indicated by positive coefficients, such as $S=0, \, I=1/2$ and $S=-1, \, I=0$. The states generated by a coupled-channel mechanism, such as those in sectors $S=-1, \, I=1$ and $S=-2, \, I=1/2$, also appear at similar energies in both models because the relevant coupled-channel coefficient has the same size, $\sqrt{2}$. The discrepancies in the widths can be traced to differences in off-diagonal coefficients involving the $\eta_c$ meson. As argued in Subsect.~\ref{models} these are a factor $\sqrt{3}$ smaller in size in the LHGA-WF model compared to the LHGA-SH(4) ones. Consequently, the LWGA-WF resonances turn out to be roughly a factor three narrower than the LWGA-SU(4) ones, because they couple strongly to states involving charmed mesons and decay mostly to states having the $\eta_c$.

Analogous results are obtained for the VB interaction, since the kernel coefficients coupling the different states are the same as those for the PB interaction, simply replacing the pseudoscalar mesons with the corresponding vector ones. In this case we omit the plot of the amplitudes and just present the properties of the poles in the bottom half of Tables~\ref{tab_WF-S0I1}-\ref{tab_WF-S2I1}.

Let us now focus on the states that can be generated from meson-baryon interactions involving B$^*$ baryons with $J^{P} = 3/2^+$, which are naturally included in the LHGA-WF method. The resulting pentaquark states from these PB$^*$ and VB$^*$ interactions differ in spin-parity assignments from those arising in the PB and VB cases. The states generated from $\text{PB}^*$ interactions in S-wave possess $J^P = 3/2^-$, while those from $\text{VB}^*$ interactions can have spin-parities $J^P = 1/2^-, 3/2^-, 5/2^-$ and appear degenerate in the present work, as no spin-dependent terms are included in the interaction potential.

 The kernel coefficients computed using the LHGA-WF method are compiled in Appendix \ref{app:C}. Following Refs.~\cite{Debastiani, Roca, Wang}, we assume that the $\text{PB}^*$ and $\text{VB}^*$ loop functions are given by Eq.~\eqref{loop_function}, which we regularize using the hybrid scheme.

\begin{figure*}[ht!]
     \hfil \begin{subfigure}[b]{0.42\textwidth}
         \includegraphics[width=\textwidth]{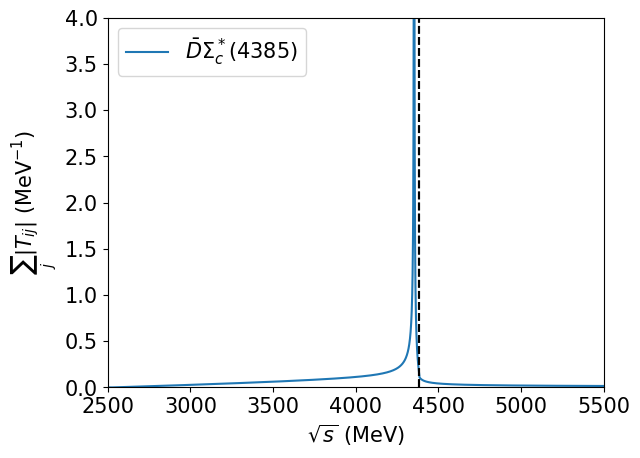}
         \caption{$S=0, I=1/2$}
         \label{PBast_S0I1}
     \end{subfigure}
     \hfill
     \begin{subfigure}[b]{0.42\textwidth}
         \includegraphics[width=\textwidth]{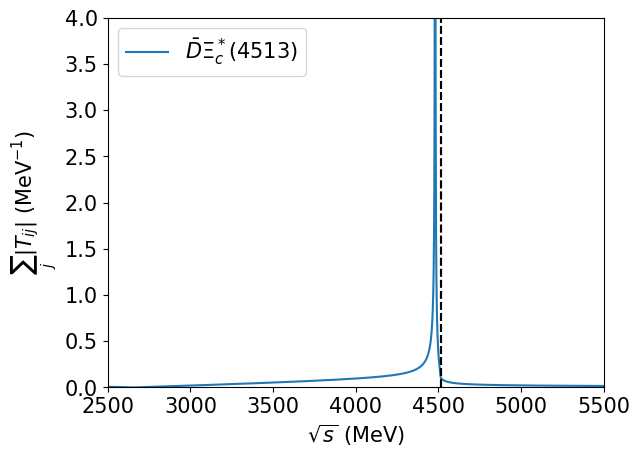}
         \caption{$S=-1, I=0$}
         \label{PBast_S1I1}
     \end{subfigure}
     \hfill
     \begin{subfigure}[b]{0.42\textwidth}
         \includegraphics[width=\textwidth]{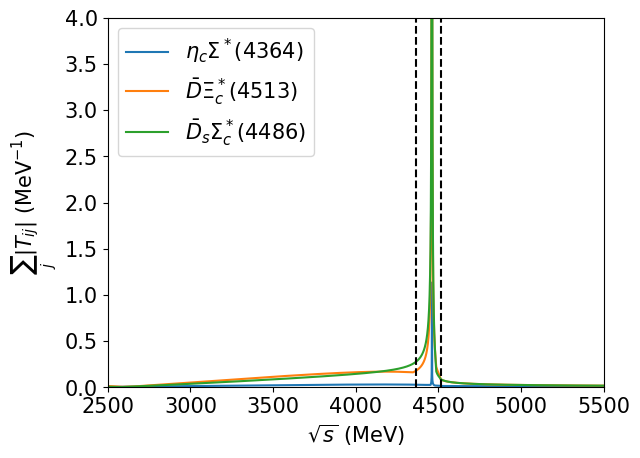}
         \caption{$S=-1, I=1$}
         \label{PBast_S1I2}
     \end{subfigure}
     \hfill
     \begin{subfigure}[b]{0.42\textwidth}
         \includegraphics[width=\textwidth]{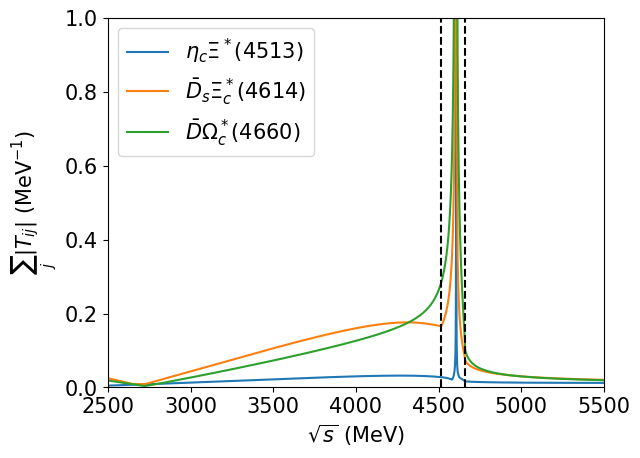}
         \caption{$S=-2, I=1/2$}
         \label{PBast_S2I1}
     \end{subfigure}
     \hfil 
        \caption{$\text{PB}^*$ scattering amplitudes for different strange and isospin sectors within the LHGA-WF method. Vertical black dashed lines indicate the thresholds of the lightest and heaviest channels, or the single channel in cases (a) and (b).}
        \label{fig_PBast}
\end{figure*}

We present in Fig.~\ref{fig_PBast} the $\text{PB}^*$ scattering amplitudes with hidden charm obtained with $q_{\max} = 600 \, \text{MeV}$ for all the strangeness and isospin sectors that produce a resonant structure. A similar plot, but shifted to higher energies, is found in the case of the $\text{VB}^*$  interaction. The corresponding pole properties and coupling strengths are collected in Tables~\ref{tab:PBast} and~\ref{tab:VBast} for $q_{\max} = 600 \, \text{MeV}$ and $q_{\max} = 800 \, \text{MeV}$. We focus our discussion on the states found in the $\text{PB}^*$ interactions, as those from the $\text{VB}^*$ interactions are analogous, just shifted to higher energies.

\begin{table}[h!]
    \begin{tabular}{c|ccccc|cc}
        \cline{1-3} \cline{6-8}
        \multicolumn{3}{c}{} & & & \multicolumn{3}{c}{ }  \\[-2.2ex] \cline{1-3} \cline{6-8}
         \multicolumn{8}{c}{ } \\[-3mm]
        \multicolumn{3}{c}{PB$^*$, $(S,I) = (0,\frac{1}{2})$} & & & \multicolumn{3}{c}{PB$^*$, $(S,I) = (-1, 0)$} \\
                 \multicolumn{8}{c}{ } \\[-3mm]
        \cline{1-3} \cline{6-8}
        \multicolumn{3}{c}{} & & & \multicolumn{3}{c}{ }  \\[-2.2ex] \cline{1-3} \cline{6-8}
          & \multicolumn{2}{c}{ } & & & & \multicolumn{2}{c}{ }  \\[-3mm]
        $q_{\max}$ (MeV) & ~~~600~~~ & 800 & & &$q_{\max}$ (MeV) & ~~~600~~~ & 800 \\ \cline{1-3} \cline{6-8}
          & \multicolumn{2}{c}{ } & & & & \multicolumn{2}{c}{ }  \\[-3mm]
        M(MeV) & 4353 & 4284 &&& M(MeV) & 4480 & 4410 \\
        $\Gamma$(MeV) & $\sim 0$ & $\sim 0$ &&& $\Gamma$(MeV) & $\sim 0$ & $\sim 0$ \\ \cline{1-3} \cline{6-8}
        & $|g_i|$& $|g_i|$ &&& & $|g_i|$& $|g_i|$ \\ 
        $\bar{D} \Sigma_c^* (4385)$ & 3.16 &  4.61 &&& $\bar{D} \Xi_c^* (4513)$ & 3.20 & 4.72 \\ \cline{1-3} \cline{6-8}
        \multicolumn{3}{c}{} & & & \multicolumn{3}{c}{ }  \\[-2.2ex] \cline{1-3} \cline{6-8}
        \multicolumn{8}{c}{} \\
        \multicolumn{8}{c}{} \\
          \cline{1-3} \cline{6-8}
        \multicolumn{3}{c}{} & & & \multicolumn{3}{c}{ }  \\[-2.2ex] \cline{1-3} \cline{6-8}
         \multicolumn{8}{c}{ } \\[-3mm]
        \multicolumn{3}{c}{PB$^*$, $(S,I) = (-1,1)$} & & & \multicolumn{3}{c}{PB$^*$, $(S,I) = (-2, \frac{1}{2})$} \\
                 \multicolumn{8}{c}{ } \\[-3mm]
        \cline{1-3} \cline{6-8}
        \multicolumn{3}{c}{} & & & \multicolumn{3}{c}{ }  \\[-2.2ex] \cline{1-3} \cline{6-8}
          & \multicolumn{2}{c}{ } & & & & \multicolumn{2}{c}{ }  \\[-3mm]
        $q_{\max}$ (MeV) & 600 & 800 & & &$q_{\max}$ (MeV) & ~~~600~~~ & 800 \\ \cline{1-3} \cline{6-8}
          & \multicolumn{2}{c}{ } & & & & \multicolumn{2}{c}{ }  \\[-3mm]
        M(MeV) & 4461 & 4390 &&& M(MeV) & 4607 & 4538\\
        $\Gamma$(MeV) & 0.03 & 0.01 &&& $\Gamma$(MeV) & 0.12 & 0.02 \\ \cline{1-3} \cline{6-8}
        & $|g_i|$& $|g_i|$ &&& & $|g_i|$& $|g_i|$ \\ 
        $\eta_c \Sigma^* (4364)$ & 0.05 & 0.03 &&& $\eta_c \Xi^* (4513)$ & 0.09 & 0.06 \\
        $\bar{D} \Xi_c^* (4513)$ & 2.04 & 2.87  &&& $\bar{D}_s \Xi_c^* (4614)$ & 1.34 & 2.54 \\
        $\bar{D}_s \Sigma_c^* (4486)$ & 2.49 & 3.86 &&& $\bar{D} \Omega_c^* (4660)$ & 2.70 & 4.01\\
        \cline{1-3} \cline{6-8}
        \multicolumn{3}{c}{} & & & \multicolumn{3}{c}{ }  \\[-2.2ex] \cline{1-3} \cline{6-8}
    \end{tabular}
    \caption{Masses, widths and coupling strengths of the poles found in the $\text{PB}^*$ interaction for different charmless sectors.}
    \label{tab:PBast}
\end{table}

The $S = 0, \, I = 1/2$ and $S = -1, \, I =0$ sectors, each involve a single channel, $\bar{D} \Sigma_c^*$ and the $\bar{D} \Xi^*$, respectively. The kernel coefficients, listed in Tables~\ref{tab:S0I1_Bast} and~\ref{tab:S1I1_Bast}, are attractive and identical, producing similar scattering amplitudes profiles and a bound state, located at at $M = 4353 \, \text{MeV}$ in the case of $S = 0, \, I = 1/2$ and at $M = 4480 \, \text{MeV}$ for $S = -1, \, I =0$.

The other two sectors, $S = -1, \, I = 1$ and $S=-2, \, I = 1/2$, display a three-coupled channel interaction kernel with identical coefficients, as can be seen from Tables~\ref{tab:S1I2_Bast} and~\ref{tab:S2I1_Bast}. The diagonal coefficients imply a repulsive interaction or a tiny attraction, indicating that the structure observed in Figs.~\ref{PBast_S1I2} and \ref{PBast_S2I1} is tied to a coupled-channel mechanism due the sizable non-diagonal coefficient of value $-\sqrt{2}$ coupling the two higher energy meson-baryon channels. This is completely analogous to what has been found for the PB interactions in the same sectors discussed in the previous LHGA-SU(4) section. As seen in Table~\ref{tab:PBast} the PB$^*$ states appear at $M =4461 \, \text{MeV}$ and $M = 4607 \, \text{MeV}$, for $S = -1, \, I = 1$ and $S=-2, \, I = 1/2$, respectively, they couple strongly to the two highest mass states, and they are very narrow, due to their small coupling to the only open decay channel involving the $\eta_c$ meson. 

The $\text{VB}^*$ interaction leads to completely analogous results, just exchanging the role of the $\eta_c$, $\overline{D}$ and $\overline{D}_s$ mesons by the $J/\psi$, $\overline{D}^*$ and $\overline{D}_s^*$ ones. The properties of the resulting pentaquarks are summarized in Table \ref{tab:VBast}.

\begin{table}[h!]
    \begin{tabular}{c|ccccc|cc}
        \cline{1-3} \cline{6-8}
        \multicolumn{3}{c}{} & & & \multicolumn{3}{c}{ }  \\[-2.2ex] \cline{1-3} \cline{6-8}
         \multicolumn{8}{c}{ } \\[-3mm]
        \multicolumn{3}{c}{VB$^*$, $(S,I) = (0,\frac{1}{2})$} & & & \multicolumn{3}{c}{VB$^*$, $(S,I) = (-1, 0)$} \\
                 \multicolumn{8}{c}{ } \\[-3mm]
        \cline{1-3} \cline{6-8}
        \multicolumn{3}{c}{} & & & \multicolumn{3}{c}{ }  \\[-2.2ex] \cline{1-3} \cline{6-8}
          & \multicolumn{2}{c}{ } & & & & \multicolumn{2}{c}{ }  \\[-3mm]
        $q_{\max}$ (MeV) & ~~~600~~~ & 800 & & &$q_{\max}$ (MeV) & ~~~600~~~ & 800 \\ \cline{1-3} \cline{6-8}
          & \multicolumn{2}{c}{ } & & & & \multicolumn{2}{c}{ }  \\[-3mm]
        M(MeV) & 4492 & 4420 &&& M(MeV) & 4619 & 4546 \\
        $\Gamma$(MeV) & $\sim 0$ & $\sim 0$ &&& $\Gamma$(MeV) & $\sim 0$ & $\sim 0$ \\ \cline{1-3} \cline{6-8}
        & $|g_i|$& $|g_i|$ &&& & $|g_i|$& $|g_i|$ \\ 
        $\bar{D}^* \Sigma_c^* (4526)$ & 3.34 & 4.93 &&& $\bar{D}^* \Xi_c^* (4654)$ & 3.35 & 4.94 \\ 
        \cline{1-3} \cline{6-8}
        \multicolumn{3}{c}{} & & & \multicolumn{3}{c}{ }  \\[-2.2ex] \cline{1-3} \cline{6-8}
        \multicolumn{8}{c}{} \\
        \multicolumn{8}{c}{} \\
              \cline{1-3} \cline{6-8}
        \multicolumn{3}{c}{} & & & \multicolumn{3}{c}{ }  \\[-2.2ex] \cline{1-3} \cline{6-8}
         \multicolumn{8}{c}{ } \\[-3mm]
        \multicolumn{3}{c}{VB$^*$, $(S,I) = (-1,1)$} & & & \multicolumn{3}{c}{VB$^*$, $(S,I) = (-2, \frac{1}{2})$} \\
                 \multicolumn{8}{c}{ } \\[-3mm]
        \cline{1-3} \cline{6-8}
        \multicolumn{3}{c}{} & & & \multicolumn{3}{c}{ }  \\[-2.2ex] \cline{1-3} \cline{6-8}
          & \multicolumn{2}{c}{ } & & & & \multicolumn{2}{c}{ }  \\[-3mm]
        $q_{\max}$ (MeV) & 600 & 800 & & &$q_{\max}$ (MeV) & ~~~600~~~ & 800 \\ \cline{1-3} \cline{6-8}
          & \multicolumn{2}{c}{ } & & & & \multicolumn{2}{c}{ }  \\[-3mm]
        M(MeV) & 4602 & 4528 &&& M(MeV) & 4747 & 4674 \\
        $\Gamma$(MeV) & 0.03 & 0.01 &&& $\Gamma$(MeV) & 0.11 & 0.02 \\ \cline{1-3} \cline{6-8}
        & $|g_i|$& $|g_i|$ &&& & $|g_i|$& $|g_i|$ \\ 
        $J/\psi \Sigma^* (4481)$ & 0.04 & 0.03 &&& $J/\psi \Xi^* (4630)$ & 0.09 & 0.05 \\
        $\bar{D}^* \Xi_c^* (4654)$ & 2.11 & 2.99  &&& $\bar{D}_s^* \Xi_c^* (4758)$ & 1.47 & 2.67 \\
        $\bar{D}_s^* \Sigma_c^* (4629)$ & 2.63 & 4.05 &&& $\bar{D}^* \Omega_c^* (4801)$ & 2.86 & 4.18 \\
        \cline{1-3} \cline{6-8}
        \multicolumn{3}{c}{} & & & \multicolumn{3}{c}{ }  \\[-2.2ex] \cline{1-3} \cline{6-8}
    \end{tabular}
    \caption{Masses, widths and coupling strengths of the poles found in the $\text{VB}^*$ interaction for different charmless sectors.}
    \label{tab:VBast}
\end{table}

We conclude this section by noting that, by including the $J^{P}=3/2^+$ baryons, the LHGA-WF method has provided two new types of interactions ($\text{PB}^*$ and $\text{VB}^*$). Unitarizing the scattering amplitudes and regularizing the loop functions within the $G^{\text{HY}}$ scheme, we have found four hidden-charm $\text{PB}^*$ states (Table~\ref{tab:PBast}) and four hidden-charm $\text{VB}^*$ states (Table~\ref{tab:VBast}) that could not be predicted within the LHGA-SU(4) model.

\subsection{Comparison with other references in the literature} 
\label{model_comp}
In this section, we compare our results within the hybrid regularization scheme with those reported in other theoretical works with interactions inspired on t-channel vector-meson exchange. In Table~\ref{LHGA-SU4WF_theo} we collect the mass (M), width ($\Gamma$) and spin-parity ($J^{P}$) of all the pentaquark states we found in Section~\ref{Gdr_co_hy_comp} using the LHGA-SU(4) model (column labeled LHGA-SU(4)) and in Section~\ref{LHGA-WF_results} using the LHGA-WF method (column labeled LHGA-WF), obtained with $q_{\max} = 600 \, \text{MeV}$. Notice that the $J^{P}$ values next to each state defines whether it is found in PB $(1/2^-)$, VB $(1/2^-,3/2^-)$ , $\text{PB}^*$ $(3/2^-)$ or $\text{VB}^*$ $(1/2^-,3/2^-,5/2^-)$ interactions. The first set of results in the LHGA-WF column correspond to PB and VB states, while the second set displays the  $\text{PB}^*$ and $\text{VB}^*$ ones. The last column in Table~\ref{LHGA-SU4WF_theo} presents the values of $M$, $\Gamma$ and $J^{P}$ from other theoretical works.




\begin{table*}
    \begin{tabular}{|c|ccc|ccc|cccc|}
        \cline{2-11}
         \multicolumn{1}{c}{}& \multicolumn{3}{|c|}{LHGA-SU(4)} & \multicolumn{3}{|c|}{LHGA-WF} & \multicolumn{4}{c|}{Other works} \\ \hline
                  & & & &   &   & &  & &  & \\[-3mm]   
        $(S, \, I)$ & $M$ & $\Gamma$ & $J^P$& $M$ & $\Gamma$ & $J^P$  & $M$ & $\Gamma$ & $J^P$ & Ref. \\ \hline 
         & & & &   &   & &  & &  & \\[-3mm]              
        \multirow{10}{*}{$(0, \, 1/2)$} & $4283$ & $21$ & $1/2^-$ & $4286$ & $9$ & $1/2^-$ & 4265 & 23 & $1/2^-$ &\multirow{2}{*}{\cite{Molina:2010}}\\
        & $4422$ & $30$ & $1/2^-,3/2^-$ & $4425$ & $10$ & $1/2^-, 3/2^-$ & 4415 & 19 & $1/2^-,3/2^-$ & \\
        & & & &   &   & & & & & \\
        & & & & 4353 & 0      & $3/2^-$  & 4306 & 15 & $1/2^-$ &\multirow{7}{*}{\cite{Xiao:2019}}\\
        & & & & 4492 & 0      & $1/2^-, 3/2^-, 5/2^-$  &     4453 & 23 & $1/2^-$ & \\
        & & & &   &   &   & 4452 & 3 & $3/2^-$ & \\
        & & & & &&&4374 & 14 & $3/2^-$ & \\ 
        & & & & &&&4520 & 22 &  $1/2^-$ & \\
        & & & & &&&4519 & 14 & $3/2^-$ & \\
        & & & & &&&4519 & 0  & $5/2^-$ & \\  \hline
         & & & &   &   & &  & &  & \\[-3mm]

        \multirow{11}{*}{$(-1, \, 0)$} & $4189$ & $0.3$ & $1/2^-$ & $4189$ & $0$ & $1/2^-$ & 4210 & 6 & $1/2^-$ &\multirow{4}{*}{\cite{Molina:2010}}\\
                     & 4328 & $0.3$ & $1/2^-,3/2^-$ & $4328$ & $0$ & $1/2^-, 3/2^-$ & 4368 & 6 & $1/2^-,3/2^-$ & \\
                     & 4405 & $31$ & $1/2^-$ & 4409 & $10$ & $1/2^-$ &4398 & 16 & $1/2^-$ &  \\
                     & 4544 & $31$ & $1/2^-,3/2^-$ & 4548 & $10$ & $1/2^-, 3/2^-$ & 4547 & 13 & $1/2^-,3/2^-$ & \\
                      & & & & & &  &  &  &  & \\
                     & & & & 4480 & $0$ & $3/2^-$& 4277 & 15 & $1/2^-$ &\multirow{10}{*}{\cite{Xiao:2019gjd}}\\
                     & & & & 4619  & $0$ & $1/2^-, 3/2^-, 5/2^-$ & 4430  & 16 & $1/2^-$ & \\
                     & & & & & &  & 4430  & 15 & $3/2^-$ & \\
                     & & & & & &  & 4437 & 2 & $1/2^-$ & \\
                     & & & & & &  & 4581 & 5 & $1/2^-$ & \\
                     & & & & & &  & 4581 & 1 & $3/2^-$ & \\
                     & & & & & &  & 4507 & 2 & $3/2^-$ & \\
                     & & & & & &  & 4651 & 5 & $1/2^-$ & \\
                     & & & & & &  & 4651 & 3 & $3/2^-$ & \\
                     & & & & & &  & 4651 & 0 & $5/2^-$ & \\
                      & & & & & &  &  &  &  & \\
                     & & & &  &  & & 4199 & 0.2 & $1/2^-$ &\multirow{6}{*}{\cite{Feijoo_Veff}}\\
                     & & & &  &  &  & 4338 & 0.2 & $1/2^-, 3/2^-$ & \\
                     & & & & & &  & 4423 & 15 & $1/2^-$ & \\
                     & &&&&&&4566 & 31 & $1/2^-, 3/2^-$ & \\
                     & &&&&&&4489 & 0 & $3/2^-$&  \\
                     & &&&&&&4627 & 0 & $1/2^-, 3/2^-, 5/2^-$& \\ \hline
         & & & &   &   & &  & &  & \\[-3mm]

        \multirow{4}{*}{$(-1, \, 1)$} & $4386$ & $27$ & $1/2^-$ & $4392$ & $8$ & $1/2^-$&&&&\\
                     & $4527$ & $28$ & $1/2^-,3/2^-$& $4533$ & $9$ & $1/2^-, 3/2^-$&&&&\\
                      & & & & & & & & & & \\  
                    & & & & $4461$ & $0$ & $3/2^-$&&&&\\
                     & & & &$4602$ & $0$ & $1/2^-, 3/2^-, 5/2^-$&&&&\\ \hline
         & & & &   &   & &  & &  & \\[-3mm]   
                     
        \multirow{6}{*}{$(-2, \, 1/2)$} & $4515$ & $33$ & $1/2^-$ &$4521$ & $9$ & $1/2^-$ & 4493 & 74 & $1/2^-$ &\multirow{2}{*}{\cite{Valera23}}\\ 
                       & $4654$ & $30$ & $1/2^-,3/2^-$ &  $4661$ & $10$ & $1/2^-, 3/2^-$ & 4633 & 80 & $1/2^-,3/2^-$ & \\   
                        & &&&  &  &  &  &  &   & \\
                       & &&& 4607 & 0 & $3/2^-$ &4535 & 9 & $1/2^-$ &\multirow{4}{*}{\cite{Roca}}\\
                       & &&&4747 & 0 & $1/2^-, 3/2^-, 5/2^-$ & 4675 & 10 & $1/2^-, 3/2^-$ &\\
                       & &&& & & & 4602 & 0 & $3/2^-$& \\
                       & &&& &&&4743 & 0 & $1/2^-, 3/2^-, 5/2^-$ & \\ \hline

    \end{tabular}
    \caption{Compilation of the predicted states using the LHGA-SU(4) and LHGA-WF approaches with the hybrid loop function and $q_{\max} = 600 \, \text{MeV}$ (columns labeled as LHGA-SU(4) and LHGA-WF, respectively). The masses ($M$) and widths ($\Gamma$) are in units of MeV. The table also includes results from other theoretical works (colum labeled as Other works).}
    \label{LHGA-SU4WF_theo}
\end{table*}

In the $S=0, \, I = 1/2$ sector, we found a PB state ($J^{P} = 1/2^-$), a twice-degenerate VB state ($J^{P} = 1/2^-,3/2^-$) and, additionally within the LHGA-WF model, a $\text{PB}^*$ state ($3/2^-$) and a triply degenerate $\text{VB}^*$ state ($J^{P} = 1/2^-,3/2^-,5/2^-$). In Refs.~\cite{Molina:2010} and~\cite{Xiao:2019}, meson-baryon interactions in the $S=0, \, I = 1/2$ sector were studied using UChPT and the dimensional regularization scheme, with $\mu = 1000 \, \text{MeV}$, but with different subtraction constants: $a_k(\mu) \sim -2.3$ and $a_k(\mu) \sim -2.1$, respectively. The two works also differ in the computation of the interaction kernel: while Ref.~\cite{Molina:2010} used the same LHGA-SU(4) model as the present work, Ref.~\cite{Xiao:2019} combined heavy quark spin and local hidden gauge symmetries within an SU(8) spin-flavor symmetry framework.
In both cases, the PB state strongly couples to the $\bar{D} \Sigma_c$ channel. Notably, our calculations using $G^{\text{HY}}$ find a similar state with $M = 4283 \, \text{MeV}$ (LHGA-SU(4)) or 
$M = 4286 \, \text{MeV}$ (LHGA-WF), lying between the values reported in Refs.~\cite{Molina:2010} and~\cite{Xiao:2019}.

The authors of Ref.~\cite{Molina:2010} also reported a state with $M=4415 \, \text{MeV}$, degenerate in spin-parity, $J^{P} = 1/2^-, 3/2^-$, and strongly coupled to the $\bar{D}^* \Sigma_c$ channel. Equivalently, Ref.~\cite{Xiao:2019} reported two states, since the degeneracy is broken in their model: one with $M = 4453 \, \text{MeV}$ and $J^{P} = 1/2^-$ , and another with $M = 4452 \, \text{MeV}$ and $J^P = 3/2^-$, both strongly coupled to the $\bar{D}^* \Sigma_c$ channel. The corresponding state found in our work has a mass of 
$4422 \, \text{MeV}$ (LHGA-SU(4)) or 
$4425 \, \text{MeV}$ (LHGA-WF) and spin-parity degeneracy ($J^P = 1/2^-, 3/2^-$). Once again, our result is between those reported in Refs.~\cite{Molina:2010} and~\cite{Xiao:2019}, lending further support to the reliability of the $G^{\text{HY}}$ scheme introduced in this work.

Finishing with the $S=0, \, I = 1/2$ sector, a $J^P = 3/2^-$ state at $4374 \, \text{MeV}$, strongly coupled to $\bar{D} \Sigma_c^*$,  was also reported in Ref.~\cite{Xiao:2019}, which is similar to our $\text{PB}^*$ state found at  $4353 \, \text{MeV}$. Additional three very close-by states coupling strongly to the $\bar{D}^* \Sigma^*$ channel, namely $4520.45 \, \text{MeV}$ $(J^{P} = 1/2^-)$, $4519.01 \, \text{MeV}$ $(J^{P} = 3/2^-)$ and $4519.23 \, \text{MeV}$ $(J^{P} = 5/2^-)$ where found in Ref.~\cite{Xiao:2019}, which would correspond to our spin-degenerate state $\text{VB}^*$ state at $4492 \, \text{MeV}$ ($J^P = 1/2^-,3/2^-, 5/2^-$).

Proceeding to the $S=-1, \, I = 0$ sector, let us start mentioning that the pioneer results of Ref.~\cite{Molina:2010} should be directly comparable to what we find with the LHGA-SU(4) method. Two states are found with the PB interaction and two more (twice degenerate) with the VB one. The minor differences found in the position and widths of theses states with respect to those in Ref.~\cite{Molina:2010} should be simply attributed to the slight differences in the loop prescriptions. 
We also display in Table~\ref{LHGA-SU4WF_theo}
the results of Ref.~\cite{Xiao:2019gjd}, where the interaction was built imposing heavy quark spin symmetry combined with the LHGA-WF method to derive the unknown low energy constants. That work reports the existence of the same ten states in this $S=-1, \, I = 0$ sector as in the present work, but with the spin degeneracy slightly broken due to the coupling of channels involving pseudoscalar mesons with related channels involving vector mesons, as well as channels involving $1/2^+$ baryons with related channels involving  $3/2^+$ baryons. It is seen that, with respect to ours, the states in Ref.~\cite{Xiao:2019gjd} appear at higher energies, which is attributed to the dimensional regularization scheme adopted there. This is especially
noteworthy for the three lower energy states, which appear above the threshold of a channel to which they couple substantially ($\bar{D}_s\Lambda_c$ or $\bar{D}_s^*\Lambda_c$) and therefore develop a substantial width. The recent results of 
Ref.~\cite{Feijoo_Veff} where obtained with the LHGA-WF model and a cut-off regularization scheme, using $q_{\max} = 600 \, \text{MeV}$. It is important to mention that, to avoid the spiked structure that appears in the range of energies explored, the real part of the $\eta_c \Lambda$ $G^{\text{CO}}$ loop was set to zero, while the imaginary part kept the unitary expression of Eq.~(\ref{Ghy}). Again, we find a one-to-one correspondence between all PB, VB, $\text{PB}^*$ and $\text{VB}^*$ states found, ours being slightly more bound. The only difference worth mentioning is the width of $\Gamma = 10$~MeV that we find for the higher energy VB state located at $M = 4548 \, \text{MeV}$, which is notably smaller than the value of 31~MeV reported in Ref.~\cite{Feijoo_Veff} for the VB state found at 4566~MeV. The discrepancy may arise from a treatment of the $J/\psi\Lambda$ loop similar to that of the $\eta_c \Lambda$ in the PB sector mentioned above, as the state lies about 350 MeV above the $J/\psi \Lambda$ threshold, although this is not explicitly mentioned in Ref.~\cite{Feijoo_Veff}.

We next compare the results in the $S=-2, \, I =1/2$ sector to other works in the literature. We recall that the existence of a dynamically resonance in this sector was predicted recently in Ref.~\cite{Valera23} employing the LHGA-SU(4) model for the kernel-coefficients and a cut-off regularization scheme. As emphasized there, and corroborated here, this state is generated by a coupled-channel mechanism that had been overlooked until then. Later, the state was also obtained in Ref.~\cite{Roca} within the LHGA-WF method and a cut-off regularization scheme.
The resonances found in Ref.~\cite{Valera23} for the PB and VB interactions are just slightly lower in mass but twice wider than the ones found in the present work with the LHGA-SU(4) method. This behavior was already discussed in Section~\ref{Gdr_co_hy_comp} and attributed to an anomaly developed by the cut-off loop function in the second Riemann sheet.
Our LHGA-WF states found in this sector are very similar, in location and width, to  those reported in Ref.~\cite{Roca}. This resemblance is surprising, given that one expects the cut-off method employed in Ref.~\cite{Roca} to yield larger width values. Although not explicitly mentioned, the results in Ref.~\cite{Roca} might have been obtained with the prescription of neglecting the real part the loop for the lower energy channels, as in Ref.~\cite{Feijoo_Veff}.

Regarding the $S=-1, \, I = 1$ sector, no recent theoretical studies have reported any state. The last results we were able to find are those of 2005 in Ref.~\cite{Lutz}, where a comprehensive study of dynamical generated pentaquark states across various charmed and charmless sectors using the dimensional regularization scheme was performed. A PB state located at $M = 3602 \, \text{MeV}$ was found in the $C=0, \, S=-1, \, I = 1$ sector, which is in clear disagreement with our state at $4392 \, \text{MeV}$. However, the low mass of the state found in Ref.~\cite{Lutz} suggests that it is most probably a fake pole rather than a physical resonance, as can easily be inferred from Fig.~\ref{fig_PB_S1I2}. 
We recall that, in spite of repulsive or tiny attractive diagonal kernel coefficients that discourage the search of resonances in this sector, a physical state appears when the coupling between the $\bar{D}_s \Sigma_c$ and $\bar{D}\Xi'_c$ channels is considered, as the corresponding kernel coefficient related to the transition $\bar{D}_s \Sigma_c \rightarrow\bar{D}\Xi'_c$, $\sqrt{2}$, is large enough to generate an attractive effective potential.  Finding this state is challenging in the commonly employed regularization schemes, as in Ref.~\cite{Molina:2010}, since the physical structure may go unnoticed in front of the dominant size of the unphysical peaks, as clearly shown in the left and middle panels of Fig.~\ref{fig_PB_S1I2}. 
 We therefore conclude that employing the hybrid loop function $G^{\text{HY}}$ is crucial for identifying physical states, as it eliminates uncertainties caused by the unphysical structures generated by the $G^{\text{DR}}$ and $G^{\text{CO}}$ schemes.

In summary, we have shown that all the relevant states reported in the literature are recovered within the hybrid prescription introduced in this work. 
 We predict new states in the $S=-1, \, I = 1$ sector, never reported before, which may have been obscured by the unphysical structures arising in the scattering amplitude when using the $G^{\text{DR}}$ and $G^{\text{CO}}$ schemes.

\subsection{Comparison with experimental data} 
\label{data_comp}

\begin{table*}
    \begin{tabular}{|c|ccc|ccc|ccccc|}
        \cline{2-12}
         \multicolumn{1}{c}{}& \multicolumn{3}{|c|}{LHGA-SU(4)} & \multicolumn{3}{|c|}{LHGA-WF} & \multicolumn{5}{c|}{EXP} \\ \hline
        $(S, \, I)$ & $M$  & $\Gamma$ & $J^P$& $M$ & $\Gamma$ & $J^P$ & State & $M$ & $\Gamma$ & $J^P$& Ref. \\ \hline               &  & & & &   &   & &  & &  & \\[-3mm]

        \multirow{5}{*}{$(0, \, 1/2)$} & $4283$ & $21$ & $1/2^-$ & $4286$ & $9$ & $1/2^-$ & $P_{c\bar{c}}(4380)^+$ & $4380\pm30$ & $210\pm 90$ & - &\cite{Aaij2015}\\
        & $4422$ & $30$ & $1/2^-,3/2^-$ &   $4425$ & $10$ & $1/2^-, 3/2^-$  & $P_{c\bar{c}}(4312)^+$ & $4312^{+7}_{-0.9}$ & $10 \pm 5$ & - &\cite{Aaij2019} \\
         & & & &   &     & & $P_{c\bar{c}}(4440)^+$ & $4440.3^{+4}_{-5}$ & $20.6\pm4.9^{+8.7}_{-10.1}$& -&\cite{Aaij2019}\\
        & & & & $4353$ & $0$      & $3/2^-$ & $P_{c\bar{c}}(4457)^+$ &$4457.3^{+4}_{-1.8}$ & $6.4\pm2^{+5.7}_{-1.9}$ & -&\cite{Aaij2019} \\
        & & & & $4492$ & $0$      & $1/2^-, 3/2^-, 5/2^-$ & $P_{c\bar{c}}(4337)^+$ & $4337^{+7}_{-4}{}^{+2}_{-2}$ & $29^{+26}_{-12}{}^{+14}_{-14}$ & -&\cite{Aaij2021}\\
         & & & &   &   & &  & &  && \\[-3mm]   
\hline                 & & & &   &   & &  & &  && \\[-3mm]

        \multirow{7}{*}{$(-1, \, 0)$} & $4189$ & $0.3$ & $1/2^-$ & $4189$ & $0$ & $1/2^-$ & ~~$P_{c\bar{c}s}(4338)^0$~~~ & $4338.2\pm0.8$ & $7.0\pm 1.8$ & $1/2^-$ &\cite{Aaij2022}\\
                     & $4328$ & $0.3$ & $1/2^-,3/2^-$ & $4328$ & $0$ & $1/2^-, 3/2^-$ & $~~P_{c\bar{c}s}(4459)^0$~~~ & $4458.8^{+6}_{-3.1}$ & $17.3 \pm 6.5^{+8}_{-5.7}$& - &\cite{Aaij2020}\\
                     & $4405$ & $31$ & $1/2^-$ & $4409$ & $10$ & $1/2^-$ &  ~~$P_{c\bar{c}s}(4459)^0$~~~ & $4471.7 \pm 4.8 \pm 0.6$ & $22 \pm 13 \pm3$ & - & \cite{Belle:2025pey}\\
                     & $4544$ & $31$ & $1/2^-,3/2^-$ & $4548$ & $10$ & $1/2^-, 3/2^-$ & & & &&\\
                     & &&&  &  &  & &&&&\\
                     & &&& $4480$ & $0$ & $3/2^-$  & &&&&\\
                     & &&&$4619$ & $0$ & $1/2^-, 3/2^-, 5/2^-$ & &&&&\\ \hline                 & & & &   &   & &  & &  && \\[-3mm]

        \multirow{5}{*}{$(-1, \, 1)$} & $4386$ & $27$ & $1/2^-$ & $4392$ & $8$ & $1/2^-$ & &&&& \\
                     & $4527$ & $28$ & $1/2^-,3/2^-$& $4533$ & $9$ & $1/2^-, 3/2^-$ & &&&&\\
                                          & &&& &  &  &&&&& \\
                     & &&&$4461$ & $0$ & $3/2^-$ &&&&& \\
                     & &&&$4602$ & $0$ & $1/2^-, 3/2^-, 5/2^-$ &&&&& \\ \hline                 & & & &   &   & &  & &  & &\\[-3mm]   
     
        \multirow{5}{*}{$(-2, \, 1/2)$} & $4515$ & $33$ & $1/2^-$ &$4521$ & $9$ & $1/2^-$ & && &&\\ 
                       & $4654$ & $30$ & $1/2^-,3/2^-$ & $4661$ & $10$ & $1/2^-, 3/2^-$ & &&&& \\
                       & &&&   &  &  &&&&& \\
                       & &&&  $4607$ & $0$ & $3/2^-$  &&&&& \\
                       & &&& $4747$ & $0$ & $1/2^-, 3/2^-, 5/2^-$ &&&&& \\ \hline
    \end{tabular}
    \caption{Compilation of the predicted states using the LHGA-SU(4) and LHGA-WF approaches with the hybrid loop function and $q_{\max} = 600 \, \text{MeV}$ (columns labeled as LHGA-SU(4) and LHGA-WF, respectively). The masses ($M$) and widths ($\Gamma$) are in units of MeV. The table also includes the experimentally observed exotic states (column labeled as EXP).}
    \label{LHGA-SU4WF_exp}
\end{table*}

Finally, we display again our results in
Table~\ref{LHGA-SU4WF_exp} together with the pentaquark states reported by the PDG \cite{PDG2024}: $P_{c\bar{c}}(4380)^+$\cite{Aaij2015}, $P_{c\bar{c}}(4312)^+$, $P_{c\bar{c}}(4440)^+$, $P_{c\bar{c}}(4457)^+$\cite{Aaij2019}, $P_{c\bar{c}s}(4459)^0$\cite{Aaij2020,Belle:2025pey} and $P_{c\bar{c}s}(4338)^0$\cite{Aaij2022}, as well as the $P_{c\bar{c}}(4337)^+$ structure seen in $J/\psi\, p$ and $J/\psi \,\bar{p}$ invariant mass distributions from the decay $B^0_s \to J/\psi \,p\,\bar{p}$ \cite{Aaij2021}. We collected the masses and widths of these states, but for most of them the $J^{P}$ assignment is unknown. No hidden charm pentaquarks in the $S=-1, \, I = 1$ and $S=-2, \, I=1/2$ sectors have been discovered yet. 

A direct inspection of Table~\ref{LHGA-SU4WF_exp} indicates the associations of some of the the $P_{c\bar{c}}^+$ states with our predictions in the $S=0,\, I=1/2$ sector. 
First, the $P_{c\bar{c}}(4380)^+$ state cannot be identified with any of our results due to its large width, although it must be noticed that the newer LHCb analysis~\cite{Aaij2019} was insensitive to the existence of such wide states. 
The state most likely to be associated with the lower energy pentaquark $P_{c\bar{c}}(4312)^+$ (with $\Gamma = 10\pm5 \, \text{MeV}$)~\cite{Aaij2019} is the PB state appearing at $M = 4283 \, \text{MeV}$ ($\Gamma = 21 \, \text{MeV}$) in the LHGA-SU(4) model or at $M = 4286 \, \text{MeV}$ ($\Gamma = 9 \, \text{MeV}$) in the LHGA-WF one, coupled strongly to $\bar{D}\Sigma_c$. Compared to experiment, the predicted masses are around $30\, \text{MeV}$ lower and the predicted widths lie within $+2\sigma$ for the LHGA-SU(4) model and $-1\sigma$ for the LHGA-WF one. 
The next two states, $P_{c\bar{c}}(4440)^+$ ($\Gamma = 20.6 \pm 4.9^{+8.7}_{-10.1} \, \text{MeV}$) and $P_{c\bar{c}}(4457)^+$ ($\Gamma = 6.4\pm2^{+5.7}_{-1.9} \, \text{MeV}$) found in~\cite{Aaij2019}, may correspond to the spin-parity degenerate VB state at $4422 \, \text{MeV}$ ($\Gamma = 30 \, \text{MeV}$) in the LHGA-SU(4) model or at $4425 \, \text{MeV}$ ($\Gamma = 10 \, \text{MeV}$) in the LHGA-WF one, coupling strongly to $\bar{D}^*\Sigma_c$. The slightly lower predicted masses could increase with a fine tuning of the $q_{\text{max}}$ parameter. As for their widths, the predictions of the LHGA-WF method are compatible with the experimental values within the uncertainties, while the LHGA-SU(4) model does not accommodate to the narrower width of the $P_{c\bar{c}}(4457)^+$.

An interesting situation arises for the $P_{c\bar c}(4337)$ state, reported in Ref.~\cite{Aaij2021} at $M=4337^{+7}_{-4}{}^{+2}_{-2}$~MeV and $\Gamma=29^{+26}_{-12}{}^{+14}_{-14}$~MeV. 
Several recent references \cite{Husken:2024rdk,Merk:2026xye}
share the opinion that a molecular interpretation of $P_{c\bar c}(4337)$ is unlikely since it lies about $40$ MeV above the $\bar{D}^* \Lambda_c$ threshold. This argument is  misleading, since we find two possible ways of being interpreted as a bound meson-baryon pair.
Firstly, the heaviest channel for the VB interaction in this sector is $\bar{D}^* \Sigma_c$, the threshold of which is at $4460$ MeV, and therefore there is absolutely no problem in generating a bound state in the range of the $P_{c\bar c}(4337)$. Indeed, our VB model for $q_{\text{max}}=800$ MeV predicts a state strongly coupled to $\bar{D}^* \Sigma_c$,  having a mass of 
$4326 \, \text{MeV}$ (LHGA-SU(4)) or 
$4347 \, \text{MeV}$ (LHGA-WF) and a width comparable to the experimental value within uncertainties in both cases.   
Secondly, the $P_{c\bar c}(4337)$ pentaquark can also be 
interpreted as a PB$^*$ state. Indeed, in the LHGA-WF model we find a resonance at $4353 \, \text{MeV}$  with $\Gamma = 0 \, \text{MeV}$ width, strongly coupled to $\bar{D} \Sigma_c^*$, the threshold of which is 4385 MeV.  We note that the
model of Ref.~\cite{Xiao:2019}, which combined heavy quark spin and local hidden gauge symmetries and permits the mixing of PB, VB, PB$^*$ and VB$^*$ states, finds a sizable coupling of this state to the $J/\psi N$ decay channel, a fact that would explain the large experimental width measured, albeit with substantial errors.

Similarly, the $P_{c\bar{c}s}^0$ pentaquarks can be associated with some of our predicted states in the $S=-1,\, I=0$ sector. The narrow $P_{c\bar{c}s}(4338)^0$ ($\Gamma = 7.0 \pm1.8 \, \text{MeV}$) state is closest in mass to the VB state appearing at $4328 \, \text{MeV}$ and a negligible width in both models, strongly coupled to the $\bar{D}_s\Lambda_c$, $\bar{D}^*\Xi_c$ channels.  The experimental width is not reproduced. Allowing $\text{VB} \rightarrow \text{PB}$ transitions may not increase the width substantially in this case, as argued in Ref.~\cite{Feijoo_Veff}, but possible three body PVB channels (out of the scope of our theoretical scheme) may provide additional decaying sources. 

Regarding the $P_{c\bar{c}s}(4459)^0$ pentaquark, there are at the moment in the literature two slightly different estimations of its mass and width, as shown in Table \ref{LHGA-SU4WF_exp}.  Following Ref.~\cite{Feijoo_Veff}  
this pentaquark could be associated to the PB state found at $4405 \, \text{MeV}$ ($\Gamma = 31\, \text{MeV}$) or $4409 \, \text{MeV}$ ($\Gamma = 10 \, \text{MeV}$) within the  LHGA-SU(4) or LHGA-WF models, respectively. 
Reducing the $q_{\text{max}}$ parameter brings the resonance closer to the $\bar{D}\Xi_c'$ threshold at 4444~MeV and, due to its strong coupling to this channel, the corresponding amplitude in the real axis develops a sizable Flatté effect there \cite{Flatte:1976xu}. 
Recall that, within a Breit-Wigner form and close to the resonance mass, the amplitude is inversely proportional to its width. Therefore, as soon as the resonance mass crosses the $\bar{D}\Xi_c'$ threshold with the lowering of $q_{\text{max}}$, there is a sudden increase of the width, which grows with energy and produces an abrupt distorsion of the amplitude: the resonance  gets apparently stuck at this threshold and acquires a peculiar line-shape which is narrower than the actual position of the pole would suggest \cite{Ramos:2002xh}. In this sense, identifying our PB state to the $P_{c\bar{c}s}(4459)^0$ LHCb pentaquark located at $4458.8 \, \text{MeV}$ \cite{Aaij2020} is rather questionable, since the model underpredicts the mass by 13~MeV in the best of the cases. This interpretation becomes even more unlikely if the heavier recent mass measurement of this state by the Belle collaboration, $M=4471.7 \, \text{MeV}$ \cite{Belle:2025pey}, is confirmed. 
 
The alternative and more probable choice in our opinion is to associate the $P_{c\bar{c}s}(4459)^0$ pentaquark to the $J^P=3/2^+$
$\text{PB}^*$ state with $M = 4480 \, \text{MeV}$ ($\Gamma = 0 \, \text{MeV}$) obtained within the LHGA-WF method, coupling strongly to $\bar{D}\Xi_c^*$. In this case, the experimental width would probably be associated to three-body decaying channels, as the couplings to other two body configurations of VB type are very small \cite{Xiao:2019gjd}.

In summary, we have shown that six out of the seven observed pentaquarks with hidden charm can be reproduced using the hybrid regularization scheme. In all cases, the predictions that could be associated with experimental states were obtained at slightly different masses, so further parameter fitting would be necessary for a more accurate description.

\section{Conclusions}
\label{conclusions}
In this work, we have done a comprehensive study of pentaquarks in the hidden charm sector predicted by theoretical models of the meson-baryon interaction based on the exchange of vector mesons. By unitarizing the scattering amplitude, we investigated the dynamically generated states arising across different strangeness and isospin sectors. The unitarization procedure requires the regularization of the meson-baryon loop function, commonly done using either a cut-off  or a dimensional regularization scheme. We, however, have introduced a novel hybrid loop function, which combines both dimensional and cut-off regularizations. This approach enables a cleaner unitarization of the scattering amplitude by avoiding the appearance of unphysical poles, while keeping the properties of the genuine dynamically generated resonances unaltered.

In Section~\ref{Gdr_co_hy_comp}, we reviewed the states generated  employing the LHGA-SU(4) model for the meson-baryon interaction. We analyzed the scattering amplitudes computed with both the cut-off and dimensional regularization methods and showed that these predict a different number of structures. Using the uncoupled approach, we were able to interpret the nature of some structures as unphysical since they result from a repulsive interaction. 

However, not all peaks in the T-matrix could be explained by means of the uncoupled approach and were found to be dynamically generated due to a strong coupling between essentially two channels. In this cases, a more complex analysis, based in the computation of an effective one-channel potential, was needed to understand the nature of these peaks. By combining both the uncoupled approach and the results from the effective potential, we were able to fully interpret the shape of the scattering amplitude, identifying and discarding unphysical structures. After this filtering, both the dimensional regularization and cut-off approaches retained the same number of physical resonances. 

Remarkably, the hybrid scheme proposed here predicts the same number of physical states without the appearance of fictitious structures. A direct comparison of the masses and widths of the generated states shows the results from the dimensional regularization and the hybrid schemes to be in good agreement. On the other hand, the cut-off scheme generates states with similar masses but with consistently larger widths, a fact that we identified with an artifact of the cut-off loop function in the second Riemann sheet. 

In Section~\ref{LHGA-WF_results} the LHGA-WF method was adopted to study pentaquarks generated from PB, VB, $\text{PB}^*$ and $\text{VB}^*$ interactions and employing the hybrid loop function. When applicable, we compared these results with those from the LHGA-SU(4) model under the same regularization scheme. The predicted masses are similar but the LHGA-WF method systematically provides reduced widths, an effect easily traceable to differences in the kernel coefficients and directly related to the SU(3) substructure retained in the LHGA-WF method as opposed to the full SU(4) symmetry of the LHGA-SU(4) approach.    

In Section~\ref{model_comp}, we compared the generated states within the LHGA-SU(4) and LHGA-WF models with other theoretical works to validate the use of the hybrid loop function. No significant discrepancies were found.
Furthermore, upon comparing in Section~\ref{data_comp} our results with the experimentally observed pentaquarks in the hidden charm sector, we conclude that the absolute majority of these states (six out of seven) could be assigned with the states predicted with the LGHA-SU(4) or LHGA-WF methods, although further parameter tuning would be necessary for a more accurate description. The state $P_{c\bar{c}}(4380)^+$ is not reproduced by any of these models due to its large width.

In the same section, we noted that no theoretical predictions can be found in the literature for the $S = -1, \, I = 1$ sector. This lack of information may stem from the fact that other research groups were discouraged upon noticing that the diagonal kernel coefficients lead to repulsive interactions. Moreover, it is also possible that the presence of unphysical structures in the scattering amplitudes obtained with both the dimensional regularization and cut-off schemes obscured the identification of the genuine states. If that is the case, the hybrid loop function proposed in this work proves to be very valuable as it substantially simplifies the shape of the scattering amplitudes by only preserving the physical poles.

In summary, by successfully eliminating unphysical artifacts, the hybrid loop function method provides consistent predictions across all possible strangeness and isospin sectors and has revealed the existence of new pentaquarks with $S=-1,\, I =1$.
We hope that the results of this work will stimulate the search of hidden charm pentaquarks by some experimental collaborations, such as LHCb at CERN or Belle and Belle II at KEK.

\section*{Acknowledgments}
 This work is supported by
 MICIU/AEI/10.13039/ 501100011033  through grants PID2023-147112NB-C21 and through the Maria de Maeztu Center of
Excellence award to the Institute of Cosmos Sciences, grant CEX2024-001451-M.

\appendix
\renewcommand{\thetable}{A\arabic{table}}
\setcounter{table}{0}
\begin{table*}[ht!]
	\begin{minipage}[t]{0.48\textwidth}
		\caption{$C^{S=0,I=\frac{1}{2}}$}
		\begin{tabular}{c}
			$
			\begin{blockarray}{ccc}
				\eta_c N & \bar{D} \Lambda_c & \bar{D}\Sigma_c\\
				\begin{block}{[ccc]}
					0 & -\sqrt{\frac{3}{2}}\kappa_c & \sqrt{\frac{3}{2}} \kappa_c \\
					& -1 + \kappa_{cc} & 0 \\
					& & 1 + \kappa_{cc} \\
				\end{block}
			\end{blockarray}
			$
		\end{tabular}
		\label{tab:C_S0I1}
	\end{minipage}
	\hfill
	\begin{minipage}[t]{0.48\textwidth}
		\caption{$D^{S=0,I=\frac{1}{2}}$}
		\begin{tabular}{c}
			$\begin{blockarray}{ccc}
				\eta_c N & \bar{D} \Lambda_c & \bar{D}\Sigma_c\\
				\begin{block}{[ccc]}
					0 & \frac{1}{\sqrt{2}} \kappa_c & \frac{1}{\sqrt{2}} \kappa_c \\
					& -1 + \kappa_{cc} & 0 \\
					&  & 1 + \kappa_{cc} \\
				\end{block}
			\end{blockarray}$
		\end{tabular}
		\label{tab:D_S0I1}
	\end{minipage}
	\hfill
\end{table*}
\begin{table*}[ht!]
	\begin{minipage}[t]{0.48\textwidth}
		\caption{$C^{S=0,I=\frac{3}{2}}$}
		\begin{tabular}{c}
			$\begin{blockarray}{c}
				\bar{D} \Sigma_c\\
				\begin{block}{[c]}
					-2+k_{cc} \\
				\end{block}
			\end{blockarray}$
		\end{tabular}
		\label{tab:C_S0I2}
	\end{minipage}
	\hfill
	\begin{minipage}[t]{0.48\textwidth}
		\caption{$D^{S=0,I=\frac{3}{2}}$}
		\begin{tabular}{c}
			$\begin{blockarray}{c}
				\bar{D} \Sigma_c\\
				\begin{block}{[c]}
					-2+k_{cc} \\
				\end{block}
			\end{blockarray}$
		\end{tabular}
		\label{tab:D_S0I2}
	\end{minipage}
	\hfill
\end{table*}
\begin{table*}[ht!]
	\begin{minipage}[t]{0.48\textwidth}
		\caption{$C^{S=-1,I=0}$}
		\begin{tabular}{c}
			$\begin{blockarray}{cccc}
				\eta_c \Lambda & \bar{D}_s \Lambda_c & \bar{D} \Xi_c & \bar{D} \Xi_c' \\
				\begin{block}{[cccc]}
					0 & \kappa_c & -\frac{1}{\sqrt{2}} \kappa_c & -\sqrt{\frac{3}{2}} \kappa_c \\
					& \kappa_{cc} & \sqrt{2} & 0 \\
					& & 1 + \kappa_{cc} & 0 \\
					& & & 1 + \kappa_{cc} \\
				\end{block}
			\end{blockarray}$
		\end{tabular}
		\label{tab:C_S1I1}
	\end{minipage}
	\hfill
	\begin{minipage}[t]{0.48\textwidth}
		\caption{$D^{S=-1,I=0}$}
		\begin{tabular}{c}
			$\begin{blockarray}{cccc}
				\eta_c \Lambda & \bar{D}_s \Lambda_c & \bar{D} \Xi_c & \bar{D} \Xi_c' \\
				\begin{block}{[cccc]}
					0 & \frac{1}{\sqrt{3}}\kappa_c & -\frac{1}{\sqrt{6}} \kappa_c & \frac{1}{\sqrt{2}} \kappa_c \\
					& \kappa_{cc} & \sqrt{2} & 0 \\
					& & 1 + \kappa_{cc} & 0 \\
					& & & 1 + \kappa_{cc} \\
				\end{block}
			\end{blockarray}$
		\end{tabular}
		\label{tab:D_S1I1}
	\end{minipage}
	\hfill
\end{table*}
\begin{table*}[ht!]
	\begin{minipage}[t]{0.48\textwidth}
		\caption{$C^{S=-1,I=1}$}
		\begin{tabular}{c}
			$\begin{blockarray}{cccc}
				\eta_c \Sigma & \bar{D} \Xi_c & \bar{D}_s \Sigma_c & \bar{D} \Xi_c'\\
				\begin{block}{[cccc]}
					0 & \sqrt{\frac{3}{2}} \kappa_c & -\kappa_c & - \frac{1}{\sqrt{2}} \kappa_c\\
					& -1+\kappa_{cc} & 0 & 0 \\
					& & \kappa_{cc} & \sqrt{2} \\
					& & & -1+\kappa_{cc} \\
				\end{block}
			\end{blockarray}$
		\end{tabular}
		\label{tab:C_S1I2}
	\end{minipage}
	\hfill
	\begin{minipage}[t]{0.48\textwidth}
		\caption{$D^{S=-1,I=1}$}
		\begin{tabular}{c}
			$\begin{blockarray}{cccc}
				\eta_c \Sigma & \bar{D} \Xi_c & \bar{D}_s \Sigma_c & \bar{D} \Xi_c'\\
				\begin{block}{[cccc]}
					0 & \frac{1}{\sqrt{2}} \kappa_c & -\frac{1}{\sqrt{3}}\kappa_c &  \frac{1}{\sqrt{6}} \kappa_c\\
					& -1+\kappa_{cc} & 0 & 0 \\
					& & \kappa_{cc} & -\sqrt{2} \\
					& & & -1+\kappa_{cc} \\
				\end{block}
			\end{blockarray}$
		\end{tabular}
		\label{tab:D_S1I2}
	\end{minipage}
	\hfill
\end{table*}
\begin{table*}[ht!]
    \begin{minipage}[t]{0.48\textwidth}
    \caption{$C^{S=-2,I=\frac{1}{2}}$}
    \begin{tabular}{c}
$\begin{blockarray}{cccc}
 \eta_c \Xi & \bar{D}_s \Xi_c & \bar{D}_s \Xi_c' & \bar{D} \Omega_c \\
\begin{block}{[cccc]}
  0 & \sqrt{\frac{3}{2}} \kappa_c & \frac{1}{\sqrt{2}} \kappa_c & -\kappa_c \\
  & -1 + \kappa_{cc} & 0 & 0 \\
  & & -1 + \kappa_{cc} & -\sqrt{2} \\
  & & & \kappa_{cc} \\
\end{block}
\end{blockarray}$
    \end{tabular}
    \label{tab:C_S2I1}
    \end{minipage}
    \hfill
    \begin{minipage}[t]{0.48\textwidth}
    \caption{$D^{S=-2,I=\frac{1}{2}}$}
    \begin{tabular}{c}
$\begin{blockarray}{cccc}
 \eta_c \Xi & \bar{D}_s \Xi_c & \bar{D}_s \Xi_c' & \bar{D} \Omega_c \\
\begin{block}{[cccc]}
  0 & \frac{1}{\sqrt{2}} \kappa_c & -\frac{1}{\sqrt{6}} \kappa_c & \frac{1}{\sqrt{3}}\kappa_c \\
  & -1 + \kappa_{cc} & 0 & 0 \\
  & & -1 + \kappa_{cc} & -\sqrt{2} \\
  & & & \kappa_{cc} \\
\end{block}
\end{blockarray}$
    \end{tabular}
    \label{tab:D_S2I1}
    \end{minipage}
    \hfill
\end{table*}
\begin{table*}[ht!]
    \begin{minipage}[t]{0.48\textwidth}
    \caption{$C^{S=-3,I=0}$}
    \begin{tabular}{c}
$\begin{blockarray}{c}
 \bar{D}_s \Omega_c\\
\begin{block}{[c]}
-2+\kappa_{cc} \\
\end{block}
\end{blockarray}$
    \end{tabular}
    \label{tab:C_S3I1}
    \end{minipage}
    \hfill
    \begin{minipage}[t]{0.48\textwidth}
    \caption{$D^{S=-3,I=0}$}
    \begin{tabular}{c}
$\begin{blockarray}{c}
 \bar{D}_s \Omega_c\\
\begin{block}{[c]}
-2+\kappa_{cc} \\
\end{block}
\end{blockarray}$
    \end{tabular}
    \label{tab:D_S3I1}
    \end{minipage}
    \hfill
\end{table*}

\section{PB and VB kernel coefficients using the LHGA-SU(4) and LHGA-WF models} \label{app:B}

We present the kernel coefficients required in Eq.~\eqref{V_SU4_swave}. The coefficients computed within the LHGA-SU(4) model are listed in odd Tables~\ref{tab:C_S0I1},~\ref{tab:C_S0I2},~\ref{tab:C_S1I1},~\ref{tab:C_S1I2},~\ref{tab:C_S2I1} and~\ref{tab:C_S3I1} (denoted as $C$ matrices), while those obtained using the LHGA-WF method are shown in even Tables~\ref{tab:D_S0I1},~\ref{tab:D_S0I2},~\ref{tab:D_S1I1},~\ref{tab:D_S1I2},~\ref{tab:D_S2I1} and~\ref{tab:D_S3I1} (denoted as $D$ matrices). The coefficients were computed for different values of strangeness ($S$) and isospin ($I$), as indicated in the caption of each table. The meson–baryon channels involved in the PB interactions are listed above each matrix. The coefficients of the $C$ matrices were extracted from Ref.~\cite{Lutz}, whereas the those of the $D$ matrices were calculated following the procedure in Ref.~\cite{Debastiani}. 

Recall that the kernel coefficients for the VB case are related to those of the PB case through the following correspondences:
    $\eta_c \rightarrow J/\psi$, $\bar{D} \rightarrow \bar{D}^*$, $\bar{D}_s \rightarrow \bar{D}_s^* $.


\newpage

\section{PB* and VB* kernel coefficients using the LHGA-WF method} \label{app:C}
\renewcommand{\thetable}{B\arabic{table}}
\setcounter{table}{0}
\begin{table*}[ht!]
    \begin{minipage}[t]{0.48\textwidth}
    \caption{$D^{S=0,I=\frac{1}{2}}$}
    \begin{tabular}{c}
$\begin{blockarray}{c}
\bar{D} \Sigma_c^* \\
\begin{block}{[c]}
  1 + \kappa_{cc} \\
\end{block}
\end{blockarray}$
    \end{tabular}
    \label{tab:S0I1_Bast}
    \end{minipage}
    \hfill
    \begin{minipage}[t]{0.48\textwidth}
    \caption{$D^{S=0,I=\frac{3}{2}}$}
    \begin{tabular}{c}
$\begin{blockarray}{cc}
 \eta_c \Delta & \bar{D} \Sigma_c^*\\
\begin{block}{[cc]}
  0 & -\kappa_c \\
    & -2+\kappa_{cc} \\
\end{block}
\end{blockarray}$
    \end{tabular}
    \label{tab:S0I2_Bast}
    \end{minipage}
    \hfill
\end{table*}
\begin{table*}[ht!]
    \begin{minipage}[t]{0.48\textwidth}
    \caption{$D^{S=-1,I=0}$}
    \begin{tabular}{c}
$\begin{blockarray}{c}
 \bar{D} \Xi_c^* \\
\begin{block}{[c]}
  1+\kappa_{cc} \\
\end{block}
\end{blockarray}$
    \end{tabular}
    \label{tab:S1I1_Bast}
    \end{minipage}
    \hfill
    \begin{minipage}[t]{0.48\textwidth}
    \caption{$D^{S=-1,I=1}$}
    \begin{tabular}{c}
$\begin{blockarray}{ccc}
 \eta_c \Sigma^* & \bar{D} \Xi_c^* & \bar{D}_s \Sigma_c^*\\
\begin{block}{[ccc]}
  0 & -\sqrt{\frac{2}{3}} \kappa_c & - \frac{1}{\sqrt{3}} \kappa_c \\
  & -1 + \kappa_{cc} & -\sqrt{2} \\
  & & \kappa_{cc} \\
\end{block}
\end{blockarray}$
    \end{tabular}
    \label{tab:S1I2_Bast}
    \end{minipage}
    \hfill
\end{table*}
\begin{table*}[ht!]
    \begin{minipage}[t]{0.48\textwidth}
    \caption{$D^{S=-2,I=\frac{1}{2}}$}
    \begin{tabular}{c}
$\begin{blockarray}{ccc}
 \eta_c \Xi^* & \bar{D}_s \Xi_c^* & \bar{D} \Omega_c^*\\
\begin{block}{[ccc]}
  0 & -\sqrt{\frac{2}{3}} \kappa_c & - \frac{1}{\sqrt{3}} \kappa_c \\
  & -1 + \kappa_{cc} & -\sqrt{2} \\
  & & \kappa_{cc} \\
\end{block}
\end{blockarray}$
    \end{tabular}
    \label{tab:S2I1_Bast}
    \end{minipage}
    \hfill
    \begin{minipage}[t]{0.48\textwidth}
    \caption{$D^{S=-3,I=0}$}
    \begin{tabular}{c}
$\begin{blockarray}{cc}
 \eta_c \Omega & \bar{D}_s \Omega_c^*\\
\begin{block}{[cc]}
  0 & -\kappa_c \\
    & -2+\kappa_{cc} \\
\end{block}
\end{blockarray}$
    \end{tabular}
    \label{tab:S3I1_Bast}
    \end{minipage}
    \hfill
\end{table*}

We present the kernel coefficients required in Eq.~\eqref{V_SU4_swave} for the interactions involving $J^P=3/2^+$ baryons. These coefficients are computed using the LHGA-WF method (described in Ref.~\cite{Debastiani}) and are shown in Tables~\ref{tab:S0I1_Bast} to~\ref{tab:S3I1_Bast} (denoted as $D$ matrices). The coefficients were computed for different values of strangeness ($S$) and isospin ($I$), as indicated in the caption of each table. The channels involved in the PB$^*$ interactions are indicated above each matrix.

Again, the kernel coefficients for the VB$^*$ case are related to those for the PB$^*$ case through the following correspondences:
 $\eta_c \rightarrow J/\psi$, $\bar{D} \rightarrow \bar{D}^*$, $\bar{D}_s \rightarrow \bar{D}_s^* $.


\bibliography{thebibliography}

\begin{thebibliography}{55}%
\makeatletter
\providecommand \@ifxundefined [1]{%
 \@ifx{#1\undefined}
}%
\providecommand \@ifnum [1]{%
 \ifnum #1\expandafter \@firstoftwo
 \else \expandafter \@secondoftwo
 \fi
}%
\providecommand \@ifx [1]{%
 \ifx #1\expandafter \@firstoftwo
 \else \expandafter \@secondoftwo
 \fi
}%
\providecommand \natexlab [1]{#1}%
\providecommand \enquote  [1]{``#1''}%
\providecommand \bibnamefont  [1]{#1}%
\providecommand \bibfnamefont [1]{#1}%
\providecommand \citenamefont [1]{#1}%
\providecommand \href@noop [0]{\@secondoftwo}%
\providecommand \href [0]{\begingroup \@sanitize@url \@href}%
\providecommand \@href[1]{\@@startlink{#1}\@@href}%
\providecommand \@@href[1]{\endgroup#1\@@endlink}%
\providecommand \@sanitize@url [0]{\catcode `\\12\catcode `\$12\catcode
  `\&12\catcode `\#12\catcode `\^12\catcode `\_12\catcode `\%12\relax}%
\providecommand \@@startlink[1]{}%
\providecommand \@@endlink[0]{}%
\providecommand \url  [0]{\begingroup\@sanitize@url \@url }%
\providecommand \@url [1]{\endgroup\@href {#1}{\urlprefix }}%
\providecommand \urlprefix  [0]{URL }%
\providecommand \Eprint [0]{\href }%
\providecommand \doibase [0]{https://doi.org/}%
\providecommand \selectlanguage [0]{\@gobble}%
\providecommand \bibinfo  [0]{\@secondoftwo}%
\providecommand \bibfield  [0]{\@secondoftwo}%
\providecommand \translation [1]{[#1]}%
\providecommand \BibitemOpen [0]{}%
\providecommand \bibitemStop [0]{}%
\providecommand \bibitemNoStop [0]{.\EOS\space}%
\providecommand \EOS [0]{\spacefactor3000\relax}%
\providecommand \BibitemShut  [1]{\csname bibitem#1\endcsname}%
\let\auto@bib@innerbib\@empty
\bibitem [{\citenamefont {Dalitz}\ and\ \citenamefont
  {Tuan}(1959)}]{Dalitz:1959dn}%
  \BibitemOpen
  \bibfield  {author} {\bibinfo {author} {\bibfnamefont {R.~H.}\ \bibnamefont
  {Dalitz}}\ and\ \bibinfo {author} {\bibfnamefont {S.~F.}\ \bibnamefont
  {Tuan}},\ }\bibfield  {title} {\bibinfo {title} {{A possible resonant state
  in pion-hyperon scattering}},\ }\href
  {https://doi.org/10.1103/PhysRevLett.2.425} {\bibfield  {journal} {\bibinfo
  {journal} {Phys. Rev. Lett.}\ }\textbf {\bibinfo {volume} {2}},\ \bibinfo
  {pages} {425} (\bibinfo {year} {1959})}\BibitemShut {NoStop}%
\bibitem [{\citenamefont {Alston}\ \emph {et~al.}(1961)\citenamefont {Alston},
  \citenamefont {Alvarez}, \citenamefont {Eberhard}, \citenamefont {Good},
  \citenamefont {Graziano}, \citenamefont {Ticho},\ and\ \citenamefont
  {Wojcicki}}]{Alston:1961zzd}%
  \BibitemOpen
  \bibfield  {author} {\bibinfo {author} {\bibfnamefont {M.~H.}\ \bibnamefont
  {Alston}}, \bibinfo {author} {\bibfnamefont {L.~W.}\ \bibnamefont {Alvarez}},
  \bibinfo {author} {\bibfnamefont {P.}~\bibnamefont {Eberhard}}, \bibinfo
  {author} {\bibfnamefont {M.~L.}\ \bibnamefont {Good}}, \bibinfo {author}
  {\bibfnamefont {W.}~\bibnamefont {Graziano}}, \bibinfo {author}
  {\bibfnamefont {H.~K.}\ \bibnamefont {Ticho}},\ and\ \bibinfo {author}
  {\bibfnamefont {S.~G.}\ \bibnamefont {Wojcicki}},\ }\bibfield  {title}
  {\bibinfo {title} {{Study of Resonances of the Sigma-pi System}},\ }\href
  {https://doi.org/10.1103/PhysRevLett.6.698} {\bibfield  {journal} {\bibinfo
  {journal} {Phys. Rev. Lett.}\ }\textbf {\bibinfo {volume} {6}},\ \bibinfo
  {pages} {698} (\bibinfo {year} {1961})}\BibitemShut {NoStop}%
\bibitem [{\citenamefont {Isgur}\ and\ \citenamefont
  {Karl}(1978)}]{Isgur:1978xj}%
  \BibitemOpen
  \bibfield  {author} {\bibinfo {author} {\bibfnamefont {N.}~\bibnamefont
  {Isgur}}\ and\ \bibinfo {author} {\bibfnamefont {G.}~\bibnamefont {Karl}},\
  }\bibfield  {title} {\bibinfo {title} {{P Wave Baryons in the Quark Model}},\
  }\href {https://doi.org/10.1103/PhysRevD.18.4187} {\bibfield  {journal}
  {\bibinfo  {journal} {Phys. Rev. D}\ }\textbf {\bibinfo {volume} {18}},\
  \bibinfo {pages} {4187} (\bibinfo {year} {1978})}\BibitemShut {NoStop}%
\bibitem [{\citenamefont {Capstick}\ and\ \citenamefont
  {Isgur}(1986)}]{Capstick:1986ter}%
  \BibitemOpen
  \bibfield  {author} {\bibinfo {author} {\bibfnamefont {S.}~\bibnamefont
  {Capstick}}\ and\ \bibinfo {author} {\bibfnamefont {N.}~\bibnamefont
  {Isgur}},\ }\bibfield  {title} {\bibinfo {title} {{Baryons in a relativized
  quark model with chromodynamics}},\ }\href
  {https://doi.org/10.1103/physrevd.34.2809} {\bibfield  {journal} {\bibinfo
  {journal} {Phys. Rev. D}\ }\textbf {\bibinfo {volume} {34}},\ \bibinfo
  {pages} {2809} (\bibinfo {year} {1986})}\BibitemShut {NoStop}%
\bibitem [{\citenamefont {Dobson}\ and\ \citenamefont
  {Mcelhaney}(1972)}]{Dobson:1972jib}%
  \BibitemOpen
  \bibfield  {author} {\bibinfo {author} {\bibfnamefont {P.~N.}\ \bibnamefont
  {Dobson}}\ and\ \bibinfo {author} {\bibfnamefont {R.}~\bibnamefont
  {Mcelhaney}},\ }\bibfield  {title} {\bibinfo {title} {{Interpretation of the
  y*0(1405) resonance}},\ }\href {https://doi.org/10.1103/PhysRevD.6.3256}
  {\bibfield  {journal} {\bibinfo  {journal} {Phys. Rev. D}\ }\textbf {\bibinfo
  {volume} {6}},\ \bibinfo {pages} {3256} (\bibinfo {year} {1972})}\BibitemShut
  {NoStop}%
\bibitem [{\citenamefont {Siegel}\ and\ \citenamefont
  {Weise}(1988)}]{Siegel:1988rq}%
  \BibitemOpen
  \bibfield  {author} {\bibinfo {author} {\bibfnamefont {P.~B.}\ \bibnamefont
  {Siegel}}\ and\ \bibinfo {author} {\bibfnamefont {W.}~\bibnamefont {Weise}},\
  }\bibfield  {title} {\bibinfo {title} {{Low-energy $K^-$ Nucleon Potentials
  and the Nature of the $\Lambda$ (1405)}},\ }\href
  {https://doi.org/10.1103/PhysRevC.38.2221} {\bibfield  {journal} {\bibinfo
  {journal} {Phys. Rev. C}\ }\textbf {\bibinfo {volume} {38}},\ \bibinfo
  {pages} {2221} (\bibinfo {year} {1988})}\BibitemShut {NoStop}%
\bibitem [{\citenamefont {Mueller-Groeling}\ \emph {et~al.}(1990)\citenamefont
  {Mueller-Groeling}, \citenamefont {Holinde},\ and\ \citenamefont
  {Speth}}]{Mueller-Groeling:1990uxr}%
  \BibitemOpen
  \bibfield  {author} {\bibinfo {author} {\bibfnamefont {A.}~\bibnamefont
  {Mueller-Groeling}}, \bibinfo {author} {\bibfnamefont {K.}~\bibnamefont
  {Holinde}},\ and\ \bibinfo {author} {\bibfnamefont {J.}~\bibnamefont
  {Speth}},\ }\bibfield  {title} {\bibinfo {title} {{K- N interaction in the
  meson exchange framework}},\ }\href
  {https://doi.org/10.1016/0375-9474(90)90398-6} {\bibfield  {journal}
  {\bibinfo  {journal} {Nucl. Phys. A}\ }\textbf {\bibinfo {volume} {513}},\
  \bibinfo {pages} {557} (\bibinfo {year} {1990})}\BibitemShut {NoStop}%
\bibitem [{\citenamefont {Glozman}\ and\ \citenamefont
  {Riska}(1996)}]{Glozman:1995fu}%
  \BibitemOpen
  \bibfield  {author} {\bibinfo {author} {\bibfnamefont {L.~Y.}\ \bibnamefont
  {Glozman}}\ and\ \bibinfo {author} {\bibfnamefont {D.~O.}\ \bibnamefont
  {Riska}},\ }\bibfield  {title} {\bibinfo {title} {{The Spectrum of the
  nucleons and the strange hyperons and chiral dynamics}},\ }\href
  {https://doi.org/10.1016/0370-1573(95)00062-3} {\bibfield  {journal}
  {\bibinfo  {journal} {Phys. Rept.}\ }\textbf {\bibinfo {volume} {268}},\
  \bibinfo {pages} {263} (\bibinfo {year} {1996})},\ \Eprint
  {https://arxiv.org/abs/hep-ph/9505422} {arXiv:hep-ph/9505422} \BibitemShut
  {NoStop}%
\bibitem [{\citenamefont {Nakamura}\ \emph {et~al.}(2010)\citenamefont
  {Nakamura} \emph {et~al.}}]{PDG2010}%
  \BibitemOpen
  \bibfield  {author} {\bibinfo {author} {\bibfnamefont {K.}~\bibnamefont
  {Nakamura}} \emph {et~al.} (\bibinfo {collaboration} {Particle Data Group}),\
  }\bibfield  {title} {\bibinfo {title} {{Review of particle physics}},\ }\href
  {https://doi.org/10.1088/0954-3899/37/7A/075021} {\bibfield  {journal}
  {\bibinfo  {journal} {J. Phys. G}\ }\textbf {\bibinfo {volume} {37}},\
  \bibinfo {pages} {075021} (\bibinfo {year} {2010})}\BibitemShut {NoStop}%
\bibitem [{\citenamefont {Gasser}\ and\ \citenamefont
  {Leutwyler}(1984)}]{Gasser:1983yg}%
  \BibitemOpen
  \bibfield  {author} {\bibinfo {author} {\bibfnamefont {J.}~\bibnamefont
  {Gasser}}\ and\ \bibinfo {author} {\bibfnamefont {H.}~\bibnamefont
  {Leutwyler}},\ }\bibfield  {title} {\bibinfo {title} {{Chiral Perturbation
  Theory to One Loop}},\ }\href {https://doi.org/10.1016/0003-4916(84)90242-2}
  {\bibfield  {journal} {\bibinfo  {journal} {Annals Phys.}\ }\textbf {\bibinfo
  {volume} {158}},\ \bibinfo {pages} {142} (\bibinfo {year}
  {1984})}\BibitemShut {NoStop}%
\bibitem [{\citenamefont {Weinberg}(1991)}]{Weinberg:1991um}%
  \BibitemOpen
  \bibfield  {author} {\bibinfo {author} {\bibfnamefont {S.}~\bibnamefont
  {Weinberg}},\ }\bibfield  {title} {\bibinfo {title} {{Effective chiral
  Lagrangians for nucleon - pion interactions and nuclear forces}},\ }\href
  {https://doi.org/10.1016/0550-3213(91)90231-L} {\bibfield  {journal}
  {\bibinfo  {journal} {Nucl. Phys. B}\ }\textbf {\bibinfo {volume} {363}},\
  \bibinfo {pages} {3} (\bibinfo {year} {1991})}\BibitemShut {NoStop}%
\bibitem [{\citenamefont {Kaiser}\ \emph {et~al.}(1995)\citenamefont {Kaiser},
  \citenamefont {Siegel},\ and\ \citenamefont {Weise}}]{Kaiser:1995eg}%
  \BibitemOpen
  \bibfield  {author} {\bibinfo {author} {\bibfnamefont {N.}~\bibnamefont
  {Kaiser}}, \bibinfo {author} {\bibfnamefont {P.~B.}\ \bibnamefont {Siegel}},\
  and\ \bibinfo {author} {\bibfnamefont {W.}~\bibnamefont {Weise}},\ }\bibfield
   {title} {\bibinfo {title} {{Chiral dynamics and the low-energy kaon -
  nucleon interaction}},\ }\href {https://doi.org/10.1016/0375-9474(95)00362-5}
  {\bibfield  {journal} {\bibinfo  {journal} {Nucl. Phys. A}\ }\textbf
  {\bibinfo {volume} {594}},\ \bibinfo {pages} {325} (\bibinfo {year}
  {1995})},\ \Eprint {https://arxiv.org/abs/nucl-th/9505043}
  {arXiv:nucl-th/9505043} \BibitemShut {NoStop}%
\bibitem [{\citenamefont {Oset}\ and\ \citenamefont
  {Ramos}(1998)}]{Oset:1997it}%
  \BibitemOpen
  \bibfield  {author} {\bibinfo {author} {\bibfnamefont {E.}~\bibnamefont
  {Oset}}\ and\ \bibinfo {author} {\bibfnamefont {A.}~\bibnamefont {Ramos}},\
  }\bibfield  {title} {\bibinfo {title} {{Nonperturbative chiral approach to s
  wave anti-K N interactions}},\ }\href
  {https://doi.org/10.1016/S0375-9474(98)00170-5} {\bibfield  {journal}
  {\bibinfo  {journal} {Nucl. Phys. A}\ }\textbf {\bibinfo {volume} {635}},\
  \bibinfo {pages} {99} (\bibinfo {year} {1998})},\ \Eprint
  {https://arxiv.org/abs/nucl-th/9711022} {arXiv:nucl-th/9711022} \BibitemShut
  {NoStop}%
\bibitem [{\citenamefont {Oller}\ \emph {et~al.}(2000)\citenamefont {Oller},
  \citenamefont {Oset},\ and\ \citenamefont {Ramos}}]{Oller:2000ma}%
  \BibitemOpen
  \bibfield  {author} {\bibinfo {author} {\bibfnamefont {J.~A.}\ \bibnamefont
  {Oller}}, \bibinfo {author} {\bibfnamefont {E.}~\bibnamefont {Oset}},\ and\
  \bibinfo {author} {\bibfnamefont {A.}~\bibnamefont {Ramos}},\ }\bibfield
  {title} {\bibinfo {title} {{Chiral unitary approach to meson meson and meson
  - baryon interactions and nuclear applications}},\ }\href
  {https://doi.org/10.1016/S0146-6410(00)00104-6} {\bibfield  {journal}
  {\bibinfo  {journal} {Prog. Part. Nucl. Phys.}\ }\textbf {\bibinfo {volume}
  {45}},\ \bibinfo {pages} {157} (\bibinfo {year} {2000})},\ \Eprint
  {https://arxiv.org/abs/hep-ph/0002193} {arXiv:hep-ph/0002193} \BibitemShut
  {NoStop}%
\bibitem [{\citenamefont {Hyodo}\ and\ \citenamefont
  {Jido}(2012)}]{Hyodo:2011ur}%
  \BibitemOpen
  \bibfield  {author} {\bibinfo {author} {\bibfnamefont {T.}~\bibnamefont
  {Hyodo}}\ and\ \bibinfo {author} {\bibfnamefont {D.}~\bibnamefont {Jido}},\
  }\bibfield  {title} {\bibinfo {title} {{The nature of the Lambda(1405)
  resonance in chiral dynamics}},\ }\href
  {https://doi.org/10.1016/j.ppnp.2011.07.002} {\bibfield  {journal} {\bibinfo
  {journal} {Prog. Part. Nucl. Phys.}\ }\textbf {\bibinfo {volume} {67}},\
  \bibinfo {pages} {55} (\bibinfo {year} {2012})},\ \Eprint
  {https://arxiv.org/abs/1104.4474} {arXiv:1104.4474 [nucl-th]} \BibitemShut
  {NoStop}%
\bibitem [{\citenamefont {Feijoo}\ \emph {et~al.}(2019)\citenamefont {Feijoo},
  \citenamefont {Magas},\ and\ \citenamefont {Ramos}}]{Feijoo19}%
  \BibitemOpen
  \bibfield  {author} {\bibinfo {author} {\bibfnamefont {A.}~\bibnamefont
  {Feijoo}}, \bibinfo {author} {\bibfnamefont {V.}~\bibnamefont {Magas}},\ and\
  \bibinfo {author} {\bibfnamefont {A.}~\bibnamefont {Ramos}},\ }\bibfield
  {title} {\bibinfo {title} {{$S$=\ensuremath{-}1 meson-baryon interaction and
  the role of isospin filtering processes}},\ }\href
  {https://doi.org/10.1103/PhysRevC.99.035211} {\bibfield  {journal} {\bibinfo
  {journal} {Phys. Rev. C}\ }\textbf {\bibinfo {volume} {99}},\ \bibinfo
  {pages} {035211} (\bibinfo {year} {2019})},\ \Eprint
  {https://arxiv.org/abs/1810.07600} {arXiv:1810.07600 [hep-ph]} \BibitemShut
  {NoStop}%
\bibitem [{\citenamefont {Oller}\ and\ \citenamefont
  {Meissner}(2001)}]{Oller:2000fj}%
  \BibitemOpen
  \bibfield  {author} {\bibinfo {author} {\bibfnamefont {J.~A.}\ \bibnamefont
  {Oller}}\ and\ \bibinfo {author} {\bibfnamefont {U.~G.}\ \bibnamefont
  {Meissner}},\ }\bibfield  {title} {\bibinfo {title} {{Chiral dynamics in the
  presence of bound states: Kaon nucleon interactions revisited}},\ }\href
  {https://doi.org/10.1016/S0370-2693(01)00078-8} {\bibfield  {journal}
  {\bibinfo  {journal} {Phys. Lett. B}\ }\textbf {\bibinfo {volume} {500}},\
  \bibinfo {pages} {263} (\bibinfo {year} {2001})},\ \Eprint
  {https://arxiv.org/abs/hep-ph/0011146} {arXiv:hep-ph/0011146} \BibitemShut
  {NoStop}%
\bibitem [{\citenamefont {Jido}\ \emph {et~al.}(2003)\citenamefont {Jido},
  \citenamefont {Oller}, \citenamefont {Oset}, \citenamefont {Ramos},\ and\
  \citenamefont {Meissner}}]{Jido:2003cb}%
  \BibitemOpen
  \bibfield  {author} {\bibinfo {author} {\bibfnamefont {D.}~\bibnamefont
  {Jido}}, \bibinfo {author} {\bibfnamefont {J.~A.}\ \bibnamefont {Oller}},
  \bibinfo {author} {\bibfnamefont {E.}~\bibnamefont {Oset}}, \bibinfo {author}
  {\bibfnamefont {A.}~\bibnamefont {Ramos}},\ and\ \bibinfo {author}
  {\bibfnamefont {U.~G.}\ \bibnamefont {Meissner}},\ }\bibfield  {title}
  {\bibinfo {title} {{Chiral dynamics of the two Lambda(1405) states}},\ }\href
  {https://doi.org/10.1016/S0375-9474(03)01598-7} {\bibfield  {journal}
  {\bibinfo  {journal} {Nucl. Phys. A}\ }\textbf {\bibinfo {volume} {725}},\
  \bibinfo {pages} {181} (\bibinfo {year} {2003})},\ \Eprint
  {https://arxiv.org/abs/nucl-th/0303062} {arXiv:nucl-th/0303062} \BibitemShut
  {NoStop}%
\bibitem [{\citenamefont {Magas}\ \emph {et~al.}(2005)\citenamefont {Magas},
  \citenamefont {Oset},\ and\ \citenamefont {Ramos}}]{Magas:2005vu}%
  \BibitemOpen
  \bibfield  {author} {\bibinfo {author} {\bibfnamefont {V.~K.}\ \bibnamefont
  {Magas}}, \bibinfo {author} {\bibfnamefont {E.}~\bibnamefont {Oset}},\ and\
  \bibinfo {author} {\bibfnamefont {A.}~\bibnamefont {Ramos}},\ }\bibfield
  {title} {\bibinfo {title} {{Evidence for the two pole structure of the
  Lambda(1405) resonance}},\ }\href
  {https://doi.org/10.1103/PhysRevLett.95.052301} {\bibfield  {journal}
  {\bibinfo  {journal} {Phys. Rev. Lett.}\ }\textbf {\bibinfo {volume} {95}},\
  \bibinfo {pages} {052301} (\bibinfo {year} {2005})},\ \Eprint
  {https://arxiv.org/abs/hep-ph/0503043} {arXiv:hep-ph/0503043} \BibitemShut
  {NoStop}%
\bibitem [{\citenamefont {Patrignani}\ \emph {et~al.}(2016)\citenamefont
  {Patrignani} \emph {et~al.}}]{PDG2016}%
  \BibitemOpen
  \bibfield  {author} {\bibinfo {author} {\bibfnamefont {C.}~\bibnamefont
  {Patrignani}} \emph {et~al.} (\bibinfo {collaboration} {Particle Data
  Group}),\ }\bibfield  {title} {\bibinfo {title} {{Review of Particle
  Physics}},\ }\href {https://doi.org/10.1088/1674-1137/40/10/100001}
  {\bibfield  {journal} {\bibinfo  {journal} {Chin. Phys. C}\ }\textbf
  {\bibinfo {volume} {40}},\ \bibinfo {pages} {100001} (\bibinfo {year}
  {2016})}\BibitemShut {NoStop}%
\bibitem [{\citenamefont {Mai}(2021)}]{Mai:2020ltx}%
  \BibitemOpen
  \bibfield  {author} {\bibinfo {author} {\bibfnamefont {M.}~\bibnamefont
  {Mai}},\ }\bibfield  {title} {\bibinfo {title} {{Review of the ${\Lambda
  }$(1405) A curious case of a strangeness resonance}},\ }\href
  {https://doi.org/10.1140/epjs/s11734-021-00144-7} {\bibfield  {journal}
  {\bibinfo  {journal} {Eur. Phys. J. ST}\ }\textbf {\bibinfo {volume} {230}},\
  \bibinfo {pages} {1593} (\bibinfo {year} {2021})},\ \Eprint
  {https://arxiv.org/abs/2010.00056} {arXiv:2010.00056 [nucl-th]} \BibitemShut
  {NoStop}%
\bibitem [{\citenamefont {Choi}\ \emph {et~al.}(2003)\citenamefont {Choi} \emph
  {et~al.}}]{Belle:2003}%
  \BibitemOpen
  \bibfield  {author} {\bibinfo {author} {\bibfnamefont {S.~K.}\ \bibnamefont
  {Choi}} \emph {et~al.} (\bibinfo {collaboration} {Belle}),\ }\bibfield
  {title} {\bibinfo {title} {{Observation of a narrow charmonium-like state in
  exclusive $B^\pm \to K^\pm \pi^+ \pi^- J/\psi$ decays}},\ }\href
  {https://doi.org/10.1103/PhysRevLett.91.262001} {\bibfield  {journal}
  {\bibinfo  {journal} {Phys. Rev. Lett.}\ }\textbf {\bibinfo {volume} {91}},\
  \bibinfo {pages} {262001} (\bibinfo {year} {2003})},\ \Eprint
  {https://arxiv.org/abs/hep-ex/0309032} {arXiv:hep-ex/0309032} \BibitemShut
  {NoStop}%
\bibitem [{\citenamefont {Aaij}\ \emph
  {et~al.}(2022{\natexlab{a}})\citenamefont {Aaij} \emph {et~al.}}]{LHCb:Tcc}%
  \BibitemOpen
  \bibfield  {author} {\bibinfo {author} {\bibfnamefont {R.}~\bibnamefont
  {Aaij}} \emph {et~al.} (\bibinfo {collaboration} {LHCb}),\ }\bibfield
  {title} {\bibinfo {title} {{Observation of an exotic narrow doubly charmed
  tetraquark}},\ }\href {https://doi.org/10.1038/s41567-022-01614-y} {\bibfield
   {journal} {\bibinfo  {journal} {Nature Phys.}\ }\textbf {\bibinfo {volume}
  {18}},\ \bibinfo {pages} {751} (\bibinfo {year} {2022}{\natexlab{a}})},\
  \Eprint {https://arxiv.org/abs/2109.01038} {arXiv:2109.01038 [hep-ex]}
  \BibitemShut {NoStop}%
\bibitem [{\citenamefont {Aaij}\ \emph
  {et~al.}(2022{\natexlab{b}})\citenamefont {Aaij} \emph
  {et~al.}}]{Aijj2022_Tcc}%
  \BibitemOpen
  \bibfield  {author} {\bibinfo {author} {\bibfnamefont {R.}~\bibnamefont
  {Aaij}} \emph {et~al.} (\bibinfo {collaboration} {LHCb}),\ }\bibfield
  {title} {\bibinfo {title} {{Study of the doubly charmed tetraquark
  $T_{cc}^{+}$}},\ }\href {https://doi.org/10.1038/s41467-022-30206-w}
  {\bibfield  {journal} {\bibinfo  {journal} {Nature Commun.}\ }\textbf
  {\bibinfo {volume} {13}},\ \bibinfo {pages} {3351} (\bibinfo {year}
  {2022}{\natexlab{b}})},\ \Eprint {https://arxiv.org/abs/2109.01056}
  {arXiv:2109.01056 [hep-ex]} \BibitemShut {NoStop}%
\bibitem [{\citenamefont {Du}\ \emph {et~al.}(2022)\citenamefont {Du},
  \citenamefont {Baru}, \citenamefont {Dong}, \citenamefont {Filin},
  \citenamefont {Guo}, \citenamefont {Hanhart}, \citenamefont {Nefediev},
  \citenamefont {Nieves},\ and\ \citenamefont {Wang}}]{Du:2021zzh}%
  \BibitemOpen
  \bibfield  {author} {\bibinfo {author} {\bibfnamefont {M.-L.}\ \bibnamefont
  {Du}}, \bibinfo {author} {\bibfnamefont {V.}~\bibnamefont {Baru}}, \bibinfo
  {author} {\bibfnamefont {X.-K.}\ \bibnamefont {Dong}}, \bibinfo {author}
  {\bibfnamefont {A.}~\bibnamefont {Filin}}, \bibinfo {author} {\bibfnamefont
  {F.-K.}\ \bibnamefont {Guo}}, \bibinfo {author} {\bibfnamefont
  {C.}~\bibnamefont {Hanhart}}, \bibinfo {author} {\bibfnamefont
  {A.}~\bibnamefont {Nefediev}}, \bibinfo {author} {\bibfnamefont
  {J.}~\bibnamefont {Nieves}},\ and\ \bibinfo {author} {\bibfnamefont
  {Q.}~\bibnamefont {Wang}},\ }\bibfield  {title} {\bibinfo {title}
  {{Coupled-channel approach to Tcc+ including three-body effects}},\ }\href
  {https://doi.org/10.1103/PhysRevD.105.014024} {\bibfield  {journal} {\bibinfo
   {journal} {Phys. Rev. D}\ }\textbf {\bibinfo {volume} {105}},\ \bibinfo
  {pages} {014024} (\bibinfo {year} {2022})},\ \Eprint
  {https://arxiv.org/abs/2110.13765} {arXiv:2110.13765 [hep-ph]} \BibitemShut
  {NoStop}%
\bibitem [{\citenamefont {Feijoo}\ \emph {et~al.}(2021)\citenamefont {Feijoo},
  \citenamefont {Liang},\ and\ \citenamefont {Oset}}]{Feijoo:2021ppq}%
  \BibitemOpen
  \bibfield  {author} {\bibinfo {author} {\bibfnamefont {A.}~\bibnamefont
  {Feijoo}}, \bibinfo {author} {\bibfnamefont {W.~H.}\ \bibnamefont {Liang}},\
  and\ \bibinfo {author} {\bibfnamefont {E.}~\bibnamefont {Oset}},\ }\bibfield
  {title} {\bibinfo {title} {{D0D0{\ensuremath{\pi}}+ mass distribution in the
  production of the Tcc exotic state}},\ }\href
  {https://doi.org/10.1103/PhysRevD.104.114015} {\bibfield  {journal} {\bibinfo
   {journal} {Phys. Rev. D}\ }\textbf {\bibinfo {volume} {104}},\ \bibinfo
  {pages} {114015} (\bibinfo {year} {2021})},\ \Eprint
  {https://arxiv.org/abs/2108.02730} {arXiv:2108.02730 [hep-ph]} \BibitemShut
  {NoStop}%
\bibitem [{\citenamefont {Meng}\ \emph {et~al.}(2021)\citenamefont {Meng},
  \citenamefont {Wang}, \citenamefont {Wang},\ and\ \citenamefont
  {Zhu}}]{Meng:2021jnw}%
  \BibitemOpen
  \bibfield  {author} {\bibinfo {author} {\bibfnamefont {L.}~\bibnamefont
  {Meng}}, \bibinfo {author} {\bibfnamefont {G.-J.}\ \bibnamefont {Wang}},
  \bibinfo {author} {\bibfnamefont {B.}~\bibnamefont {Wang}},\ and\ \bibinfo
  {author} {\bibfnamefont {S.-L.}\ \bibnamefont {Zhu}},\ }\bibfield  {title}
  {\bibinfo {title} {{Probing the long-range structure of the Tcc+ with the
  strong and electromagnetic decays}},\ }\href
  {https://doi.org/10.1103/PhysRevD.104.L051502} {\bibfield  {journal}
  {\bibinfo  {journal} {Phys. Rev. D}\ }\textbf {\bibinfo {volume} {104}},\
  \bibinfo {pages} {051502} (\bibinfo {year} {2021})},\ \Eprint
  {https://arxiv.org/abs/2107.14784} {arXiv:2107.14784 [hep-ph]} \BibitemShut
  {NoStop}%
\bibitem [{\citenamefont {Albaladejo}(2022)}]{Albaladejo:2021vln}%
  \BibitemOpen
  \bibfield  {author} {\bibinfo {author} {\bibfnamefont {M.}~\bibnamefont
  {Albaladejo}},\ }\bibfield  {title} {\bibinfo {title} {{Tcc+ coupled channel
  analysis and predictions}},\ }\href
  {https://doi.org/10.1016/j.physletb.2022.137052} {\bibfield  {journal}
  {\bibinfo  {journal} {Phys. Lett. B}\ }\textbf {\bibinfo {volume} {829}},\
  \bibinfo {pages} {137052} (\bibinfo {year} {2022})},\ \Eprint
  {https://arxiv.org/abs/2110.02944} {arXiv:2110.02944 [hep-ph]} \BibitemShut
  {NoStop}%
\bibitem [{\citenamefont {Aaij}\ \emph {et~al.}(2015)\citenamefont {Aaij} \emph
  {et~al.}}]{Aaij2015}%
  \BibitemOpen
  \bibfield  {author} {\bibinfo {author} {\bibfnamefont {R.}~\bibnamefont
  {Aaij}} \emph {et~al.} (\bibinfo {collaboration} {LHCb}),\ }\bibfield
  {title} {\bibinfo {title} {{Observation of $J/\psi p$ Resonances Consistent
  with Pentaquark States in $\Lambda_b^0 \to J/\psi K^- p$ Decays}},\ }\href
  {https://doi.org/10.1103/PhysRevLett.115.072001} {\bibfield  {journal}
  {\bibinfo  {journal} {Phys. Rev. Lett.}\ }\textbf {\bibinfo {volume} {115}},\
  \bibinfo {pages} {072001} (\bibinfo {year} {2015})},\ \Eprint
  {https://arxiv.org/abs/1507.03414} {arXiv:1507.03414 [hep-ex]} \BibitemShut
  {NoStop}%
\bibitem [{\citenamefont {Aaij}\ \emph {et~al.}(2019)\citenamefont {Aaij} \emph
  {et~al.}}]{Aaij2019}%
  \BibitemOpen
  \bibfield  {author} {\bibinfo {author} {\bibfnamefont {R.}~\bibnamefont
  {Aaij}} \emph {et~al.} (\bibinfo {collaboration} {LHCb}),\ }\bibfield
  {title} {\bibinfo {title} {{Observation of a narrow pentaquark state,
  $P_c(4312)^+$, and of two-peak structure of the $P_c(4450)^+$}},\ }\href
  {https://doi.org/10.1103/PhysRevLett.122.222001} {\bibfield  {journal}
  {\bibinfo  {journal} {Phys. Rev. Lett.}\ }\textbf {\bibinfo {volume} {122}},\
  \bibinfo {pages} {222001} (\bibinfo {year} {2019})},\ \Eprint
  {https://arxiv.org/abs/1904.03947} {arXiv:1904.03947 [hep-ex]} \BibitemShut
  {NoStop}%
\bibitem [{\citenamefont {Aaij}\ \emph {et~al.}(2021)\citenamefont {Aaij} \emph
  {et~al.}}]{Aaij2020}%
  \BibitemOpen
  \bibfield  {author} {\bibinfo {author} {\bibfnamefont {R.}~\bibnamefont
  {Aaij}} \emph {et~al.} (\bibinfo {collaboration} {LHCb}),\ }\bibfield
  {title} {\bibinfo {title} {{Evidence of a $J/\psi\Lambda$ structure and
  observation of excited $\Xi^-$ states in the $\Xi^-_b \to J/\psi\Lambda K^-$
  decay}},\ }\href {https://doi.org/10.1016/j.scib.2021.02.030} {\bibfield
  {journal} {\bibinfo  {journal} {Sci. Bull.}\ }\textbf {\bibinfo {volume}
  {66}},\ \bibinfo {pages} {1278} (\bibinfo {year} {2021})},\ \Eprint
  {https://arxiv.org/abs/2012.10380} {arXiv:2012.10380 [hep-ex]} \BibitemShut
  {NoStop}%
\bibitem [{\citenamefont {Aaij}\ \emph {et~al.}(2023)\citenamefont {Aaij} \emph
  {et~al.}}]{Aaij2022}%
  \BibitemOpen
  \bibfield  {author} {\bibinfo {author} {\bibfnamefont {R.}~\bibnamefont
  {Aaij}} \emph {et~al.} (\bibinfo {collaboration} {LHCb}),\ }\bibfield
  {title} {\bibinfo {title} {{Observation of a
  J/\ensuremath{\psi}\ensuremath{\Lambda} Resonance Consistent with a Strange
  Pentaquark Candidate in
  B-\textrightarrow{}J/\ensuremath{\psi}\ensuremath{\Lambda}p\textasciimacron{}
  Decays}},\ }\href {https://doi.org/10.1103/PhysRevLett.131.031901} {\bibfield
   {journal} {\bibinfo  {journal} {Phys. Rev. Lett.}\ }\textbf {\bibinfo
  {volume} {131}},\ \bibinfo {pages} {031901} (\bibinfo {year} {2023})},\
  \Eprint {https://arxiv.org/abs/2210.10346} {arXiv:2210.10346 [hep-ex]}
  \BibitemShut {NoStop}%
\bibitem [{\citenamefont {Adachi}\ \emph {et~al.}(2025)\citenamefont {Adachi}
  \emph {et~al.}}]{Belle:2025pey}%
  \BibitemOpen
  \bibfield  {author} {\bibinfo {author} {\bibfnamefont {I.}~\bibnamefont
  {Adachi}} \emph {et~al.} (\bibinfo {collaboration} {Belle, Belle II}),\
  }\bibfield  {title} {\bibinfo {title} {{Search for Pcs(4459) and Pcs(4338) in
  Upsilon(1S,2S) inclusive decays at Belle}},\ }\href@noop {} {\  (\bibinfo
  {year} {2025})},\ \Eprint {https://arxiv.org/abs/2502.09951}
  {arXiv:2502.09951 [hep-ex]} \BibitemShut {NoStop}%
\bibitem [{\citenamefont {Aaij}\ \emph
  {et~al.}(2022{\natexlab{c}})\citenamefont {Aaij} \emph {et~al.}}]{Aaij2021}%
  \BibitemOpen
  \bibfield  {author} {\bibinfo {author} {\bibfnamefont {R.}~\bibnamefont
  {Aaij}} \emph {et~al.} (\bibinfo {collaboration} {LHCb}),\ }\bibfield
  {title} {\bibinfo {title} {{Evidence for a new structure in the $J/\psi p$
  and $J/\psi \bar{p}$ systems in $B_s^0 \to J/\psi p \bar{p}$ decays}},\
  }\href {https://doi.org/10.1103/PhysRevLett.128.062001} {\bibfield  {journal}
  {\bibinfo  {journal} {Phys. Rev. Lett.}\ }\textbf {\bibinfo {volume} {128}},\
  \bibinfo {pages} {062001} (\bibinfo {year} {2022}{\natexlab{c}})},\ \Eprint
  {https://arxiv.org/abs/2108.04720} {arXiv:2108.04720 [hep-ex]} \BibitemShut
  {NoStop}%
\bibitem [{\citenamefont {Hofmann}\ and\ \citenamefont {Lutz}(2005)}]{Lutz}%
  \BibitemOpen
  \bibfield  {author} {\bibinfo {author} {\bibfnamefont {J.}~\bibnamefont
  {Hofmann}}\ and\ \bibinfo {author} {\bibfnamefont {M.~F.~M.}\ \bibnamefont
  {Lutz}},\ }\bibfield  {title} {\bibinfo {title} {{Coupled-channel study of
  crypto-exotic baryons with charm}},\ }\href
  {https://doi.org/10.1016/j.nuclphysa.2005.08.022} {\bibfield  {journal}
  {\bibinfo  {journal} {Nucl. Phys. A}\ }\textbf {\bibinfo {volume} {763}},\
  \bibinfo {pages} {90} (\bibinfo {year} {2005})},\ \Eprint
  {https://arxiv.org/abs/hep-ph/0507071} {arXiv:hep-ph/0507071} \BibitemShut
  {NoStop}%
\bibitem [{\citenamefont {Wu}\ \emph {et~al.}(2011)\citenamefont {Wu},
  \citenamefont {Molina}, \citenamefont {Oset},\ and\ \citenamefont
  {Zou}}]{Molina:2010}%
  \BibitemOpen
  \bibfield  {author} {\bibinfo {author} {\bibfnamefont {J.-J.}\ \bibnamefont
  {Wu}}, \bibinfo {author} {\bibfnamefont {R.}~\bibnamefont {Molina}}, \bibinfo
  {author} {\bibfnamefont {E.}~\bibnamefont {Oset}},\ and\ \bibinfo {author}
  {\bibfnamefont {B.~S.}\ \bibnamefont {Zou}},\ }\bibfield  {title} {\bibinfo
  {title} {{Dynamically generated $N^{*}$ and $\Lambda^*$ resonances in the
  hidden charm sector around 4.3 GeV}},\ }\href
  {https://doi.org/10.1103/PhysRevC.84.015202} {\bibfield  {journal} {\bibinfo
  {journal} {Phys. Rev. C}\ }\textbf {\bibinfo {volume} {84}},\ \bibinfo
  {pages} {015202} (\bibinfo {year} {2011})},\ \Eprint
  {https://arxiv.org/abs/1011.2399} {arXiv:1011.2399 [nucl-th]} \BibitemShut
  {NoStop}%
\bibitem [{\citenamefont {Monta\~na}\ \emph {et~al.}(2018)\citenamefont
  {Monta\~na}, \citenamefont {Feijoo},\ and\ \citenamefont {Ramos}}]{Gloria}%
  \BibitemOpen
  \bibfield  {author} {\bibinfo {author} {\bibfnamefont {G.}~\bibnamefont
  {Monta\~na}}, \bibinfo {author} {\bibfnamefont {A.}~\bibnamefont {Feijoo}},\
  and\ \bibinfo {author} {\bibfnamefont {A.}~\bibnamefont {Ramos}},\ }\bibfield
   {title} {\bibinfo {title} {{A meson-baryon molecular interpretation for some
  $\Omega_{c}$ excited states}},\ }\href
  {https://doi.org/10.1140/epja/i2018-12498-1} {\bibfield  {journal} {\bibinfo
  {journal} {Eur. Phys. J. A}\ }\textbf {\bibinfo {volume} {54}},\ \bibinfo
  {pages} {64} (\bibinfo {year} {2018})},\ \Eprint
  {https://arxiv.org/abs/1709.08737} {arXiv:1709.08737 [hep-ph]} \BibitemShut
  {NoStop}%
\bibitem [{\citenamefont {Debastiani}\ \emph {et~al.}(2018)\citenamefont
  {Debastiani}, \citenamefont {Dias}, \citenamefont {Liang},\ and\
  \citenamefont {Oset}}]{Debastiani}%
  \BibitemOpen
  \bibfield  {author} {\bibinfo {author} {\bibfnamefont {V.~R.}\ \bibnamefont
  {Debastiani}}, \bibinfo {author} {\bibfnamefont {J.~M.}\ \bibnamefont
  {Dias}}, \bibinfo {author} {\bibfnamefont {W.~H.}\ \bibnamefont {Liang}},\
  and\ \bibinfo {author} {\bibfnamefont {E.}~\bibnamefont {Oset}},\ }\bibfield
  {title} {\bibinfo {title} {{Molecular $\Omega_c$ states generated from
  coupled meson-baryon channels}},\ }\href
  {https://doi.org/10.1103/PhysRevD.97.094035} {\bibfield  {journal} {\bibinfo
  {journal} {Phys. Rev. D}\ }\textbf {\bibinfo {volume} {97}},\ \bibinfo
  {pages} {094035} (\bibinfo {year} {2018})},\ \Eprint
  {https://arxiv.org/abs/1710.04231} {arXiv:1710.04231 [hep-ph]} \BibitemShut
  {NoStop}%
\bibitem [{\citenamefont {Mars\'e-Valera}\ \emph {et~al.}(2023)\citenamefont
  {Mars\'e-Valera}, \citenamefont {Magas},\ and\ \citenamefont
  {Ramos}}]{Valera23}%
  \BibitemOpen
  \bibfield  {author} {\bibinfo {author} {\bibfnamefont {J.~A.}\ \bibnamefont
  {Mars\'e-Valera}}, \bibinfo {author} {\bibfnamefont {V.~K.}\ \bibnamefont
  {Magas}},\ and\ \bibinfo {author} {\bibfnamefont {A.}~\bibnamefont {Ramos}},\
  }\bibfield  {title} {\bibinfo {title} {{Double-Strangeness Molecular-Type
  Pentaquarks from Coupled-Channel Dynamics}},\ }\href
  {https://doi.org/10.1103/PhysRevLett.130.091903} {\bibfield  {journal}
  {\bibinfo  {journal} {Phys. Rev. Lett.}\ }\textbf {\bibinfo {volume} {130}},\
  \bibinfo {pages} {091903} (\bibinfo {year} {2023})},\ \Eprint
  {https://arxiv.org/abs/2210.02792} {arXiv:2210.02792 [hep-ph]} \BibitemShut
  {NoStop}%
\bibitem [{\citenamefont {Roca}\ \emph {et~al.}(2024)\citenamefont {Roca},
  \citenamefont {Song},\ and\ \citenamefont {Oset}}]{Roca}%
  \BibitemOpen
  \bibfield  {author} {\bibinfo {author} {\bibfnamefont {L.}~\bibnamefont
  {Roca}}, \bibinfo {author} {\bibfnamefont {J.}~\bibnamefont {Song}},\ and\
  \bibinfo {author} {\bibfnamefont {E.}~\bibnamefont {Oset}},\ }\bibfield
  {title} {\bibinfo {title} {{Molecular pentaquarks with hidden charm and
  double strangeness}},\ }\href {https://doi.org/10.1103/PhysRevD.109.094005}
  {\bibfield  {journal} {\bibinfo  {journal} {Phys. Rev. D}\ }\textbf {\bibinfo
  {volume} {109}},\ \bibinfo {pages} {094005} (\bibinfo {year} {2024})},\
  \Eprint {https://arxiv.org/abs/2403.08732} {arXiv:2403.08732 [hep-ph]}
  \BibitemShut {NoStop}%
\bibitem [{\citenamefont {Yu}\ \emph {et~al.}(2019)\citenamefont {Yu},
  \citenamefont {Pavao}, \citenamefont {Debastiani},\ and\ \citenamefont
  {Oset}}]{Yu:2018yxl}%
  \BibitemOpen
  \bibfield  {author} {\bibinfo {author} {\bibfnamefont {Q.~X.}\ \bibnamefont
  {Yu}}, \bibinfo {author} {\bibfnamefont {R.}~\bibnamefont {Pavao}}, \bibinfo
  {author} {\bibfnamefont {V.~R.}\ \bibnamefont {Debastiani}},\ and\ \bibinfo
  {author} {\bibfnamefont {E.}~\bibnamefont {Oset}},\ }\bibfield  {title}
  {\bibinfo {title} {{Description of the $\Xi _c$ and $\Xi _b$ states as
  molecular states}},\ }\href {https://doi.org/10.1140/epjc/s10052-019-6665-z}
  {\bibfield  {journal} {\bibinfo  {journal} {Eur. Phys. J. C}\ }\textbf
  {\bibinfo {volume} {79}},\ \bibinfo {pages} {167} (\bibinfo {year} {2019})},\
  \Eprint {https://arxiv.org/abs/1811.11738} {arXiv:1811.11738 [hep-ph]}
  \BibitemShut {NoStop}%
\bibitem [{\citenamefont {Bando}\ \emph {et~al.}(1985)\citenamefont {Bando},
  \citenamefont {Kugo}, \citenamefont {Uehara}, \citenamefont {Yamawaki},\ and\
  \citenamefont {Yanagida}}]{Bando}%
  \BibitemOpen
  \bibfield  {author} {\bibinfo {author} {\bibfnamefont {M.}~\bibnamefont
  {Bando}}, \bibinfo {author} {\bibfnamefont {T.}~\bibnamefont {Kugo}},
  \bibinfo {author} {\bibfnamefont {S.}~\bibnamefont {Uehara}}, \bibinfo
  {author} {\bibfnamefont {K.}~\bibnamefont {Yamawaki}},\ and\ \bibinfo
  {author} {\bibfnamefont {T.}~\bibnamefont {Yanagida}},\ }\bibfield  {title}
  {\bibinfo {title} {{Is rho Meson a Dynamical Gauge Boson of Hidden Local
  Symmetry?}},\ }\href {https://doi.org/10.1103/PhysRevLett.54.1215} {\bibfield
   {journal} {\bibinfo  {journal} {Phys. Rev. Lett.}\ }\textbf {\bibinfo
  {volume} {54}},\ \bibinfo {pages} {1215} (\bibinfo {year}
  {1985})}\BibitemShut {NoStop}%
\bibitem [{\citenamefont {Oset}\ and\ \citenamefont
  {Ramos}(2010)}]{OsetRamos10}%
  \BibitemOpen
  \bibfield  {author} {\bibinfo {author} {\bibfnamefont {E.}~\bibnamefont
  {Oset}}\ and\ \bibinfo {author} {\bibfnamefont {A.}~\bibnamefont {Ramos}},\
  }\bibfield  {title} {\bibinfo {title} {{Dynamically generated resonances from
  the vector octet-baryon octet interaction}},\ }\href
  {https://doi.org/10.1140/epja/i2010-10957-3} {\bibfield  {journal} {\bibinfo
  {journal} {Eur. Phys. J. A}\ }\textbf {\bibinfo {volume} {44}},\ \bibinfo
  {pages} {445} (\bibinfo {year} {2010})},\ \Eprint
  {https://arxiv.org/abs/0905.0973} {arXiv:0905.0973 [hep-ph]} \BibitemShut
  {NoStop}%
\bibitem [{\citenamefont {Oller}\ and\ \citenamefont
  {Oset}(1997)}]{Oller:1997ti}%
  \BibitemOpen
  \bibfield  {author} {\bibinfo {author} {\bibfnamefont {J.~A.}\ \bibnamefont
  {Oller}}\ and\ \bibinfo {author} {\bibfnamefont {E.}~\bibnamefont {Oset}},\
  }\bibfield  {title} {\bibinfo {title} {{Chiral symmetry amplitudes in the S
  wave isoscalar and isovector channels and the $\sigma$, f$_0$(980),
  a$_0$(980) scalar mesons}},\ }\href
  {https://doi.org/10.1016/S0375-9474(97)00160-7} {\bibfield  {journal}
  {\bibinfo  {journal} {Nucl. Phys. A}\ }\textbf {\bibinfo {volume} {620}},\
  \bibinfo {pages} {438} (\bibinfo {year} {1997})},\ \bibinfo {note} {[Erratum:
  Nucl.Phys.A 652, 407--409 (1999)]},\ \Eprint
  {https://arxiv.org/abs/hep-ph/9702314} {arXiv:hep-ph/9702314} \BibitemShut
  {NoStop}%
\bibitem [{\citenamefont {Roca}\ \emph {et~al.}(2005)\citenamefont {Roca},
  \citenamefont {Oset},\ and\ \citenamefont {Singh}}]{Roca2005}%
  \BibitemOpen
  \bibfield  {author} {\bibinfo {author} {\bibfnamefont {L.}~\bibnamefont
  {Roca}}, \bibinfo {author} {\bibfnamefont {E.}~\bibnamefont {Oset}},\ and\
  \bibinfo {author} {\bibfnamefont {J.}~\bibnamefont {Singh}},\ }\bibfield
  {title} {\bibinfo {title} {{Low lying axial-vector mesons as dynamically
  generated resonances}},\ }\href {https://doi.org/10.1103/PhysRevD.72.014002}
  {\bibfield  {journal} {\bibinfo  {journal} {Phys. Rev. D}\ }\textbf {\bibinfo
  {volume} {72}},\ \bibinfo {pages} {014002} (\bibinfo {year} {2005})},\
  \Eprint {https://arxiv.org/abs/hep-ph/0503273} {arXiv:hep-ph/0503273}
  \BibitemShut {NoStop}%
\bibitem [{\citenamefont {Oller}\ \emph {et~al.}(1999)\citenamefont {Oller},
  \citenamefont {Oset},\ and\ \citenamefont {Pelaez}}]{OllerOset99}%
  \BibitemOpen
  \bibfield  {author} {\bibinfo {author} {\bibfnamefont {J.~A.}\ \bibnamefont
  {Oller}}, \bibinfo {author} {\bibfnamefont {E.}~\bibnamefont {Oset}},\ and\
  \bibinfo {author} {\bibfnamefont {J.~R.}\ \bibnamefont {Pelaez}},\ }\bibfield
   {title} {\bibinfo {title} {{Meson meson interaction in a nonperturbative
  chiral approach}},\ }\href {https://doi.org/10.1103/PhysRevD.59.074001}
  {\bibfield  {journal} {\bibinfo  {journal} {Phys. Rev. D}\ }\textbf {\bibinfo
  {volume} {59}},\ \bibinfo {pages} {074001} (\bibinfo {year} {1999})},\
  \bibinfo {note} {[Erratum: Phys.Rev.D 60, 099906 (1999), Erratum: Phys.Rev.D
  75, 099903 (2007)]},\ \Eprint {https://arxiv.org/abs/hep-ph/9804209}
  {arXiv:hep-ph/9804209} \BibitemShut {NoStop}%
\bibitem [{\citenamefont {Wang}\ \emph {et~al.}(2022)\citenamefont {Wang},
  \citenamefont {Feijoo}, \citenamefont {Song},\ and\ \citenamefont
  {Oset}}]{Wang}%
  \BibitemOpen
  \bibfield  {author} {\bibinfo {author} {\bibfnamefont {W.-F.}\ \bibnamefont
  {Wang}}, \bibinfo {author} {\bibfnamefont {A.}~\bibnamefont {Feijoo}},
  \bibinfo {author} {\bibfnamefont {J.}~\bibnamefont {Song}},\ and\ \bibinfo
  {author} {\bibfnamefont {E.}~\bibnamefont {Oset}},\ }\bibfield  {title}
  {\bibinfo {title} {{Molecular \ensuremath{\Omega}cc, \ensuremath{\Omega}bb,
  and \ensuremath{\Omega}bc states}},\ }\href
  {https://doi.org/10.1103/PhysRevD.106.116004} {\bibfield  {journal} {\bibinfo
   {journal} {Phys. Rev. D}\ }\textbf {\bibinfo {volume} {106}},\ \bibinfo
  {pages} {116004} (\bibinfo {year} {2022})},\ \Eprint
  {https://arxiv.org/abs/2208.14858} {arXiv:2208.14858 [hep-ph]} \BibitemShut
  {NoStop}%
\bibitem [{\citenamefont {Xiao}\ \emph
  {et~al.}(2019{\natexlab{a}})\citenamefont {Xiao}, \citenamefont {Nieves},\
  and\ \citenamefont {Oset}}]{Xiao:2019}%
  \BibitemOpen
  \bibfield  {author} {\bibinfo {author} {\bibfnamefont {C.~W.}\ \bibnamefont
  {Xiao}}, \bibinfo {author} {\bibfnamefont {J.}~\bibnamefont {Nieves}},\ and\
  \bibinfo {author} {\bibfnamefont {E.}~\bibnamefont {Oset}},\ }\bibfield
  {title} {\bibinfo {title} {{Heavy quark spin symmetric molecular states from
  ${\bar D}^{(*)}\Sigma_c^{(*)}$ and other coupled channels in the light of the
  recent LHCb pentaquarks}},\ }\href
  {https://doi.org/10.1103/PhysRevD.100.014021} {\bibfield  {journal} {\bibinfo
   {journal} {Phys. Rev. D}\ }\textbf {\bibinfo {volume} {100}},\ \bibinfo
  {pages} {014021} (\bibinfo {year} {2019}{\natexlab{a}})},\ \Eprint
  {https://arxiv.org/abs/1904.01296} {arXiv:1904.01296 [hep-ph]} \BibitemShut
  {NoStop}%
\bibitem [{\citenamefont {Xiao}\ \emph
  {et~al.}(2019{\natexlab{b}})\citenamefont {Xiao}, \citenamefont {Nieves},\
  and\ \citenamefont {Oset}}]{Xiao:2019gjd}%
  \BibitemOpen
  \bibfield  {author} {\bibinfo {author} {\bibfnamefont {C.~W.}\ \bibnamefont
  {Xiao}}, \bibinfo {author} {\bibfnamefont {J.}~\bibnamefont {Nieves}},\ and\
  \bibinfo {author} {\bibfnamefont {E.}~\bibnamefont {Oset}},\ }\bibfield
  {title} {\bibinfo {title} {{Prediction of hidden charm strange molecular
  baryon states with heavy quark spin symmetry}},\ }\href
  {https://doi.org/10.1016/j.physletb.2019.135051} {\bibfield  {journal}
  {\bibinfo  {journal} {Phys. Lett. B}\ }\textbf {\bibinfo {volume} {799}},\
  \bibinfo {pages} {135051} (\bibinfo {year} {2019}{\natexlab{b}})},\ \Eprint
  {https://arxiv.org/abs/1906.09010} {arXiv:1906.09010 [hep-ph]} \BibitemShut
  {NoStop}%
\bibitem [{\citenamefont {Feijoo}\ \emph {et~al.}(2023)\citenamefont {Feijoo},
  \citenamefont {Wang}, \citenamefont {Xiao}, \citenamefont {Wu}, \citenamefont
  {Oset}, \citenamefont {Nieves},\ and\ \citenamefont {Zou}}]{Feijoo_Veff}%
  \BibitemOpen
  \bibfield  {author} {\bibinfo {author} {\bibfnamefont {A.}~\bibnamefont
  {Feijoo}}, \bibinfo {author} {\bibfnamefont {W.-F.}\ \bibnamefont {Wang}},
  \bibinfo {author} {\bibfnamefont {C.-W.}\ \bibnamefont {Xiao}}, \bibinfo
  {author} {\bibfnamefont {J.-J.}\ \bibnamefont {Wu}}, \bibinfo {author}
  {\bibfnamefont {E.}~\bibnamefont {Oset}}, \bibinfo {author} {\bibfnamefont
  {J.}~\bibnamefont {Nieves}},\ and\ \bibinfo {author} {\bibfnamefont {B.-S.}\
  \bibnamefont {Zou}},\ }\bibfield  {title} {\bibinfo {title} {{A new look at
  the Pcs states from a molecular perspective}},\ }\href
  {https://doi.org/10.1016/j.physletb.2023.137760} {\bibfield  {journal}
  {\bibinfo  {journal} {Phys. Lett. B}\ }\textbf {\bibinfo {volume} {839}},\
  \bibinfo {pages} {137760} (\bibinfo {year} {2023})},\ \Eprint
  {https://arxiv.org/abs/2212.12223} {arXiv:2212.12223 [hep-ph]} \BibitemShut
  {NoStop}%
\bibitem [{\citenamefont {Navas}\ \emph {et~al.}(2024)\citenamefont {Navas}
  \emph {et~al.}}]{PDG2024}%
  \BibitemOpen
  \bibfield  {author} {\bibinfo {author} {\bibfnamefont {S.}~\bibnamefont
  {Navas}} \emph {et~al.} (\bibinfo {collaboration} {Particle Data Group}),\
  }\bibfield  {title} {\bibinfo {title} {{Review of particle physics}},\ }\href
  {https://doi.org/10.1103/PhysRevD.110.030001} {\bibfield  {journal} {\bibinfo
   {journal} {Phys. Rev. D}\ }\textbf {\bibinfo {volume} {110}},\ \bibinfo
  {pages} {030001} (\bibinfo {year} {2024})}\BibitemShut {NoStop}%
\bibitem [{\citenamefont {H{\"u}sken}\ \emph {et~al.}(2025)\citenamefont
  {H{\"u}sken}, \citenamefont {Norella},\ and\ \citenamefont
  {Polyakov}}]{Husken:2024rdk}%
  \BibitemOpen
  \bibfield  {author} {\bibinfo {author} {\bibfnamefont {N.}~\bibnamefont
  {H{\"u}sken}}, \bibinfo {author} {\bibfnamefont {E.~S.}\ \bibnamefont
  {Norella}},\ and\ \bibinfo {author} {\bibfnamefont {I.}~\bibnamefont
  {Polyakov}},\ }\bibfield  {title} {\bibinfo {title} {{A brief guide to exotic
  hadrons}},\ }\href {https://doi.org/10.1142/S0217732325300022} {\bibfield
  {journal} {\bibinfo  {journal} {Mod. Phys. Lett. A}\ }\textbf {\bibinfo
  {volume} {40}},\ \bibinfo {pages} {2530002} (\bibinfo {year} {2025})},\
  \Eprint {https://arxiv.org/abs/2410.06923} {arXiv:2410.06923 [hep-ph]}
  \BibitemShut {NoStop}%
\bibitem [{\citenamefont {Merk}\ and\ \citenamefont
  {Tuning}(2026)}]{Merk:2026xye}%
  \BibitemOpen
  \bibfield  {author} {\bibinfo {author} {\bibfnamefont {M.~H.~M.}\
  \bibnamefont {Merk}}\ and\ \bibinfo {author} {\bibfnamefont {N.}~\bibnamefont
  {Tuning}},\ }\bibfield  {title} {\bibinfo {title} {{The LHCb Experiment}},\
  }\href@noop {} {\  (\bibinfo {year} {2026})},\ \Eprint
  {https://arxiv.org/abs/2605.03745} {arXiv:2605.03745 [hep-ex]} \BibitemShut
  {NoStop}%
\bibitem [{\citenamefont {Flatte}(1976)}]{Flatte:1976xu}%
  \BibitemOpen
  \bibfield  {author} {\bibinfo {author} {\bibfnamefont {S.~M.}\ \bibnamefont
  {Flatte}},\ }\bibfield  {title} {\bibinfo {title} {{Coupled - Channel
  Analysis of the pi eta and K anti-K Systems Near K anti-K Threshold}},\
  }\href {https://doi.org/10.1016/0370-2693(76)90654-7} {\bibfield  {journal}
  {\bibinfo  {journal} {Phys. Lett. B}\ }\textbf {\bibinfo {volume} {63}},\
  \bibinfo {pages} {224} (\bibinfo {year} {1976})}\BibitemShut {NoStop}%
\bibitem [{\citenamefont {Ramos}\ \emph {et~al.}(2002)\citenamefont {Ramos},
  \citenamefont {Oset},\ and\ \citenamefont {Bennhold}}]{Ramos:2002xh}%
  \BibitemOpen
  \bibfield  {author} {\bibinfo {author} {\bibfnamefont {A.}~\bibnamefont
  {Ramos}}, \bibinfo {author} {\bibfnamefont {E.}~\bibnamefont {Oset}},\ and\
  \bibinfo {author} {\bibfnamefont {C.}~\bibnamefont {Bennhold}},\ }\bibfield
  {title} {\bibinfo {title} {{On the spin, parity and nature of the Xi(1620)
  resonance}},\ }\href {https://doi.org/10.1103/PhysRevLett.89.252001}
  {\bibfield  {journal} {\bibinfo  {journal} {Phys. Rev. Lett.}\ }\textbf
  {\bibinfo {volume} {89}},\ \bibinfo {pages} {252001} (\bibinfo {year}
  {2002})},\ \Eprint {https://arxiv.org/abs/nucl-th/0204044}
  {arXiv:nucl-th/0204044} \BibitemShut {NoStop}%
\end{thebibliography}%

\end{document}